\documentclass[amssymb,nobibnotes,aps,prd,preprintnumbers]{revtex4}

 \usepackage{here}

\usepackage[dvipdfmx]{graphicx}
\usepackage{amssymb,amsfonts,amsmath,bm,color,multirow,cases,empheq,hyperref}
\usepackage{mathrsfs}
\definecolor{darkergreen}{rgb}{0,0.5,0}

\usepackage{enumerate}
\usepackage{ulem}
\usepackage{afterpage}

\hypersetup{%
 setpagesize=false,
 bookmarksnumbered=true,%
 bookmarksopen=true,%
 colorlinks=true,%
 linkcolor=blue,%
 citecolor=red}

\newcommand{\fr}[1]{\frac{1}{#1}}

\newcommand{\cD}{{\mathcal D}}

\newcommand{\nonum}{\nonumber\\ }

\newcommand{\cout}[1]{}

\newcommand{\arrayL}[1]{\left(\begin{array}{#1}}
\newcommand{\arrayR}{\end{array}\right)}
\newcommand{\arrayLb}[1]{\left[\begin{array}{#1}}
\newcommand{\arrayRb}{\end{array}\right]}

\begin{document}
\title{Solution Generation of a Capped Black Hole }

\author{Ryotaku Suzuki}
\email{sryotaku@toyota-ti.ac.jp}
\author{Shinya Tomizawa}
\email{tomizawa@toyota-ti.ac.jp}
\affiliation{\vspace{3mm}Mathematical Physics Laboratory, Toyota Technological Institute\vspace{2mm}\\Hisakata 2-12-1, Tempaku-ku, Nagoya, Japan 468-8511\vspace{3mm}}

\begin{abstract}

Utilizing the electric Harrison transformation developed in five-dimensional minimal supergravity, we construct an exact solution characterizing non-BPS charged rotating black holes with a horizon cross-section of a lens space $L(n;1)$.
Among these solutions, only the ones corresponding to $n=0$ and $n=1$ do not have any curvature singularities, conical singularities, Dirac-Misner string singularities, and orbifold singularities both on and outside the horizon; additionally, it is free from closed timelike curves.
The solution for $n=0$ corresponds to the charged dipole black ring that we constructed  in the previous paper.
The specific solution for $n=1$, referred to as the ``capped black hole," was introduced in our previous letter.
This provides the first example of a non-BPS exact solution, representing an asymptotically flat, stationary spherical black hole with a domain of outer communication (DOC) having a nontrivial topology in five-dimensional minimal supergravity. 
We demonstrate that the DOC on a timeslice has the topology of $[{\mathbb R}^4 \# {\mathbb CP}^2 ]\setminus {\mathbb B}^4$.
Differing from the well-known Myers-Perry and Cveti\v{c}-Youm black holes describing a spherical horizon topology  and a DOC with a trivial topology of  ${\mathbb R}^4 \setminus {\mathbb B}^4$ on a timeslice, the capped black hole's horizon is capped by a disc-shaped bubble.
We explicitly demonstrate that the capped black hole carries mass, two angular momenta, an electric charge, and a magnetic flux, with only three of these quantities being independent.
Furthermore, we reveal that this black hole can possess identical conserved charges as the Cveti\v{c}-Youm black hole.
The existence of this solution challenges black hole uniqueness beyond both the black ring and the BPS spherical black hole.
Moreover, within specific parameter regions, the capped black hole can exhibit a larger entropy than the Cveti\v{c}-Youm black hole.

\end{abstract}

\date{\today}
\preprint{TTI-MATHPHYS-27}

\maketitle

\section{Introduction}
In the domain of string theory and its associated disciplines, higher-dimensional black holes and other extended black objects have played a pivotal role in our comprehension of these higher-dimensional theories over the past two decades~\cite{Emparan:2008eg,Emparan:2006mm}. Of particular interest is the physics of black holes within the framework of five-dimensional minimal supergravity, which is recognized as a low-energy approximation of string theory. This theory shares similarities with eleven-dimensional supergravity, particularly in terms of its Lagrangians, where the three-form field in eleven-dimensional supergravity is replaced by Maxwell's $U(1)$ gauge field. The correspondence between five-dimensional minimal supergravity and eleven-dimensional supergravity has been previously investigated~\cite{Mizoguchi:1998wv,Mizoguchi:1999fu}. Furthermore, the formulation of five-dimensional supergravity can be derived through a truncated toroidal compactification of eleven-dimensional supergravity by identifying three vector fields and freezing out the moduli~\cite{Cremmer:1997ct,Cremmer:1998px}. This highlights the significance of discovering and classifying all exact solutions of black holes within the framework of five-dimensional minimal supergravity, as it contributes significantly to our understanding of string theory. Despite ongoing efforts, achieving this goal remains elusive, although various exact solutions of black holes within this theory have been generated through recent advancements in solution-generation techniques~\cite{Bouchareb:2007ax,Ford:2007th,Galtsov:2008pbf,Galtsov:2008bmt,Galtsov:2008jjb,Compere:2009zh,Figueras:2009mc,Mizoguchi:2011zj}.

\medskip
It is now well-established that even within vacuum Einstein gravity, there exists a diverse kind of  black hole solutions in higher dimensions~\cite{Myers:1986un,Emparan:2001wn}. 
However, the classification of asymptotically flat and stationary black holes remains a significant open problem. 
For instance, according to the topology theorem of a stationary black hole in five dimensions~\cite{Hollands:2007aj},
the allowed topology of the cross-section of the event horizon is restricted to either a sphere $S^3$, a ring $S^1\times S^2$, or lens spaces $L(n;m)$, given the spacetime is asymptotically flat and allows two commuting axial Killing vector fields.
Emparan and Reall~\cite{Emparan:2001wn} first showed that five-dimensional vacuum Einstein theory allows for the existence of a $S^1$-rotating spherical black hole and two rotating black rings with identical conserved charges, thus explicitly illustrating the non-uniqueness property in higher dimensions. 
The $S^2$-rotating black ring was initially derived independently by Mishima and Iguchi~\cite{Mishima:2005id} and Figueras~\cite{Figueras:2005zp}, although it exhibited conical singularities. Subsequently, Pomerasnky and Sen'kov~\cite{Pomeransky:2006bd} succeeded in constructing the general black ring solution with rotations in both $S^1$ and $S^2$.
Numerous efforts have been made by various authors to discover an asymptotically flat black lens solution to the five-dimensional vacuum Einstein equations. 
However, regrettably, all such endeavors have ended in failure~\cite{Evslin:2008gx,Chen:2008fa,Tomizawa:2019acu,Lucietti:2020phh}. 
The main obstacle lies in the fact that the resulting solutions obtained are always marred by naked singularities.
Several black objects have been extensively studied within the context of asymptotically flat supersymmetric solutions in five-dimensional minimal supergravity, leveraging techniques pioneered by Gauntlett et al.~\cite{Gauntlett:2002nw}. 
Reall demonstrated that the possible topologies of these supersymmetric black holes are limited to $S^3$, $S^1\times S^2$, $T^3$, or quotients thereof~\cite{Reall:2002bh}. 
For the $S^3$ case, Breckenridge et al.~\cite{Breckenridge:1996is} constructed a black hole solution with spherical topology featuring equal angular momenta, commonly referred to as the Breckenridge-Myers-Peet-Vafa (BMPV) black hole. 
Elvang et al. discovered a black ring solution in the $S^1\times S^2$ case~\cite{Elvang:2004rt}. 
The black ring exhibits only $U(1)\times U(1)$ spatial symmetry and does not allow for a configuration with equal angular momenta, distinguishing it from the BMPV black hole. 
Furthermore, Kunduri and Lucietti constructed an asymptotically flat supersymmetric black lens solution with topology $L(2;1)=S^3/{\mathbb Z}_2$~\cite{Kunduri:2014kja}, which was later extended to more general black lens solutions with topology $L(n;1)=S^3/{\mathbb Z}_n$ $(n\geq 3)$ in Ref.~\cite{Tomizawa:2016kjh,Breunholder:2017ubu}. 
 So far, exact solutions for bi-axisymmetric BPS black holes have been classified~\cite{Breunholder:2017ubu}, but ones for non-BPS black holes with a single $U(1)$ symmetry, or even $U(1)\times U(1)$, remain elusive.

\medskip

In recent years, many researchers have focused on horizon topologies when constructing new exact solutions of black holes. However, it has recently become evident that different types of black holes can exist even when the horizon topology is spherical.
According to the uniqueness theorem for charged rotating black holes in the bosonic sector of five-dimensional minimal supergravity~\cite{Tomizawa:2009ua}, assuming the existence of two commuting axial isometries and a spherical topology of horizon cross-sections, an asymptotically flat, stationary charged rotating black hole with a non-extremal horizon is uniquely characterized by its mass, charge, and two independent angular momenta, and is therefore described by the five-dimensional Cveti\v{c}-Youm  solution  ~\cite{Cvetic:1996xz}. Consequently, it appears that there are no other spherical black holes in the class of asymptotically flat, regular solutions with no closed timelike curves (CTCs).

However, the topological censorship theorem proved by Friedman~\cite{Friedman:1993ty} gives us the possibility of another black hole with spherical topology since in the uniqueness theorem~\cite{Tomizawa:2009ua},  the exterior region of a black hole is assumed to have the trivial topology of ${\mathbb R}^4\setminus {\mathbb B}^4$, where ${\mathbb B}^4$ represents the black hole region.
This theorem asserts that under the averaged null energy condition, the domain of outer communication (DOC) in an asymptotically flat spacetime must be simply connected. 
In four dimensions, this implies that the topology of the intersection of a black hole's exterior region with a timeslice $\Sigma$ is limited to a trivial structure of ${\mathbb R}^3\setminus {\mathbb B}^3$, where ${\mathbb B}^3$ represents the black hole region. 
However, in higher dimensions, the DOC can exhibit non-trivial topologies, meaning that ${\rm DOC}\cap \Sigma$ can possess homology groups with ranks higher than one. 
Based on the topological censorship theorem, it was shown in Ref.~\cite{Hollands:2010qy} that in five dimensions,  the region ${\rm DOC}\cap \Sigma$ can have the non-trivial topology of $[{\mathbb R}^4\#n(S^2\times S^2)\# m (\pm {\mathbb C}P^2)]\setminus {\mathbb B}^4 $.
In static asymptotically flat spacetimes, the uniqueness theorems~\cite{Gibbons:2002bh,Gibbons:2002av} establish that the higher-dimensional Schwarzschild solution~\cite{Tangherlini:1963bw}  and the higher-dimensional Reissner-Nordstr\"om solution are the only vacuum and charged black hole solutions, respectively. 
Consequently, any solutions with a non-trivial DOC---if they exist---must belong to a class of solutions that are not static rather stationary. 
Kunduri and Lucietti~\cite{Kunduri:2014iga} have constructed four parameter family of supersymmetric black hole solutions with spherical horizon topology and a 2-cycle in the exterior in  five-dimensional minimal supergravity, which indicates a charged spherical black hole such that ${\rm DOC}\cap \Sigma$ has the topology of $[{\mathbb R}^4 \# S^2\times S^2]\setminus {\mathbb B}^4$.
The presence of such a solution indicates the existence of black holes within this family that possess conserved charges identical to those of the BMPV black hole~\cite{Breckenridge:1996is}, highlighting the violation of uniqueness among black holes within a certain class of BPS spherical black holes.

\medskip
It is well-known that dimensionally reduced gravity theories and supergravity exhibit a global symmetry known as ``hidden symmetry," which often proves to be a powerful tool in discovering new solutions.
New solutions can be obtained by applying this group transformation to a known solution within the same theory, referred to as a "seed solution" (see Refs.~\cite{Ehlers,Harrison,exact} for four-dimensional Einstein gravity).
The dimensional reduction of five-dimensional minimal supergravity to four dimensions, as explored in Refs.~\cite{Chamseddine:1980mpx,Mizoguchi:1998wv}, reveals precisely an $SL(2, {\Bbb R})$ symmetry, arising from the dimensional reduction of eleven-dimensional supergravity~\cite{Cremmer:1979up}.
The new solution-generation technique utilizing this $SL(2,{\Bbb R})$ symmetry~\cite{Mizoguchi:2011zj} has successfully produced the Kaluza-Klein black hole solutions~\cite{Mizoguchi:2012vg,Tomizawa:2012nk}.
First explored by Mizoguchi and Ohta~\cite{Mizoguchi:1998wv,Mizoguchi:1999fu} in five-dimensional minimal supergravity, the presence of two commuting Killing vector fields reduces the theory to a three-dimensional non-linear sigma model with a $G_{2(2)}$ target space symmetry.
With two spacelike commuting Killing vector fields, it is described by the $G_{2(2)}/SO(4)$ sigma model coupled to gravity, while if one of the two commuting Killing vector fields is timelike, the symmetry becomes $G_{2(2)}/[SL(2,{\Bbb R})\times SL(2,{\Bbb R}))]$.
Utilizing this $G_{2(2)}$ symmetry, Bouchareb et al.~\cite{Bouchareb:2007ax} developed a solution-generation technique involving an electric Harrison transformation, capable of transforming a five-dimensional vacuum solution into an electrically charged solution in five-dimensional minimal supergravity.
By representing the coset in terms of a $7\times 7$ matrix, this transformation applied to the five-dimensional vacuum rotating black hole (the Myers-Perry solution~\cite{Myers:1986un}) yields the five-dimensional charged rotating black hole (the Cveti\v{c}-Youm solution~\cite{Cvetic:1996xz}).
However, applying this transformation to the vacuum doubly rotating black ring (the Pomeransky-Sen'kov solution~\cite{Pomeransky:2006bd}) fails to produce a regular charged doubly spinning black ring solution, as the resulting solution inevitably suffers from a Dirac-Misner string singularity. 
In Ref.~\cite{Tomizawa:2008qr}, this transformation is also applied to the Rasheed solution~\cite{Rasheed:1995zv} producing 
the rotating generalization of the static charged Kaluza-Klein black hole found by Ishihara and Matsuno~\cite{Ishihara:2005dp}.

\medskip

In our prior research~\cite{Suzuki:2024coe}, we employed the electric Harrison transformation to derive an exact solution for a non-BPS charged rotating black ring with a dipole charge within the bosonic sector of five-dimensional minimal supergravity.
To achieve this solution, we employed a vacuum solution of a rotating black ring that inherently contained a Dirac-Misner string singularity as the seed solution for the Harrison transformation. Subsequently, we adjusted the parameters appropriately to eliminate the Dirac-Misner string singularity inside the black ring.
To procure a vacuum seed solution having a Dirac-Misner string singularity, the inverse scattering method (ISM) proves invaluable. 
In Ref.~\cite{Suzuki:2024coe}, we successfully constructed such a vacuum solution, which serves as the foundational seed for the Harrison transformation. 
The ISM stands out as one of the most valuable tools for obtaining exact solutions of the vacuum Einstein equations with $D-2$ Killing isometries ($D:$ spacetime dimension).
This method enables the systematic derivation of new solutions with the same isometries through the soliton transformation from a known simple solution.
While the original ISM, as formulated by Belinski and Zakharov~\cite{Belinski:2001ph,Belinsky:1979mh}, typically yields singular solutions when applied directly to higher dimensions, Pomeransky modified the ISM to generate regular solutions even in higher dimensions~\cite{Pomeransky:2005sj}.
Notably, when combined with the rod structure~\cite{Emparan:2001wk,Harmark:2004rm}, this modified ISM has been highly successful, particularly in the context of five-dimensional vacuum black hole solutions~\cite{
Mishima:2005id,
Tomizawa:2005wv,Tomizawa:2006jz,Tomizawa:2006vp,
Iguchi:2006rd,
Elvang:2007rd,
Tomizawa:2007mz,Iguchi:2007xs,
Pomeransky:2006bd,
Iguchi:2007is,Evslin:2007fv,
Elvang:2007hs,Izumi:2007qx,
Chen:2011jb,
Chen:2008fa,
Chen:2012kd,
Rocha:2011vv,
Rocha:2012vs,
Chen:2015iex,
Chen:2012zb,
Lucietti:2020ltw,
Lucietti:2020phh,
Morisawa:2007di,Evslin:2008gx,
Feldman:2012vd,
Chen:2010ih,
Iguchi:2011qi,
Tomizawa:2019acu,
Tomizawa:2022qyd,
Suzuki:2023nqf,
Figueras:2009mc}.
The first example of this success was the re-derivation of the five-dimensional Myers-Perry black hole solution~\cite{Pomeransky:2005sj}.
Subsequently, the $S^2$-rotating black ring was re-derived from the Minkowski seed~\cite{Tomizawa:2005wv}, though the generation of the $S^1$ rotating black ring presented a more delicate problem due to the choice of seed leading to singular solutions.
The appropriate seed for deriving the black ring with $S^1$ rotation was first considered in~\cite{Iguchi:2006rd,Tomizawa:2006vp}, culminating in the construction of the regular black ring solution with both $S^1$ and $S^2$ rotations by Pomeransky and Sen'kov~\cite{Pomeransky:2006bd}.
In attempts to construct asymptotically flat black lens solutions in five-dimensional vacuum Einstein equations, several authors have employed the ISM.
For instance, Evslin~\cite{Evslin:2008gx} attempted to construct a static black lens with lens space topology $L(n^2+1;1)$, only to find that while orbifold singularities could be eliminated, curvature singularities remained unavoidable.
Similarly, Chen and Teo~\cite{Chen:2008fa} constructed a black lens solution with horizon topology $L(n;1)=S^3/{\mathbb Z}_n$ by the ISM, but encountered either conical singularities or naked curvature singularities.
The primary obstacle in constructing black lens solutions has thus been the presence of naked singularities.
However, breakthroughs in this regard have emerged from supersymmetric solutions.
Building upon the framework developed by Gauntlett {\it et al.} for supersymmetric solutions in the bosonic sector of five-dimensional minimal supergravity~\cite{Gauntlett:2002nw}, Kundhuri and Lucietti~\cite{Kunduri:2014kja} derived the first regular exact solution of an asymptotically flat black lens with horizon topology $L(2;1)=S^3/{\mathbb Z}_2$.
This solution was subsequently extended to more general supersymmetric black lens solutions with horizon topology $L(n;1)=S^3/{\mathbb Z}_n$ $(n\ge 3)$ in the same theory~\cite{Tomizawa:2016kjh}. 
Thus, non-BPS black lens solutions, including vacuum solutions, have not yet been discovered.

\medskip

In this paper, we derive an exact solution representing an asymptotically flat, stationary, non-BPS black hole characterized by a horizon cross-section with trivial topology $S^3$ and a DOC exhibiting non-trivial topology, within the bosonic sector of five-dimensional minimal supergravity. 
To begin, we employ the ISM to construct a vacuum black lens harboring a Dirac-Misner string singularity. 
Subsequently, employing the electric Harrison transformation on this vacuum solution, we derive a charged rotating black lens solution characterized by a horizon topology of lens space $L(n;1)$, still retaining the Dirac-Misner string singularity. 
Finally, we adjust the solution's parameters to eliminate the Dirac-Misner string singularity, ensuring its regularity.
Among these solutions, only those corresponding to $n=0$ and $n=1$ exhibit regularity, the absence of curvature, conical, Dirac-Misner string, or orbifold singularities both inside and outside the horizon, and additionally CTCs.
The $n=0$ solution corresponds to the charged dipole black ring previously constructed in our earlier work~\cite{Suzuki:2024coe}.
Specifically, the $n=1$ solution, termed the ``capped black hole," was introduced in our preceding work~\cite{Suzuki:2023nqf}.
This presents the first instance of a non-BPS exact solution, delineating an asymptotically flat, stationary spherical black hole with a nontrivially topological DOC within five-dimensional minimal supergravity.
In contrast to the familiar Cveti\v{c}-Youm solution with a spherical horizon topology, the capped black hole's horizon is capped by a disc-shaped bubble.
Additionally, we demonstrate the existence of spherical black holes possessing the same conserved charges as the Cveti\v{c}-Youm solution, which implies the violation of the uniqueness for a spherical black hole.

\medskip
The remainder of this paper is structured as follows: In Sec.~\ref{sec:setup}, we provide an overview of the setup and formalism employed in our analysis. Sec.~\ref{sec:neutral} is dedicated to the construction of the neutral metric using the soliton transformation. Following this, in Sec.~\ref{sec:charging}, we detail the application of the electric Harrison transformation to the neutral metric, resulting in the derivation of the charged metric and gauge field. Furthermore, we demonstrate that the only regular charged solution corresponds to a black ring and a black hole with a disc-like bubble. Subsequently, in Sec.~\ref{sec:phase}, we delve into an examination of the physical properties of the regular black hole solution. Finally, we encapsulate our findings and conclusions in Sec,~~\ref{sec:sum}.

\section{Preliminary}\label{sec:setup}

Let us start by explainning the fundamental framework for asymptotically flat, stationary, and bi-axisymmetric solutions within the bosonic sector of five-dimensional minimal ungauged supergravity (Einstein-Maxwell-Chern-Simons theory). The action governing this theory is given by:
\begin{eqnarray}
S=\frac{1}{16 \pi G_5}  \left[ 
        \int d^5x \sqrt{-g}\left(R-\frac{1}{4}F^2\right) 
       -\frac{1}{3\sqrt{3}} \int F\wedge F\wedge A 
  \right] \,, 
\label{action} 
\end{eqnarray} 
where $F=dA$. 
The field equations governing the dynamics of the system consist of the Einstein equation and the Maxwell equation with a Chern-Simons term. They are expressed as
\begin{eqnarray}
 R_{\mu \nu } -\frac{1}{2} R g_{\mu \nu } 
 = \frac{1}{2} \left( F_{\mu \lambda } F_\nu^{ ~ \lambda } 
  - \frac{1}{4} g_{\mu \nu } F_{\rho \sigma } F^{\rho \sigma } \right) \,, 
 \label{Eineq}
\end{eqnarray}
and 
\begin{eqnarray}
 d\star F+\frac{1}{\sqrt{3}}F\wedge F=0 \,. 
\label{Maxeq}
\end{eqnarray}

\subsection{Five-dimensional minimal supergravity with symmetry}

By assuming the presence of one timelike Killing vector $\xi_0 = \partial/\partial t$ and one spacelike axial Killing vector $\xi_1=\partial/\partial\psi$, the theory reduces to the $G_{2(2)}/SL(2,{\Bbb R})\times SL(2,{\Bbb R})$ non-linear sigma models coupled to three-dimensional gravity~\cite{Mizoguchi:1998wv,Mizoguchi:1999fu}.
Further, the assumption of the existence of a third spacelike axial Killing vector $\xi_2=\partial/\partial\phi$, implying the presence of three mutually commuting Killing vectors, reduces the theory to a two-dimensional non-linear sigma model, and additionally ensures the integrability conditions discussed in Ref.~\cite{Emparan:2001wk,Harmark:2004rm}, as a result,  the metric can be expressed in the Weyl-Papapetrou form:
\begin{eqnarray}
ds^2 &=& \lambda_{ab}(dx^a + a_\phi^a d\phi)(dx^b + a_\phi^b d\phi) + \tau^{-1}\rho^2 d\phi^2 + \tau^{-1}e^{2\sigma}(d\rho^2 + dz^2),
\label{eq:WPform}
\end{eqnarray}
and the gauge potential is given by:
\begin{eqnarray}
A = \sqrt{3}\psi_a dx^a + A_\phi d\phi,
\label{pote:gauge}
\end{eqnarray}
where the coordinates $x^a = (t,\psi)$ ($a=0,1$) represent the Killing coordinates, and thus all functions $\lambda_{ab}$, $\tau := -\mathrm{det}(\lambda_{ab})$, $a^a$, $\sigma$, and $(\psi_a, A_\phi)$ are independent of $\phi$ and $x^a$.
Additionally, as shown in Appendix of Ref.~\cite{Tomizawa:2009ua},  one can always set $A_{\rho}=A_{z}=0$, using the gauge transformation.
It is important to note that the coordinates $(\rho, z)$, spanning a two-dimensional base space $\Sigma = {(\rho, z) | \rho \geq 0, -\infty < z < \infty}$, are globally well-defined, harmonic, and mutually conjugate on $\Sigma$.

\medskip
The magnetic potential $\mu$ and twist potentials $\omega_a$ can be introduced using Eqs.~(\ref{Eineq}) and (\ref{Maxeq}), as discussed in Ref.~\cite{Tomizawa:2009ua}, expressed as
\begin{eqnarray}
d\mu &=& \frac{1}{\sqrt{3}} \star (\xi_0 \wedge \xi_1 \wedge F) - \epsilon^{ab} \psi_a d\psi_b ,
\label{eq:mu} \\
d\omega_a &=& \star (\xi_0 \wedge \xi_1 \wedge d\xi_a) + \psi_a(3d\mu+\epsilon^{bc} \psi_b d\psi_c) ,
\label{eq:twistpotential}
\end{eqnarray}
where $\epsilon^{01}=-\epsilon^{10}=1$, and $\xi_a$ ($a=0,1$) are written as Killing one-forms.
Thus, as a consequence of the existence of isometries $\xi_a$, 
we have eight scalar fields $\lambda_{ab},\omega_a,\psi_a,\mu$, which we denote collectively by coordinates 
$\Phi^A=(\lambda_{ab},\omega_a,\psi_a,\mu)$ $(a=0,1)$ and then, the action~(\ref{action}) reduces to the following nonlinear sigma model for the eight scalar functions $\Phi^A$  invariant under the $G_{2(2)}$-transformation:
\begin{eqnarray}
S&=&\int d\rho dz \rho\left[ G_{AB}(\partial\Phi^A)(\partial\Phi^B)\right] 
\nonumber \\
 &=& \int d\rho dz \rho 
     \biggl[\:  
             \frac{1}{4}{\rm Tr}(\lambda^{-1}\partial\lambda\lambda^{-1}
                                             \partial\lambda )
           + \frac{1}{4}\tau^{-2}\partial\tau^2 
           + \frac{3}{2}\partial\psi^T \lambda^{-1}\partial\psi 
\nonumber \\
& &{} \qquad \qquad 
           - \frac{1}{2}\tau^{-1}v^T\lambda^{-1}v 
           - \frac{3}{2}\tau^{-1}(\partial\mu 
                                   +\epsilon^{ab}\psi_a\partial\psi_b)^2 
     \biggr]  \,,
\label{eq:action} 
\end{eqnarray} 
where $v=\partial \omega-\psi(3\partial \mu+\epsilon^{bc}\psi_b\partial \psi_c)$.
In this coordinate system, $\Phi^A=(\lambda_{ab},\omega_a,\psi_a,\mu)$ are determined by the equations of motion 
\begin{eqnarray}
\Delta_\gamma \Phi^A + 
\Gamma^A_{BC}[\Phi^B_{,\rho}\Phi^C_{,\rho}+\Phi^C_{,z}\Phi^C_{,z}]=0 \,, 
\label{eq:scalar}
\end{eqnarray}
where $\Delta_\gamma$ is the Laplacian with respect to the abstract 
three-dimensional metric $\gamma=d\rho^2+dz^2+\rho^2d\varphi^2$, 
and $\Gamma_{BC}^A$ is the Christoffel symbol  with respect to the target space metric $G_{AB}$.

\medskip
On the other hand, once $\Phi^A$ are given, one can completely determine 
$\sigma$, $a^t{}_\phi$, $a^\psi{}_\phi$, $A_i$. 
In fact, the function $\sigma$ is determined by 
\begin{eqnarray}
\frac{2}{\rho}\sigma_{,\rho}
 =G_{AB}[\Phi^A_{,\rho}\Phi^B_{,\rho}-\Phi^A_{,z}\Phi^B_{,z}] \,,\quad
   \frac{1}{\rho}\sigma_{,z}
                            =G_{AB}\Phi^A_{,\rho}\Phi^B_{,z} \,.\label{eq:eq-sigma}
\end{eqnarray}
The integrability $\sigma_{,\rho z}=\sigma_{,z \rho }$ is assured by  
eq.~(\ref{eq:scalar}). From eq.(\ref{eq:twistpotential}), the metric functions $a^a{_\phi}$ are determined by 

\begin{align}
\begin{split}\label{eq:a0phi}
&a^a{}_{\phi,\rho}=-\rho \tau^{-1} \lambda^{ab}(\omega_{b,z}-3\psi_b\mu_{,z}-\psi_b\epsilon^{cd}\psi_c\psi_{d,z}) \\
&a^a{}_{\phi,z}=\rho\tau^{-1}\lambda^{ab}(\omega_{b,\rho}-3\psi_b\mu_{,\rho}-\psi_b\epsilon^{cd}\psi_c\psi_{d,\rho}), 
\end{split}
\end{align}
where we have set 
\begin{align}
\epsilon_{012\rho z}=1.\label{eq:orient}
\end{align}
Therefore it follows from eq.~(\ref{eq:mu}) that the gauge potential $A_\phi$ is determined by 
\begin{eqnarray}
A_{\phi,\rho}&=&\sqrt{3}\left[a^a{}_\phi\psi_{a,\rho}-\rho \tau^{-1}(\mu_{,z}+\epsilon^{bc}\psi_b\psi_{c,z})\right],\\
A_{\phi,z}&=&\sqrt{3}\left[a^a{}_\phi\psi_{a,z}+\rho\tau^{-1}(\mu_{,\rho}+\epsilon^{bc}\psi_b\psi_{c,\rho})\right].
\end{eqnarray}
Thus, once $\Phi^A= (\lambda_{ab},\omega_a,\psi_a,\mu)$ are determined, one can determine the solutions of the system given by the action~(\ref{action}).

\medskip 
Following Ref.~\cite{Bouchareb:2007ax}, we introduce the $G_{2(2)}/[SL(2,R)\times SL(2,R)]$ coset 
matrix, $M$, which is defined by 
\begin{eqnarray}
M= \left(
  \begin{array}{ccc}
  \hat A&\hat B&\sqrt{2}\hat U\\
  \hat B^T&\hat C&\sqrt{2}\hat V\\
  \sqrt{2}\hat U^T&\sqrt{2}\hat V^T&\hat S\\
  \end{array}
 \right) \,,
\end{eqnarray}
where $\hat A$ and $\hat C$ are symmetric $3\times 3$ matrices, $\hat B$ is a $3\times 3$ 
matrix, $\hat U$ and $\hat V$ are 3-component column matrices, and $\hat S$ is a scalar, 
defined, respectively, by  
\begin{eqnarray}
&&\hat A=\left(
  \begin{array}{ccc}
  [(1-y)\lambda+(2+x)\psi \psi^T-\tau^{-1}\tilde\omega\tilde\omega^T+\mu(\psi \psi^T\lambda^{-1}\hat J-\hat J\lambda^{-1}\psi\psi^T)]&\tau^{-1}\tilde\omega\\
 \tau^{-1}\tilde\omega^T& -\tau^{-1}
  \end{array}
 \right) \,,\nonumber\\
&&\hat B=\left(
  \begin{array}{ccc}
  (\psi\psi^T-\mu \hat J)\lambda^{-1}-\tau^{-1}\tilde\omega \psi^T \hat J&[(-(1+y)\lambda \hat J-(2+x)\mu+\psi^T\lambda^{-1}\tilde\omega)\psi+(z-\mu \hat J\lambda^{-1}\tilde)\omega] \\
  \tau^{-1}\psi^T \hat J&-z\\
  \end{array}
 \right) \,, \nonumber\\
&&\hat C=\left(
  \begin{array}{ccc}
 (1+x)\lambda^{-1}-\lambda^{-1}\psi\psi^T\lambda^{-1}&\lambda^{-1}\tilde\omega-\hat J(z-\mu \hat J \lambda^{-1})\psi\\
 \tilde\omega^T\lambda^{-1}+\psi^T(z+\mu \lambda^{-1}\hat J)\hat J&[\tilde\omega^T\lambda^{-1}\tilde\omega-2\mu\psi^T\lambda^{-1}\tilde\omega-\tau(1+x-2y-xy+z^2)]\\
  \end{array}
 \right) \,,\nonumber\\
&&\hat U=
\left(
  \begin{array}{cc}
  (1+x-\mu \hat J\lambda^{-1})\psi-\mu \tau^{-1}\tilde\omega\\
  \mu\tau^{-1}\\
  \end{array}
 \right) \,,\nonumber\\
&&\hat V=\left(
  \begin{array}{ccc}
  (\lambda^{-1}+\mu\tau^{-1}\hat J)\psi\\
  \psi^T\lambda^{-1}\tilde\omega-\mu(1+x-z)\nonumber\\
  \end{array}
 \right) \,,\nonumber\\
&&\hat S=1+2(x-y) \,, \nonumber
\end{eqnarray}
with 
\begin{eqnarray}
&&\tilde \omega=\omega-\mu\psi \,,
\end{eqnarray}
\begin{eqnarray}
&&x=\psi^T\lambda^{-1}\psi,\quad y 
   =\tau^{-1}\mu^2,\quad z=y-\tau^{-1}\psi^T\hat J\tilde\omega \,,
\end{eqnarray}
and the $2\times 2$ matrix, 
\begin{eqnarray}
\hat J= \left(
  \begin{array}{ccccccc}
   0&1\\
   -1&0\\
  \end{array}
 \right) \,.  
\end{eqnarray}
We note that this $7\times7$ matrix $M$ is symmetric, $M^T=M$, 
and unimodular, $\det(M)= 1$. 
\medskip
We define a current matrix as
\begin{eqnarray}
 J_i = M^{-1} \partial_i M \,,
\end{eqnarray}
which is conserved if the scalar fields are the solutions of the equation of motion derived by the action (\ref{eq:action}). Then, the action (\ref{eq:action}) can be written in terms of $J$ and $M$ as follows
\begin{eqnarray}
S&=&\frac{1}{4}\int d\rho dz \rho {\rm tr}(J_iJ^i) \nonumber \\
 &=&\frac{1}{4}\int d\rho dz 
    \rho {\rm tr}(M^{-1}\partial_iMM^{-1}\partial^iM)\,. \label{eq:action2}
\end{eqnarray}
Thus, the matrix $M$ completely specify the solutions to our system.   
Therefore, one can find that the equation of motion~(\ref{eq:scalar}) can be written as
\begin{align}
& \partial_\rho( \rho \partial_\rho M M^{-1}) + \partial_z( \rho \partial_z M M^{-1}) = 0.
\end{align}
The action~(\ref{eq:action2}) is invariant under the $G_{2(2)}$ transformation.

\subsection{Electric Harrison transformation}

In particular, utilizing the $G_{2(2)}$ symmetry, Ref.~\cite{Bouchareb:2007ax} constructed the electric Harrison transformation preserving asymptotic flatness that transform a five-dimensional vacuum solution $\Phi^{A}=\{\lambda_{ab},\omega_a,\psi_a=0,\mu=0\}$ into a charged solution $\Phi^{\prime A}=\{\lambda'_{ab},\omega'_a,\psi'_a,\mu'\}$ in  five-dimensional minimal supergravity, which is given by
\begin{align}
\begin{split}\label{eq:ctrans-metric}
&\tau' = D^{-1} \tau,\quad
\lambda'_{00} = D^{-2} \lambda_{00},\quad
 \lambda'_{01} =D^{-2} (c^3 \lambda_{01}+s^3 \lambda_{00} \omega_0),\\
& \lambda'_{11} = -\frac{\tau D}{\lambda_{00}} + \frac{(c^3 \lambda_{01}+s^3 \omega_0 \lambda_{00})^2}{D^2 \lambda_{00}},\\
&\omega'_0 = D^{-2}\left[c^3(c^2+s^2+2s^2 \lambda_{00})\omega_0-s^3 (2c^2+(c^2+s^2)\lambda_{00}) \lambda_{01}\right],\\
& \omega'_1 = \omega_1 + D^{-2} s^3 \left[-c^3 \lambda_{01}^2+s(2c^2-\lambda_{00})\lambda_{01} \omega_{0}-c^3 \omega_0^2\right],\\
& \psi'_0 = D^{-1} sc\, (1+\lambda_{00}),\quad
\psi'_1 = D^{-1} sc\, (c\lambda_{01}-s \omega_0),\\
&\mu' = D^{-1}sc\,(c\,\omega_0-s\lambda_{01}),
\end{split}
\end{align}
with
\begin{align}
 D  = c^2+ s^2 \lambda_{00} = 1 + s^2 (1+\lambda_{00}),
\end{align}
where the new parameter $\alpha$ in $(c,s):=(\cosh\alpha,\sinh\alpha)$ is related to the electric charge.
The functions $a^{\prime a}{}_{\phi}\ (a=0,1)$ and the component $A'_\phi$ for the charged solution are determined by the eight scalar functions $\{\lambda'_{ab},\omega'_a,\psi'_a,\mu'\}$ from  Eqs.~(\ref{eq:mu}) and (\ref{eq:twistpotential}) after the replacement of $\{\lambda_{ab},\omega_a,\psi_a,\mu\}$ with $\{\lambda'_{ab},\omega'_a,\psi'_a,\mu'\}$:
First, the functions $a'{}^a{}_\phi $ $(a=0,1)$ are determined by
\begin{align}
\begin{split} \label{eq:a0phi-charged}
& \partial_\rho a'{}^a{}_{\phi} = -\frac{\rho }{\tau'} \lambda'{}^{ab}(\partial_z \omega'_{b}
 -3 \psi'_b \partial_z\mu'-\psi'_b \epsilon^{c d} \psi'_c \partial_z\psi'_{d}),\\
&  \partial_z a'{}^a{}_\phi =\frac{\rho }{\tau'} \lambda'{}^{ab}(\partial_\rho\omega'_{b}
 -3 \psi'_b \partial_\rho\mu'-\psi'_b \epsilon^{cd} \psi'_c \partial_\rho\psi'_{d}). 
\end{split}
\end{align}
We can show from Eq.~(\ref{eq:ctrans-metric}) that  Eq.(\ref{eq:a0phi-charged}) can be written as
\begin{align}
\partial_\rho (a'{}^1{}_\phi{} -a^1{}_\phi) = 0 ,\quad \partial_z (a'{}^1{}_\phi -a^1{}_\phi)=0,
\end{align}
 Hence,  $a'^1{}_\phi$ can be obtained up to a constant as
\begin{align}
 \quad a'^1{}_\phi = a^1{}_\phi, \label{eq:a1-diff}
\end{align}
Furthermore,  Eq.(\ref{eq:a0phi-charged}) can be written as
\begin{align}
\begin{split}\label{eq:a0phi-diff}
& \partial_\rho (a'{}^0{}_\phi-c^3 a{}^0{}_\phi)
 =  s^3  (-  \omega_0 \partial_\rho a^1{}_\phi  + \rho\tau^{-1} \epsilon^{cd} \lambda_{0c} \partial_z \lambda_{0d}),\\
 &  \partial_z (a'{}^0{}_\phi-c^3 a{}^0{}_\phi)
=s^3 ( - \omega_0 \partial_z a^1{}_\phi
 - \rho \tau^{-1}\epsilon^{cd} \lambda_{0c} \partial_\rho \lambda_{0d}),
  \end{split}
\end{align}
where we have used Eq.~(\ref{eq:a0phi}) corresponding to the vacuum seed solution before the Harrison transformation: 
\begin{align}
\partial_\rho a^a{}_\phi =- \frac{\rho }{\tau} \lambda^{ab}\partial_z \omega_{b},\quad \partial_z a^a{}_\phi= \frac{\rho }{\tau} \lambda^{ab}\partial_\rho \omega_{b}.\label{eq:a0phi-neutral}
\end{align}
Similarly, the gauge potential $A'{}_\phi $ are determined by
\begin{align}
\begin{split} \label{eq:A0phi-charged}
& \partial_\rho A'_\phi = \sqrt{3}( a'^a{}_\phi \partial_\rho \psi'_a  - \rho \tau'^{-1} (\partial_z \mu' + \epsilon^{ab} \psi'_a \partial_z \psi'_b)), \\
& \partial_z A'_\phi = \sqrt{3}( a'^a{}_\phi \partial_z \psi'_a + \rho \tau'^{-1} (\partial_\rho \mu' + \epsilon^{ab} \psi'_a \partial_\rho \psi'_b)).
\end{split}
\end{align}
 One can rewrite the first equation of~(\ref{eq:A0phi-charged}) as follows
\begin{align}
\partial_\rho ( A_\phi' -\sqrt{3} a'{}^a{}_\phi \psi'_a) &= \sqrt{3} (-\psi'_a \partial_\rho a'{}^a{}_\phi - \rho \tau'{}^{-1} (\partial_z \mu+ \epsilon^{ab} \psi'_a \partial_z \psi'_b))\nonum
&= \sqrt{3} \rho \tau'{}^{-1} \left[ \lambda'{}^{ab} \psi'_a \partial_z \omega_b' - (1+3 \lambda'{}^{ab}\psi'_a \psi'_b)\partial_z \mu' - (1+\lambda'^{ab}\psi'_a \psi'_b)\epsilon^{cd} \psi'_c \partial_z \psi_d'\right].
\end{align}
where we have used Eq.~(\ref{eq:a0phi-charged}) to eliminate $\partial_\rho a'{}^a{}_\phi$ in the second line.
The similar expression is obtained for the second equation in Eq.~(\ref{eq:A0phi-charged}).
Moreover, substituting Eq.~(\ref{eq:ctrans-metric}) into the right hand sides, we can expressed only in terms of the quantities of the vacuum seed solution
\begin{align}
\begin{split}
&\partial_\rho (A'_\phi-\sqrt{3} a'{}^a{}_\phi \psi_a') =  \sqrt{3}  \left[cs^2 ( \omega_0 \partial_\rho a^1{}_\phi 
-\rho \tau^{-1} \epsilon^{ab} \lambda_{0a}\partial_z \lambda_{b0})
-c^2 s \partial_\rho a^0{}_\phi\right],\\
&\partial_z (A'_\phi-\sqrt{3} a'{}^a{}_\phi \psi_a') =  \sqrt{3}  \left[cs^2 (\omega_0 \partial_z a^1{}_\phi
+\rho \tau^{-1} \epsilon^{ab} \lambda_{0a}\partial_\rho \lambda_{b0})
-c^2 s \partial_z a^0{}_\phi \right],
\end{split}
\end{align}
where we used Eq.~(\ref{eq:a0phi-neutral}) to eliminate the derivatives of $\omega_a$ on the right hand side.
Comparing with the right hand side in Eq.~(\ref{eq:a0phi-diff}),
one can write this in the total derivative form, which results in
\begin{align}
A'_\phi  = \sqrt{3} \left(a'{}^a{}_\phi \psi'_a-\frac{c}{s} a'^{0}{}_\phi +\frac{c^2}{s} a^0{}_\phi\right).\label{eq:sol-A2}
\end{align}
Hence, integrating eq.~(\ref{eq:a0phi-diff}) is the only nontrivial task to obtain the charged metric. 
As will be seen later, this can be easily integrated in the $C$-metric, as in Ref.~\cite{Bouchareb:2007ax}.

\medskip
Therefore, one can derive the new metric and gauge potential describing the charged solution from Eqs.(\ref{Eineq}) and (\ref{Maxeq}). This transformation adds an electric charge to a vacuum solution while preserving asymptotic flatness and Killing isometries. 
However, as noted in Ref.~\cite{Bouchareb:2007ax}, performing the Harrison transformation on a regular vacuum black ring, such as the Pomeransky-Sen'kov solution, unavoidably leads to a Dirac-Misner string singularity appearing on the disc inside the ring. Conversely, while the transformation can produce the regular Cveti\v{c}-Youm charged black hole from the vacuum black hole, such as the Myers-Perry solution, it poses challenges for black rings. 
In our previous work~\cite{Suzuki:2024coe}, we solved this problem by utilizing a vacuum rotating black ring with a Dirac-Misner string singularity as the seed for the Harrison transformation and subsequently eliminating it appropriately by controlling the post-transformation parameters. As a result, we have obtained a regular exact solution for a non-BPS charged rotating black ring with a dipole charge.
In the subsequent section, we will detail the procedure for utilizing a vacuum seed that includes a Dirac-Misner string singularity to derive a capped black hole solution.

\section{Construction of vacuum seed for Harrison transformation}\label{sec:neutral}

Pomeransky's pioneering work~\cite{Pomeransky:2005sj} marked the beginning of utilizing the ISM (Inverse Scattering Method)~\cite{Belinsky:1979mh,Belinski:2001ph} for constructing diverse vacuum solutions of five-dimensional black holes. 
This approach has since been employed in numerous studies~\cite{
Mishima:2005id,
Tomizawa:2005wv,Tomizawa:2006jz,Tomizawa:2006vp,
Iguchi:2006rd,
Elvang:2007rd,
Tomizawa:2007mz,Iguchi:2007xs,
Pomeransky:2006bd,
Iguchi:2007is,Evslin:2007fv,
Elvang:2007hs,Izumi:2007qx,
Chen:2011jb,
Chen:2008fa,
Chen:2012kd,
Rocha:2011vv,
Rocha:2012vs,
Chen:2015iex,
Chen:2012zb,
Lucietti:2020ltw,
Lucietti:2020phh,
Morisawa:2007di,Evslin:2008gx,
Feldman:2012vd,
Chen:2010ih,
Iguchi:2011qi,
Tomizawa:2019acu,
Tomizawa:2022qyd,
Suzuki:2023nqf,
Figueras:2009mc}, by using the rod structure~\cite{Harmark:2004rm}.
In this section, we employ the ISM to craft the five-dimensional vacuum seed solution utilized for the electric Harrison transformation detailed in the subsequent section. This solution describes a vacuum rotating black lens,  comprising a rotating black ring and a rotating black hole, with a horizon cross-section of $L(n;1)=S^3/{\mathbb Z}_n$ topology with a Dirac-Misner singularity.

\subsection{ISM construction of the vacuum seed}

As a vacuum seed for the Harrison transformation, we choose the vacuum rotating black lens with a horizon cross-section of lens space $L(n;1)$ $(n=0,1,2,\ldots)$, which initially possesses a Dirac-Misner string singularity. This singularity will be eliminated by appropriately adjusting the parameters of the solution after the Harrison transformation. 
To construct this vacuum solution, we follow the procedure outlined for the vacuum rotating black lens by Chen and Teo~\cite{Chen:2008fa}. The key distinction lies in the treatment of the Dirac-Misner string singularity: while they remove it, we retain it in our solution.

\medskip

\begin{figure}
\begin{center}
\includegraphics[width=8cm]{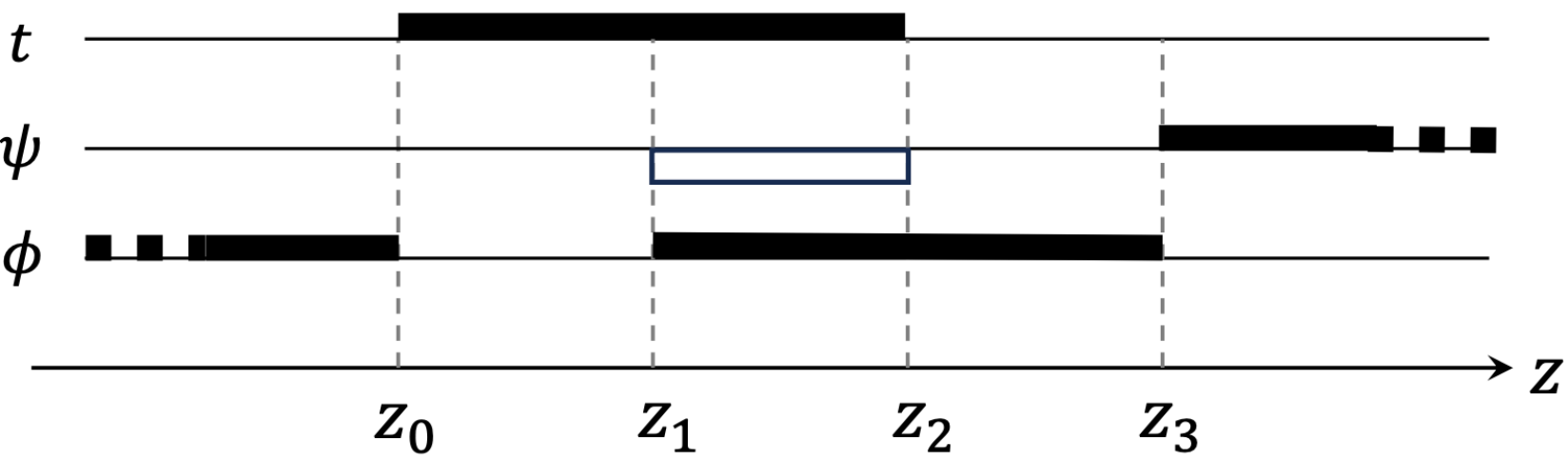}
\caption{Rod structure of the diagonal seed for the ISM. \label{fig:rodstructure}}
\end{center}
\end{figure}

To use the ISM, we rewrite the Weyl-Papapetrou form~(\ref{eq:WPform}) as 
\begin{align}
  ds^2 = G_{ij} dx^i dx^j + f (d\rho^2+dz^2),\label{eq:WPform-vac}
\end{align}
where $(x^i) = (t,\psi,\phi)$ $(i=0,1,2)$ and a $3\times 3$ matrix $G_{ij}$ and $f$ are the functions of $\rho$ and $z$, with the constraint ${\rm det}(G_{ij})=-\rho^2$.
 We begin with the diagonal metric given by:
\begin{align}
G_0 = {\rm diag}\left( -\frac{\mu_0}{\mu_2},\frac{\mu_2\mu_3}{\mu_1},\frac{\rho^2 \mu_1}{\mu_0\mu_3}\right),\quad
f_0 = \frac{C_f\mu_2 \mu_3 R_{01}R_{02}R_{12} R_{13}^2}{\mu_1 R_{00} R_{03}R_{11} R_{22} R_{23}R_{33}},
\end{align}
where $\mu_i:=\sqrt{\rho^2+(z-z_i)^2}-z+z_i$ and $R_{ij} := \rho^2 + \mu_i \mu_j$.
The rod structure is displayed in Fig.~\ref{fig:rodstructure}, and  the constant $C_f$ is consistently set to $1$ throughout this paper.
The $\psi\psi$-component of the metric diverges as $g_{\psi\psi} \sim \mathcal{O}(\rho^{-2})$ as $\rho \to 0$ for $z_1<z<z_2$. This divergence indicates naked curvature singularities on the negative rod $\rho=0, z_1<z<z_2$, as discussed in Ref.~\cite{Harmark:2004rm}.

\medskip

Then, following Pomeransky's procedure~\cite{Pomeransky:2005sj,Chen:2008fa}, we first remove three trivial solitons from the points $z=z_0,z_2,z_3$ with vectors $(0,0,1)$, $(1,0,0)$, and $(0,0,1)$, respectively.
Next, we add back three non-trivial solitons at the same points $z=z_0,z_2,z_3$ with vectors $m_{0,0}=(C_0,0,1)$, $m_{2,0}=(1,C_2,0)$, and $m_{3,0}=(0,C_3,1)$, respectively.
For the diagonal metric, removing a trivial soliton corresponds to multiplying $-\mu_k^2/\rho^2$ to the $kk$-component of $G_0$, where $k$ is the index of the nonzero component in the vector.
\begin{align}
\tilde{G}_0 &= {\rm diag} \left(-\frac{\mu_2^2}{\rho^2},1,\frac{\mu_0^2}{\rho^2}\frac{\mu_3^2}{\rho^2}\right)G_0\nonum
&=   {\rm diag}\left( \frac{\mu_0\mu_2}{\rho^2},\frac{\mu_2\mu_3}{\mu_1},\frac{\mu_0\mu_1\mu_3}{\rho^2}\right)\nonum
&=   {\rm diag}\left( \frac{\rho^2}{\bar{\mu}_0\bar{\mu}_2},\frac{\rho^4}{\mu_1\bar{\mu}_2\bar{\mu}_3},\frac{\rho^2\mu_1}{\bar{\mu}_0\bar{\mu}_3}\right),
\end{align}
where $\bar{\mu}_i := -\sqrt{\rho^2+(z-z_i)^2}-z+z_i$ and we used $\mu_i \bar{\mu}_i = -\rho^2$ in the last line for the later use.
The three-soliton solution is obtained from the modified metric $\tilde{G}_0$ and the vectors $m_i$ as
\begin{align}
G_3 = \tilde{G}_0 - \sum_{i,j=0,2,3} (\Gamma^{-1})_{ij} \frac{ (m_i \tilde{G}_0) \otimes (m_j\tilde{G}_0)}{\mu_i \mu_j},
\end{align}
where the $3\times 3$ matrix $\Gamma_{ij}$ is given by
\begin{align}
\Gamma_{ij} := \frac{m_i \tilde{G}_0 m_j}{R_{ij}} ,\quad m_i := m_{i,0} \Psi_0^{-1}(\lambda = \mu_i,\rho,z), (i,j=0,2,3),
\end{align}
with the generating matrix made from $\tilde{G}_0$ by the replacement $\mu_i \to \mu_i-\lambda,\ \bar{\mu}_i \to \bar{\mu}_i-\lambda,\ \rho^2 \to \rho^2 -2\lambda z-\lambda^2$
\begin{align}
\Psi_0(\lambda,\rho,z)={\rm diag}\left(\frac{\rho ^2-2 \lambda  z-\lambda ^2}{\left(\lambda -\bar{\mu }_0\right)
   \left(\lambda -\bar{\mu }_2\right)},
   \frac{\left(\rho ^2-2 \lambda  z-\lambda ^2\right)^2}{\left(\mu _1-\lambda \right) \left(\bar{\mu }_2-\lambda \right)
   \left(\bar{\mu }_3-\lambda \right)},
   \frac{\left(\lambda -\mu _1\right)
   \left(\rho ^2-2 \lambda  z-\lambda ^2\right)}{\left(\lambda -\bar{\mu }_0\right)
   \left(\lambda -\bar{\mu }_3\right)}\right).
\end{align}
The metric function $f$ can be obtained as
\begin{align}
f_3 = \frac{{\rm det}(\Gamma_{ij})}{{\rm det}(\Gamma_{ij})|_{C_0,C_2,C_3\to0}} f_0.
\end{align}
To remove the divergence of the metric on $\rho=0,\ z_1<z<z_2$, we set
\begin{align}
C_2 =\sqrt{ \frac{2 z_{20}}{z_{21}z_{32}}} z_2,\label{eq:no-negative}
\end{align}
where $z_{ij} := z_i-z_j$. Under this condition~(\ref{eq:no-negative}), 
the rod vectors on two rods $\{(\rho,z)|\rho=0,\ z_1<z<z_2\}$ and $\{(\rho,z)|\rho=0,\ z_2<z<z_3\}$ become parallel, merging these rods into a single rod. Note that under this condition, the point $(\rho,z)=(0,z_2)$ no longer becomes an endpoint of the rods but a mere regular point, often referred to as a phantom point.

\medskip
For later convenience, we introduce the following parameters:
\begin{align}
 b := C_0 \frac{z_{10}}{z_{30}}  \sqrt{\frac{2z_{20}z_{32}}{z_{21}}},\quad
 a:= C_3\frac{z_{31}^2}{z_3 z_{30}} -C_0 \frac{z_{10}}{z_{30}}\sqrt{\frac{2z_{20}z_{21}}{z_{32}}}.\label{eq:def-newBZ}
\end{align}
The resulting solution becomes asymptotically flat at $\sqrt{\rho^2+z^2}\to\infty$ if and only if
\begin{align}
-1 < a < 1, \label{eq:constQ}
\end{align}
or otherwise the spacetime is not Lorentzian, since the spatial metric $(G_3)_{IJ}$ ($I,J=\psi,\phi$) approaches a semi-positive definite metric multiplied by the factor $(1-a^2)^{-1}$ at infinity.
Together with this condition, the solution turns out to asymptote to the standard Minkowski metric under the global rotation, which is expressed as the coordinate change:
\begin{eqnarray}
x^i \to \Lambda^i{}_j x^j, \label{eq:globalrot}
\end{eqnarray}
where
\begin{align}
 \Lambda = \arrayL{ccc}  1 & - \Gamma_1 \Gamma_2 &b\Gamma_1\Gamma_2\\
 0 & \Gamma_1 & -a \Gamma_1  \\
 0 & -a \Gamma_1  & \Gamma_1  \arrayR,\quad \Gamma_1 := \fr{\sqrt{1-a^2}},\quad
 \Gamma_2 := \sqrt{\frac{2z_{20}z_{21}}{z_{32}}}.
\end{align}

\subsection{Vacuum seed solution for Harrison transformation}
Under the condition~(\ref{eq:no-negative}), the metric can be written in the simpler form without square root terms if we introduce  the $C$-metric coordinates $(x,y)$~\cite{Harmark:2004rm}, which are defined as
\begin{align}
 \rho = \frac{2 \ell^2 \sqrt{-G(x)G(y)}}{(x-y)^2},\quad z = \frac{\ell^2(1-xy)(2+\nu(x+y))}{(x-y)^2},
\end{align}
with the cubic function
\begin{align}
G(u) = (1-u^2)(1+\nu u).
\end{align}
The endpoints $z_i$ $(i=0,1,2,3)$ of the rods are replaced by the new parameters $\ell,\nu,\gamma$
\begin{align}
 z_0 = - \nu\ell^2,\quad z_1 = \nu \ell^2,\quad z_2 = \gamma \ell^2,\quad z_3 = \ell^2,
 \label{eq:newparamz}
 \end{align}
 where 
\begin{align}
\ell>0,\quad  \nu <  \gamma < 1.\label{eq:nugamrange}
\end{align}
The functions including nasty square roots $\mu_0$, $\mu_1$ and $\mu_3$ are written as  rational functions of $x,y$,
\begin{align}
 \mu_0 =- \frac{2\ell^2(1-x)(1+y)(1+\nu y)}{(x-y)^2},\quad
 \mu_1 = - \frac{2\ell^2(1-x)(1+\nu x)(1+y)}{(x-y)^2}\quad
 \mu_3 = \frac{2\ell^2 (1+\nu x)(y^2-1)}{(x-y)^2}.
\end{align}
By the use of Eq.~(\ref{eq:no-negative}), the square root $\sqrt{\rho^2+(z-z_2)^2}$ in $\mu_2$ can be removed from the metric, and hence the point $z=z_2$ is referred to as a phantom point.

\medskip
Finally,  the metric of the vacuum solution in the $C$-metric form can be written as
\begin{align}
 ds^2 &= -\frac{H(y,x)}{H(x,y)}(dt + \Omega_\psi(x,y) d\psi + \Omega_\phi(x,y) d\phi)^2
 + \frac{F(y,x)}{H(y,x)}d\psi^2 - \frac{2J(x,y)}{H(y,x)} d\psi d\phi- \frac{F(x,y)}{H(y,x)}d\phi^2\nonum
 &+ \frac{\ell^2 H(x,y)}{4(1-\gamma)^3(1-\nu)^2(1-a^2)(x-y)^2}\left( \frac{dx^2}{G(x)}-\frac{dy^2}{G(y)}\right),\label{eq:neutral-sol-cmetric}
\end{align}
where
\begin{align}
H(x,y)& = 2 d_1(1-\gamma ) (1-\nu ) (2+ \nu  (1+x+y-x y) )\nonum
&\times (\gamma (1+y)  (1+\nu x  )-2
   -\nu  (3 x+\nu +y (2+x+\nu +2 x \nu ))) \nonum
&+d_1 c _3(1+\nu ) (\gamma + \gamma  \nu x -\nu  (x+\nu ))(1+x) (1+y)^2  \nonum
&+(1-\gamma ) (1-\nu )^2 (x+y+\nu + \nu x y  ) \biggr[
  2  \left((1-\gamma ) (1-\nu ) (\gamma +\nu )-2  d_2\right)(2+\nu(1+x+y-x y)  )\nonum
&  +\left[2 (\gamma -\nu ) (2+(x+y) \nu )+(1-x y) ((3-\nu ) \nu +\gamma 
   (1+\nu ))- (1-\nu ) (\gamma +\nu )(x-y)\right] c _3\biggr],\label{eq:H1}\\
 F(x,y) &=\frac{2 \ell^2 (1+x)}{(1-a^2)(x-y)^2} \biggr[
4 \left[\left(1-a^2\right)^2 (y-1) (1-\gamma )^3 (1-\nu )^3-d_1^2(1+y) \right](1+\nu y  ) G(x)\nonum
   & +4\left[(1-\nu ) c _2-(1-a b) (\gamma -\nu ) (1+\nu ) c _1\right]^2 (1+\nu x  )  (1+x)G(y)\nonum
  &+ \nu^{-1} (1-\nu )^3 (\gamma -\nu ) (d_3^2(1-x^2)G(y) - c _3^2(1-y^2)G(x) )\nonum
& +\frac{G(x)G(y)\left[\left(1-a^2\right) (1-\gamma )
   d_4-(a-b)^2 y (1-\gamma )^2 (\gamma -\nu ) \nu  c _2^2+x (\gamma -\nu ) \nu  \left(b d_1-c _1 c _3\right){}^2\right]}{\nu(1-\gamma) }
\biggr],\\
J(x,y)& =  \frac{2\ell^2(1+x)(1+y)}{(1-a^2)(x-y)}
\biggr[4  d_1 
   \left((a-b) (1-\gamma ) (\gamma -\nu ) (1+\nu )-a d_2\right)(1+ \nu x) (1+ \nu y)\nonum
 &  -d_3 c _3 (1-\nu
   )^3 (\gamma -\nu ) (1-x) (1-y)-(a-b) (\gamma -\nu )  c _2 \left(c _1 c _3-b d _1\right)(1-x) (1-y) (1+\nu x  ) (1+\nu y  )\biggr],
   \end{align}
     \begin{align}
  \Omega_\psi(x,y) &= \frac{v_0 \ell (1+y)(1-\nu)} {\nu H(y,x)}\biggr[
c _2 \left(c _1 c_3-b d_1\right)(1-x) (1+\nu x ) (1+\nu y  ) \nonum
&- (1-\nu )^2 d_2 c _3 (1-x)+(1+x \nu ) d_1 \left(2 \nu(1-a b)  (1-\gamma ) (1+\nu )(1+x)+(1-3   \nu -x (1+\nu )) c _3\right)
\biggr],\\
 \Omega_\phi(x,y)& =\frac{v_0 \ell (1+x)}{H(y,x)}\biggr[b (1+x) d_1 \left(d_2(1+y) (1+\nu y  ) +\nu  c _3\left(1-y^2\right) (1-\nu ) \right)\nonum
&+\frac{2 (a-b) (1-\gamma)^2 \left(2 d_1(1+\nu x  )   (1+ \nu y  )^2 
-(1-\nu )^2 \nu c _3(1-y)  (x+y+\nu +\nu x y  ) \right)}{1+\nu }\biggr],\label{eq:Omega_psi}
\end{align}
and the coefficients are given by
\begin{align}\label{eq:def-cfs}
\begin{split}
&v_0 := \sqrt{\frac{2(\gamma^2-\nu^2)}{(1-a^2)(1-\gamma)}},\\
&c _1 :=(1-\gamma) a+(\gamma-\nu) b ,\\
&c _2 := 2 a (1-\gamma ) \nu +b (\gamma -\nu ) (1+\nu ),\\
&c _3 := 2   (1-\gamma ) \nu +b^2 (\gamma -\nu ) (1+\nu ),\\
&d_1 := (\nu +1) c _1^2-(1-\gamma ) (1-\nu )^2,\\
&d_2:= b (\nu +1) c _1 (\gamma -\nu )+2\nu (1-\gamma )   (1-\nu ) ,\\
&  d_3 := \left(a^2-1\right) b (\gamma -1) (\nu +1)-a c _3,\\
&d_4 := b^2 (\gamma -\nu )   \left[(\nu +1)^2 c _1^2 \left(-3 (1-\gamma ) \nu -\nu ^2+1\right)-(1-\gamma ) (1-\nu )^4 (2 \nu +1)\right]\\
&\quad +(1-\gamma   ) \left[\left((1-\nu ) c _2-2 \nu ^2 c _1\right)^2-4 \nu ^2 c _1^2 \left(-\gamma  (\nu +2)+3 \nu
   ^2+1\right)\right].
   \end{split}
\end{align}
We assume the ranges of the coordinates as
\begin{align}
 -\infty < t < \infty,\quad 0 \leq \psi \leq 2\pi,\quad 0\leq \phi \leq 2\pi,\label{eq:rangetpsiphi}
\end{align}
and
\begin{align}
 -1 \leq x \leq 1,\quad -1/\nu \leq y \leq -1,\label{eq:rangexy}
\end{align}
where the boundary of the coordinate $(x,y)$ corresponds to the rods plus infinity:
\begin{enumerate}[(i)]
\item
$\phi$-rotational axis : 
$\partial \Sigma_\phi=\{(x,y)|x=-1,-1<y<-1/\nu \}$ with the rod vector 
$v_\phi:=(0,0,1)$, where in the choice of $C_f=1$, the periodicity $\phi\sim \phi +2\pi$ from the coordinate ranges (\ref{eq:rangetpsiphi}) ensures the absence of 
the conical singularities, 
\item Horizon: 
$\partial \Sigma_{\cal H}=\{(x,y)|-1<x<1,y=-1/\nu \}$ with the rod vector $
 v_{\cal H} := (1, \omega^{\rm vac}_\psi, \omega^{\rm vac}_\phi),
$ 
where
\begin{align}
\begin{split}
&\omega^{\rm vac}_\psi =\frac{v_0 \left(1-a^2\right) (1-\gamma )}{2 \ell (\gamma +\nu ) \left(1-a^2+a (a-b) \gamma -(1-a b) \nu \right)},\\
&\omega^{\rm vac}_\phi = \omega^{\rm vac}_\psi \frac{2 a (1-\gamma ) \nu -\left(1-a^2\right) b (1-\gamma ) (1+\nu )+a b^2 (\gamma -\nu ) (1+\nu )}{2 (1-\gamma ) \nu +b^2 (\gamma -\nu )   (1+\nu )}.\label{eq:vH-n}
   \end{split}
\end{align}

\item
Inner axis: 
$\partial \Sigma_{\rm in}=\{(x,y)|x=1,-1<y<-1/\nu \}$ 
with the rod vector 
\begin{align}
 v_{\rm in}:= (v_0 \ell (a-b),  n , 1),\label{eq:rodvector-3}
\end{align}
with
\begin{align} 
n := \frac{a d_1+(1-\gamma)(1+\nu)(1-a^2)c_1}{d_1},\label{eq:lensN}
\end{align}
where we note that the presence of the $t$-component denotes the existence of the Dirac-Misner string singularity~\cite{Misner:1963fr},

\item
$\psi$-rotational axis: 
$\partial \Sigma_\psi=\{(x,y)|-1<x<1,y=-1 \}$
with the rod vector $v_\psi :=(0,1,0)$, 
where the periodicity $\psi\sim \psi +2\pi$ from the coordinate ranges (\ref{eq:rangetpsiphi}) also ensures the absence of the conical singularities, 

\item Infinity:  
$\partial \Sigma_\infty 
= \{(\rho,z)|\sqrt{\rho^2+z^2}\to 
\infty\ {\rm with}\ z/\sqrt{\rho^2+z^2}\ {\rm finite}  \}$$=\{(x,y)|x\to y \to -1 \}$
\end{enumerate}

\medskip

As studied earlier in Ref.~\cite{Chen:2008fa}, setting $a=b$, in which the obtained metric exactly describes the rotating black lens in~\cite{Chen:2008fa}, allows us to remove the Dirac-Misner string singularity~\cite{Misner:1963fr} on the inner axis $\partial \Sigma_{\rm in}$. 
However, similar to the approach taken for the charged dipole black ring in Ref.~\cite{Suzuki:2024coe}, we choose to use the vacuum black lens possessing a Dirac-Misner string singularity as the seed for the Harrison transformation. Hence, we do not assume its absence ($a \neq b$) before the Harrison transformation. 
In the following section, we will eliminate it after the transformation by appropriately adjusting the solution's parameters.

\section{Charged solution from Harrison transformation}\label{sec:charging}

Now, let us utilize the electric Harrison transformation~(\ref{eq:ctrans-metric}) on the vacuum solution~(\ref{eq:neutral-sol-cmetric}) derived in the previous section. 
The procedure to obtain the charged solution follows a similar approach to that used for charged  black ring~\cite{Bouchareb:2007ax}. 
First, we express the vacuum solution~(\ref{eq:neutral-sol-cmetric}) in terms of eight scalar potentials $\Phi^A=(\lambda_{ab},\omega_a,\psi_a,\mu)$. 
Since the vacuum solution possesses two axial Killing vectors $\partial/\partial\psi$ and $\partial/\partial\phi$, there are two possible ways to express the solution in terms of the potentials, depending on the following choice of the Killing vectors: 
(i) $\xi_0 = \partial_t, \ \xi_1 =\partial/\partial\psi,\ \xi_2=\partial/\partial\phi$ and (ii) $\xi_0 = \partial_t, \ \xi_1 =\partial/\partial\phi,\ \xi_2=\partial/\partial\psi$.

\medskip
In case (i), from the metric~(\ref{eq:neutral-sol-cmetric}), we can  extract $\lambda_{ab}$, $a^a{}_\phi$ and $\tau$, expressed as   
\begin{align}
\begin{split}
&\lambda_{00} = - \frac{H(y,x)}{H(x,y)},\quad \lambda_{01} =- \frac{H(y,x)}{H(x,y)}\Omega_\psi(x,y),\quad
\lambda_{11} = \frac{F(y,x)}{H(y,x)} -\frac{H(y,x)}{H(x,y)}\Omega_\psi^2(x,y).\\
&   a^0{}_\phi = \Omega_\phi(x,y) + \frac{J(x,y)}{F(y,x)} \Omega_\psi(x,y),\quad a^1{}_\phi = -\frac{J(x,y)}{F(y,x)}, \quad \tau = \frac{F(y,x)}{H(x,y)}.
   \end{split}
\end{align}
The twist potential $\omega_a$ can be obtained by directly integrating Eq.~(\ref{eq:a0phi-neutral}). This yields
\begin{align}
\omega_0& = -\Omega_\phi(y,x),\\
\omega_1&=-\frac{J(x,y)+K(x,y)}{H(x,y)},\label{eq:sol-omega1}
\end{align}
where $K(x,y)$ is a polynomial of $x$ and $y$: 
\begin{align}
&K(x,y)=\frac{\ell^2(1+y)}{1-a^2}\biggr[k_1 \left(1-x^2\right) (1+y)+k_2 \left\{(1+x) (1-y) (1+\nu x)+(1-x) (1+y) (1+\nu y)\right\}\nonum
&+k_3 (x-y) (\nu (xy-x-y-1)-2)+k_4 (1-x) (1-y) (\nu (x-y-1)+1)\nonum
&+k_5 (1-x) (1-y)+k_6 \left\{(\nu+1)^2 (x+1) (y+1)-4 (1+\nu x) (1+\nu y)\right\}+k_7 (1+\nu x) (1+\nu  y)\biggr],\label{eq:def-Kxy}
\end{align}
in which the coefficients $d_i$ $(i=1,\ldots,7)$ are given in Appendix~\ref{sec:Kxy}.
In case (ii), we denote the corresponding quantities with ``hats", which can be expressed as:
\begin{align}
\begin{split}
& \hat{\lambda}_{00}= - \frac{H(y,x)}{H(x,y)}, \quad \hat{\lambda}_{01} = \frac{H(y,x)}{H(x,y)} \Omega_\phi(x,y), \quad
\hat{\lambda}_{11} = -\frac{F(x,y)}{H(y,x)}-\frac{H(y,x)}{H(x,y)}\Omega_\phi^2(x,y), \\
& \hat{a}^0{}_\psi = \Omega_\psi(x,y) - \frac{J(x,y)}{F(x,y)} \Omega_\phi(x,y), \quad \hat{a}^1{}_\psi = \frac{J(x,y)}{F(x,y)}, \quad \hat\tau = -\frac{F(x,y)}{H(x,y)},
\end{split}
\end{align}
which yields
\begin{align}
\hat{\omega}_0 &= \Omega_\psi(y,x),\\
\hat{\omega}_1 &= \frac{J(x,y)+\hat{K}(x,y)}{H(x,y)},\label{eq:sol-hatomega1}
\end{align}
where $\hat{K}(x,y)$ can be expressed, in terms of $K(x,y)$, as
\begin{align}
\hat{K}(x,y) = -K(y,x) - \frac{2 \ell^2 c_3 d_1 (1-\nu ) (x+1)^2 (\gamma +\nu ) \left(2 a (1-\gamma ) (1-\nu )+c_2 (y+1)\right)}{(1-a^2)(1-\gamma) }.
\end{align}

\medskip

When deriving the metric and gauge potential after the Harrison transformation, the most non-trivial aspect lies in determining $a'{}^0{}_\phi$ or $\hat{a}'{}^0{}_\psi$ through the integration of Eq.~(\ref{eq:a0phi-diff}), resulting in
\begin{align}
 a'^0{}_\phi &=c^3 a^0{}_\phi(x,y) - s^3\hat{a}^0{}_\psi(y,x)
,\label{eq:ac0sol}
\end{align}
\begin{align}
\hat{a}'^0{}_\psi = c^3 \hat{a}^0{}_\psi(x,y) + s^3 a^0{}_\phi(y,x),
\end{align}
which also has the same form as that of the vacuum doubly rotating black ring.
The metric functions $a^0{}_\phi$ and $\hat a^0{}_\psi$ in cases (i) and (ii) are intertwined in the transformed metric. This highlights that the two Harrison transformations of (i) and (ii) are connected through sign flips: $s\to -s$, $t\to -t$, $\psi\to -\psi$, and $\phi\to -\phi$. 
Therefore, in what follows,  we consider only the Harrison transformation in case (i).

Since Eq.~(\ref{eq:eq-sigma}) is invariant under the transformation, the function $\sigma$ is also invariant. 
Then, from the change of $\tau$ in Eq.~(\ref{eq:ctrans-metric}), one can see that the two-dimensional conformal factor in Eq.~(\ref{eq:WPform}) is multiplied by $D$, which leads to
\begin{align}
g_{xx}' = D \, g_{xx} ,\quad g_{yy}' = D\,  g_{yy}.
\end{align}

\begin{figure}
\includegraphics[width= 7cm]{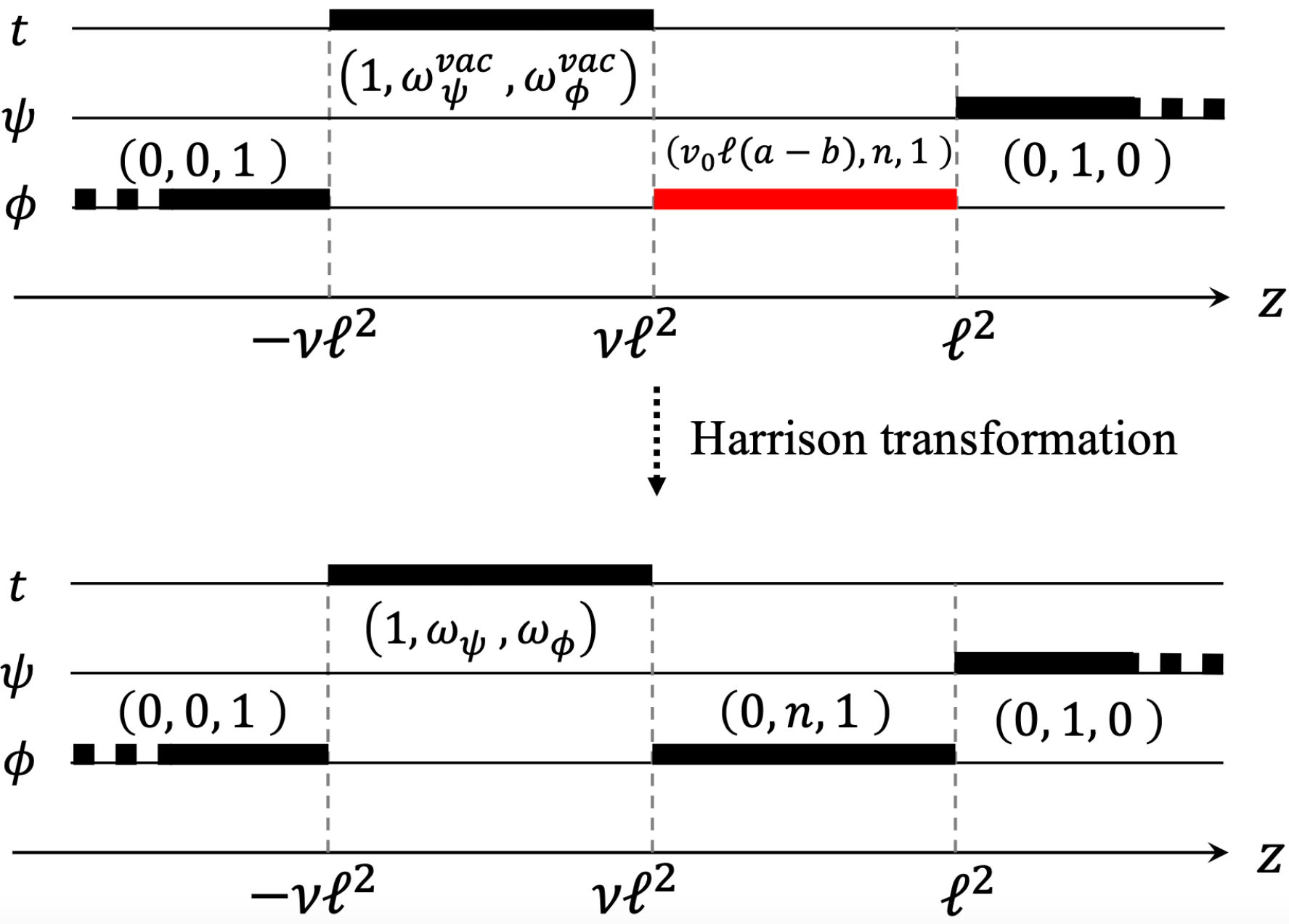}
\caption{
The rod structures before and after applying the Harrison transformation, with the condition~(\ref{eq:nodms}) imposed on the latter.
 \label{fig:rodtrans}}
\end{figure}

\subsection{Charged solution}

 In case (i), 
  the metric and gauge potential for a charged solution in five-dimensional minimal supergravity after the electric Harrison transformation can be written as
\begin{eqnarray}
ds^2& =& -\frac{H(y,x)}{D^2 H(x,y)}(dt + \Omega')^2
 +D \left[ \frac{F(y,x)}{H(y,x)}d\psi^2 - \frac{2J(x,y)}{H(y,x)} d\psi d\phi- \frac{F(x,y)}{H(y,x)}d\phi^2\right] \notag\\
 &&+ \frac{\ell^2 D H(x,y)}{4(1-\gamma)^3(1-\nu)^2(1-a^2)(x-y)^2}\left( \frac{dx^2}{G(x)}-\frac{dy^2}{G(y)}\right),\label{eq:charged-metric}\\
 A &=& \frac{\sqrt{3}cs}{D H(x,y)}\biggr[ (H(x,y)-H(y,x))dt - (c H(y,x) \Omega_\psi(x,y)-s H(x,y) \Omega_\phi(y,x))d\psi \nonum
&& - ( c H(y,x)\Omega_\phi(x,y)-s H(x,y)\Omega_\psi(y,x))d\phi \biggr], 
\label{eq:gauge-sol}
\end{eqnarray}

where the functions $D$, $\Omega_\psi' $ and $\Omega_\phi' $ are given by
\begin{align}
&D=  \frac{c^2 H(x,y)  - s^2 H(y,x)}{H(x,y)},\\
&\Omega' = (c^3 \Omega_\psi(x,y) -s^3 \Omega_\phi(y,x))d\psi+(c^3 \Omega_\phi(x,y)-s^3 \Omega_\psi(y,x))d\phi.
\end{align}

The Harrison transformation changes the rod structure of the vacuum solution~(\ref{eq:ctrans-metric}) as follows (see Fig.~\ref{fig:rodtrans} about the rod diagram):
\begin{enumerate}[(i)]
\item
$\phi$-rotational axis: 
$\partial \Sigma_\phi=\{(x,y)|x=-1,-1/\nu<y<-1 \}$ with the rod vector
$v_\phi =(0,0,1)$. The periodicity of $\phi\sim \phi+2\pi$ still leaves the absence of conical singularities.

\item Horizon: 
$\partial \Sigma_{\cal H}=\{(x,y)|-1<x<1,y=-1/\nu \}$ with the rod vector 
$ v_{\cal H} = (1, \omega_\psi , \omega_\phi)$,
with
\begin{align}
&\omega_\psi=\frac{v_0\left(1-a^2\right)  (1-\gamma )}{2 \ell (\gamma +\nu ) \left[c^3 \left(1-\nu -a c_1\right)-s^3 (a-b) (\gamma -\nu )\right]},\quad
\omega_\phi=- \frac{d_3 \omega_\psi}{c_3}.
\end{align}

\item 
Inner axis: 
$\partial \Sigma_{\rm in}=\{(x,y)|x=1,-1/\nu<y<-1\}$ with the rod vector 
\begin{align}
v_{\rm in}= ( v_0 \ell (c^3 (a-b)-s^3(1-ab)), n, 1),
\end{align}
 where $n$ is given by Eq.~(\ref{eq:lensN}).

\item[(iv)] 
$\psi$-rotational axis: 
$\partial \Sigma_\psi=\{(x,y)|-1<x<1,y=-1 \}$ with the rod vector $v_\psi=(0,1,0)$.  
The periodicity of $\psi\sim \psi+2\pi$ still leaves the absence of conical singularities,

\item[(v)] Infinity:  
$\partial \Sigma_\infty =\{(x,y)|x\to y \to -1 \}$.
\end{enumerate} 

One can see that the Harrison transformation preserves both the positions and the regularity of the $\phi$ and $\psi$-rotational axes at $x=-1$ and $y=-1$, respectively.
The event horizon remains to be at $y=-1/\nu$ but the horizon velocities are changes.
The rod vector of the inner axis $\partial \Sigma_{\rm in}$ changes from $(v_0 \ell(a-b),n,1)$ to $(v_0 \ell(c^3(a-b)-s^3(1-ab)),n,1)$, which enables one to eliminate the Dirac-Misner string singularity by setting
\begin{align}
   \tanh^3 \alpha = \frac{a-b}{1-ab},\label{eq:nodms}
\end{align}
where the vacuum case corresponds to $a=b$. 
As seen in Ref.~\cite{Suzuki:2024coe}, the Dirac-Misner string singularity is unavoidable if we transform the vacuum seed not possessing the Dirac-Misner string singularity (the vacuum seed corresponding to $a=b$). 
From the range~(\ref{eq:constQ}) of $a$ and $-1<\tanh\alpha<1$, the parameter $b$ runs the range
\begin{align}
 -1 < b < 1.
\end{align}

Under the condition~(\ref{eq:nodms}) for the absence of the Dirac-Misner string singularity, the rod vector on $\partial \Sigma_{\rm in}$ becomes $\partial_{\phi'} = \partial_\phi + n\partial_\psi$, 
and the absence of conical singularities requires
\begin{align}
\left( \frac{\Delta \phi'}{2\pi}\right)^2 =\frac{d_1^2}{\left(1-a^2\right)^2 (1-\gamma )^3
  (1-\nu )^2 (1+\nu )}=1.\label{eq:conifree}
\end{align}
The topology condition for $\partial\Sigma_{\cal H}$ requires 
\begin{align}
{\rm det\ }(\hat v_{\rm in},\hat v_\psi)=
 \frac{a d_1+(1-\gamma)(1+\nu)(1-a^2)c_1}{d_1}=n \in {\mathbb Z},\label{eq:lenscon}
\end{align}
where each hatted vector $\hat v$ denotes a two-dimensional vector made from $\psi$ and $\phi$ components of each rod vector $v$.
As proved in Ref.~\cite{Hollands:2007aj}, the horizon cross-section has the topology of $S^2 \times S^1$ for $n=0$, $S^3$ for $n=\pm1$ and $L(n;1)$ for $|n| \geq 2$. To study all possibilities, we do not fix the value of $n$ here.

\medskip

To summarize, the charged solution has six parameters $(\ell,a,b,\gamma,\nu,\alpha)$, with the following ranges:
\begin{align}
 \ell>0,\quad -1<a<1,\quad -1<b<1,\quad 0<\nu<\gamma<1,\quad -\infty < \alpha < \infty. \label{eq:paramrange}
\end{align}
The regularity of the metric at each boundary imposes the  conditions~(\ref{eq:nodms}), (\ref{eq:conifree}), and (\ref{eq:lenscon}), which reduce the independent parameters of the solution from six to three.
Moreover, the solution and the conditions are invariant under transformations $n \to - n,\quad a \to - a, \quad b\to-b,\quad \alpha \to -\alpha$, and hence we may assume $n \geq 0$ without loss of generality.
In the following, under the conditions~(\ref{eq:nodms}), (\ref{eq:conifree}), and (\ref{eq:lenscon}), we investigate whether the charged solution has curvature singularities and CTCs for each value of $n$.

\subsection{Regularity at the coordinate boundaries}

Curvature singularities on and outside the horizon may arise at points where the metric and its inverse appear to diverge in the range~(\ref{eq:rangexy}). 
This occurs on the surfaces $H(x,y)=0$ and $D=0$, as well as on the boundary of the $C$-metric coordinates at $x=\pm 1$ and $y=-1/\nu,-1$, where $G(x)=0$ or $G(y)=0$.
From $H(-1,-1)=8(1-\gamma)^3(1-\nu)^4(1-a^2)>0$ and  $D=1$ at infinity $x\to y\to -1$. 
Hence, the necessary and sufficient condition for the absence of surfaces $H(x,y)=0$ and $D=0$ is that $H(x,y)$ and $D$ are positive everywhere in the range~(\ref{eq:rangexy}).
Since the discussion regarding $H(x,y)>0$ and $D>0$ depends on the value of $n$, we will adrress this  in the next subsection.   
Here, we demonstrate the absence of curvature singularities at the coordinate boundaries, $x=\pm 1$, $y=-1,-1/\nu$, by assuming $H(x,y)>0$ and $D>0$ 
Additionally, we notice that despite its appearance in the metric \eqref{eq:charged-metric}, the surface $H(y,x)=0$ does not cause a divergence in the metric and its inverse. This is because the factor $H^{-1}(y,x)$ does not appear in each component of $g_{\mu\nu}$ and $g^{\mu\nu}$.

\medskip

\begin{enumerate}

\item 
The limit $x\to y \to -1$ corresponds to the asymptotic infinity. 
In terms of the standard spherical coordinates $(r,\theta)$, defined as, 
\begin{align}
 x=-1+4(1-\nu)\ell^2 r^{-2} \cos^2\theta ,\quad
  y=-1-4(1-\nu)\ell^2 r^{-2} \sin^2\theta, \label{eq:aslim}
\end{align}
we find that the metric at $r\to\infty$ ($x\to y \to -1$) behaves as the Minkowski metric:
\begin{align}
ds^2 \simeq -dt^2+dr^2 + r^2 (d\theta^2+\sin^2\theta d\psi^2+\cos^2\theta d\phi^2).\label{eq:aslimds}
\end{align}
Hence, the charged metric~(\ref{eq:charged-metric}) describes an asymptotically flat spacetime.

\item 
The point $(x,y)=(1,-1)$ corresponds to a center of the spacetime i.e., the intersection of the $\psi$-rotational axis and inner rotational axis.
 Using  the coordinates $(r,\theta)$  introduced by
\begin{align}
x = 1 - \frac{(1+\nu)(1-\gamma)^2(1-a^2) r^2 \cos^2\theta }{|d_1| \ell^2 (1+c^2 \nu+s^2\gamma)},\quad
y = -1 - \frac{(1-\nu)(1-\gamma)^2(1-a^2) r^2 \sin^2\theta }{|d_1| \ell^2 (1+c^2 \nu+s^2\gamma)},
\end{align}
we can show that the metric at $r\to 0$ ($(x,y)\to (1,-1)$) behaves as the origin of the Minkowski spacetime written in the spherical coordinates if $d_1<0$,
\begin{align}
ds^2 \simeq -dt'^2+(-d_1/|d_1|) \left[ dr^2 +r^2(d\theta^2+\sin^2 \theta d\psi'^2  + \cos^2 \theta d\phi^2)\right],
\end{align}
where $t' := \sqrt{(1-\gamma)(1+\nu)}t/(1+c^2 \nu+s^2 \gamma )$ and $\psi' = \psi-n\phi$. 

If $d_1>0$, the metric is not Lorentzian around this point but the negativity of $d_1$ is ensured by the positivity of $H(x,y)$ at this point, since 
\begin{eqnarray}
H(1,-1) = - 8 d_1(1-\gamma)(1-\nu)(1-\nu^2) > 0 \Longleftrightarrow d_1<0. \label{eq:d1neg}
\end{eqnarray}
Under this condition,  the point $(x,y)=(1,-1)$ is regular.

\item
The boundary $x=-1$ and $x=1$ correspond to the $\phi$-rotational axis and inner rotational axes, respectively.
Introducing the radial coordinate $r$ by  $x = \pm 1 \mp C_\pm r^2$ with positive constants $C_\pm$ for $x=\pm 1$,
we can see that, with the use of Eqs.~(\ref{eq:lenscon}) and (\ref{eq:conifree}), the metric at $r\to0$ ($x\to \pm 1$) behaves as
\begin{align}
 ds^2 \simeq \gamma^{\pm}_{tt}(y) dt^2+2 \gamma^{\pm}_{t\psi}(y) dt d\psi_\pm+ \gamma^{\pm}_{\psi\psi}(y) d\psi_\pm^2 + \alpha_\pm(y) ( dr^2+r^2 d\phi_\pm^2-G^{-1}(y) dy^2),
\end{align}
where
\begin{align}
&\gamma_{tt}^{\pm} =-\frac{D|_{x=\pm1}H(y,\pm1)}{H(\pm1,y)},\quad
\gamma_{t\psi}^{\pm} = -\frac{H(y,\pm1)[c^3 \Omega_\phi(\pm1,y)-s^3 \Omega_\phi(y,\pm1)]}{D^2|_{x=\pm1}H(\pm1,y)},\nonum
&\gamma_{\psi\psi}^{\pm} =\frac{D|_{x=\pm1} F(y,\pm1)}{H(y,\pm1)}-\frac{H(y,\pm1)[c^3\Omega_\psi(\pm1,y)-s^3 \Omega_\phi(y,\pm1)]^2}{D^2|_{x=\pm} H(\pm1,y)},\nonum
& \alpha_{\pm} = \frac{C_\pm \ell^2 D |_{x=\pm 1} H(\pm 1,y)}{2(1-\gamma)^3(1\pm \nu)(1-\nu)^2(1-a^2)(1\mp y)^2},
\end{align}
and
\begin{align}
(\psi_{-},\phi_{-}):=(\psi,\phi),\
(\psi_{+},\phi_{+}):=(\psi-n\phi,\phi).\label{eq:defpsipm}
\end{align}
Under the assumptions $H(x,y)>0$ and $D>0$, we can also show that  $\alpha_{\pm}>0$,  and 
\begin{align}
 &{\rm det} \, (\gamma^{+})
 = \frac{16\ell^2 d_1^2 (1+\nu) ( 1+y)(1+\nu y)}{(1-a^2)(1-y)D |_{x=  1} H( 1,y)}<0,\\
&{\rm det} \, (\gamma^{-})
 = \frac{16\ell^2(1-\gamma)^3(1-\nu)^4(1-a^2)( 1-y)(1+\nu y)}{(1+y)D |_{x= - 1} H( -1,y)}<0,
\end{align}
hence $\gamma^{\pm}$ is a nonsingular and non-degenerate matrix for $-1/\nu<y<-1$. 
Thus, the metric is regular at $x=\pm1$.

\item 
The boundary $y=-1$ corresponds to the $\psi$-rotational axis. 
Introducing the radial coordinate $r$ by $y=-1-C_0 r^2$ with a positive constant $C_0$, 
we can see that the metric at $r \to 0$ ($y\to -1$) behaves as
\begin{align}
 ds^2 \simeq \gamma^0_{tt}(x)dt^2 + 2 \gamma^0_{t\phi}(x) dt d\phi + \gamma^0_{\phi\phi}(x)d\phi^2 + \alpha_0(x)
 (dr^2+r^2 d\psi^2+G^{-1}(x)dx^2),
\end{align}
where 
\begin{align}
&\gamma_{tt}^{0} =-\frac{D|_{y=-1}H(-1,x)}{H(x,-1)},\quad
\gamma_{t\phi}^{0} = -\frac{H(-1,x)[c^3 \Omega_\phi(x,-1)-s^3 \Omega_\psi(-1,x)]}{D^2|_{y=-1}H(x,-1)},\nonum
&\gamma_{\phi\phi}^{0} =-\frac{D|_{y=-1} F(x,-1)}{H(-1,x)}-\frac{H(-1,x)[c^3\Omega_\phi(x,-1)-s^3 \Omega_\psi(-1,x)]^2}{D^2|_{y=-1} H(x,-1)},\quad
 \alpha_0 = \frac{ C_0 \ell^2 D|_{y=-1} H(x,-1)}{2(1-\gamma)^3(1-\nu)^3(1-a^2)(1+x)^2}.
\end{align}
Under the assumptions of $H(x,y)>0$ and $D>0$,  we can also show that $\alpha_0>0$ and 
\begin{align}
{\rm det}\,(\gamma^0)
=- \frac{16\ell^2 (1-\gamma)^3(1 - \nu)^4(1-a^2)(1-x)(1+\nu x)}{(1+x) D |_{y=-1} H(x, -1)}<0,
\end{align}
and hence $\gamma^{0}$ is a nonsingular and non-degenerate matrix for $-1<x<1$. Therefore, the metric is also regular at $y=-1$.

\item 
The boundary $y=-1/\nu$ corresponds to a Killing horizon with the surface gravity
\begin{align}
\kappa = \frac{(1-a^2)^{3/2} (1-\gamma )^2 \sqrt{\nu  (\nu   +1)(\gamma +\nu)^{-1}}}{c_3 \ell  \left(c^3 \left(1-\nu-a c_1 \right)+s^3 (b-a) (\gamma -\nu   )\right)},\label{eq:kappa}
\end{align}
and the null generator is denoted by $v_{\cal H} = \partial/\partial t + \omega_\psi \partial/\partial\psi + \omega_\phi \partial/\partial \phi$ with 
\begin{align}
(\omega_\psi,\omega_\phi) = 
\frac{ \kappa }{ \left(1-a^2\right) }
 \sqrt{\frac{\gamma-\nu}{2\nu(1-\gamma )^3 \left(1+
   +\nu \right)}}   (c_3,-d_3).\label{eq:omegai}
\end{align}
We can show that both the metric and gauge potential are regular at $y=-1/\nu$, introducing the ingoing/outgoing Eddington-Finkelstein coordinates by
\begin{align}
 dx^i = dx'^i \pm v_{\cal H}^i \frac{(1-\nu^2)}{2 \nu \kappa G(y)} dy, 
\end{align}
where $(x^i)=(t,\psi,\phi)$ $(i=0,1,2)$ and the metric near $y=-1/\nu$ behaves as
\begin{align}
ds^2 &\simeq \alpha_H(x) \left(\frac{4 \nu^2 \kappa^2 G(y) }{(1-\nu^2)^2}dt'^2 \pm \frac{4\nu \kappa}{1-\nu^2} dt' dy+\frac{dx^2}{G(x)} \right)\nonum
&+ \gamma^H_{\psi\psi}(x)(d\psi'-\omega_\psi dt')^2+ 2\gamma^H_{\psi\phi}(x)(d\psi'-\omega_\psi dt')(d\phi'-\omega_\phi dt')
+ \gamma^H_{\phi\phi}(x)(d\phi'-\omega_\phi dt')^2,
\end{align}
with
\begin{align}
\begin{split}
&\gamma_{\psi\psi}^{H} =\frac{D|_{y=-1/\nu} F(-1/\nu,x)}{H(-1/\nu,x)}
-\frac{H(-1/\nu,x)[c^3\Omega_\psi(x,-1/\nu)-s^3 \Omega_\phi(-1/\nu,x)]^2}{D^2|_{y=-1/\nu} H(x,-1/\nu)},\\
&\gamma_{\psi\phi}^{H} =-\frac{D|_{y=-1/\nu} J(x,-1/\nu)}{H(-1/\nu,x)}-\frac{H(-1/\nu,x)[c^3\Omega_\psi(x,-1/\nu)-s^3 \Omega_\phi(-1/\nu,x)][c^3\Omega_\phi(x,-1/\nu)-s^3 \Omega_\psi(-1/\nu,x)]}{D^2|_{y=-1/\nu} H(x,-1/\nu)},\\
&\gamma_{\phi\phi}^{H} =-\frac{D|_{y=-1/\nu} F(x,-1/\nu)}{H(-1/\nu,x)}-\frac{H(-1/\nu,x)[c^3\Omega_\phi(x,-1/\nu)-s^3 \Omega_\psi(-1/\nu,x)]^2}{D^2|_{y=-1/\nu} H(x,-1/\nu)},\\
& \alpha_H =\frac{ \ell^2 \nu^2 D|_{y=-1/\nu}  H\left(x,-1/\nu \right)}{4(1-\gamma)^3(1-\nu)^2\left(1-a^2\right) (1+\nu x)^2},
\end{split}
\end{align}
and hence, under the assumptions $H(x,y)>0$ and $D>0$, we can show $\alpha_H>0$ and
\begin{align}
{\rm det}\, (\gamma^H)
= \frac{4 \ell^4 c_3^2  (1-\nu )^4 (\nu +1) \left(1-x^2\right) (\gamma
   +\nu ) \left[c^3 \left(a c_1+\nu -1\right)-s^3 (b-a) (\gamma
   -\nu )\right]{}^2}{\left(1-a^2\right)^2 (1-\gamma ) D|_{y=-1/\nu}
   \nu  (\nu  x+1) H(x,-1/\nu)}>0
\end{align}
and thus $\gamma^{H}$ is a nonsingular and non-degenerate matrix for $-1<x<1$.
Hence, the metric is smoothly continued to $-\infty<y<-1/\nu$ across the horizon $y=-1/\nu$. 
Moreover, in the Eddington-Finkelstein coordinate, the gauge potential also remains regular at the horizon $y=-1/\nu$ under the gauge transformation
\begin{align}
 A' =  A \pm d\left( \frac{(1-\nu^2)\Phi_e}{2\nu \kappa} \int \frac{dy}{G(y)}\right),
\end{align}
where $\Phi_e$ is the electric potential defined by
\begin{align}
\Phi_e :=  - (A_t+A_\psi \omega_\psi+A_\phi \omega_\phi)\biggr|_{y=-1} = -\frac{\sqrt{3} c s \left((\gamma -1) (\nu +1) s (b-a) (\gamma
   -\nu )+c \left(d_1-d_2\right)\right)}{(\gamma -1) (\nu +1) s^3
   (b-a) (\gamma -\nu )+c^3 \left(d_1-d_2\right)}.
\label{eq:phi-e}
\end{align}

\item 
At $(x,y)=(-1,-1/\nu)$ and $(x,y)=(1,-1/\nu)$ correspond to the intersecting points with the rotational axes.
By introducing the coordinates $(r,\theta)$ for $(x,y)=(\pm 1,-1/\nu)$ as
\begin{align}
  x = \pm 1 \mp c_\pm r^2 \sin^2\theta,\quad y =-\fr{\nu} \left(1- \frac{(1\mp\nu)c_\pm}{2} r^2 \cos^2\theta\right),\quad c_\pm : {\rm positive \ constants},
\end{align}
and with the use of Eqs.~(\ref{eq:conifree}) and (\ref{eq:lenscon}),
the metric at $r\to 0$ ($(x,y)\to(\pm1,-1/\nu)$) behaves as
\begin{align}
 ds^2 \simeq dr^2 + r^2 d\theta^2 + r^2 \sin^2\theta (d\phi_\pm-\omega_\phi dt)^2  - r^2 \kappa^2 \cos^2 \theta dt^2 + R_{\pm}^2 (d\psi_\pm-\omega_\psi^\pm dt)^2,
\end{align}
where $\omega_\psi^+=\omega_\psi-n\omega_\phi$
, $\omega_\psi^- =\omega_\psi$
, and $(\psi_\pm, \phi_\pm)$ are defined in Eq.~(\ref{eq:defpsipm}).
$R_{\pm}:=\sqrt{g_{\psi\psi}}|_{(x,y)=(\pm1,-1/\nu)}$ are given by
\begin{align}
& R_+ =\frac{2 \ell (-d_1)  (\gamma -\nu ) \sqrt{\nu  (\gamma +\nu )} \left(
c^3
   d_2+b
   s^3 \left(d_2-(\gamma -1) (\nu -1) (\gamma +\nu )\right)\right)}{\sqrt{\left(a^2-1\right) (\gamma -1)} \left(c^2
   d_2^2 (\gamma -\nu )+2 \nu  s^2 \left(d_2-(\gamma -1) (\nu -1)
   (\gamma +\nu )\right){}^2\right)},\label{eq:defRplus}\\
&R_- = \frac{2 \ell \sqrt{\nu  (\gamma +\nu )} \left(c^3
   \left(d_2-d_1\right)-s^3(\gamma -1) (\nu +1)  (b-a) (\gamma
   -\nu )\right)}{(1-\gamma ) \sqrt{\left(1-a^2\right) (\nu +1)}
   \left((\nu +1) s^2 (\gamma -\nu )-2 c^2 (\gamma -1) \nu
   \right)}.   \label{eq:defRminus}
\end{align}
We also set $c_{\pm}$ as 
\begin{align}
 c_+ = \frac{\left(1-a^2\right) \kappa  ((1-\gamma ) (\nu +1))^{3/2}
   R_{+} }{2 (-d_1) \nu \ell^2 },\quad
   c_- = \frac{\kappa  (1-\nu ) R_{-}}{2  \nu  (\nu +1)\ell^2}.
\end{align}
Note that the negativity~(\ref{eq:d1neg}) of $d_1$ ensures that the metric is  Lorentzian at $(x,y)=(1,-1/\nu)$.
In the Cartesian coordinates $(T, X, Y, Z,W)=(\kappa t, r\cos\theta, r\sin\theta \cos(\phi_\pm-\omega_\phi t), r\sin\theta \sin(\phi_\pm-\omega_\phi t),R_{\pm} (\psi_\pm-\omega^\pm_\psi t) )$, the above asymptotic metric becomes
\begin{align}
 ds^2 \simeq - X^2 dT^2 + dX^2 + dY^2 + dZ^2 +dW^2,
 \end{align}
 where the Rindler horizon lines at $X=0$.
Therefore, the metric is regular at $(x,y)=(\pm1,-1/\nu)$.

\end{enumerate}

\subsection{Parameter regions for regularity}\label{sec:paramrange}

Since in the previous subsection, we have shown that there are no curvature singularities at the boundaries $x=\pm 1,y=-1,y=-1/\nu$ of the $C$-metric coordinates under the assumptions of $H(x,y)>0$ and $D>0$ in the coordinate ranges~(\ref{eq:rangexy}),  now we investigate whether they can indeed be positive in the ranges in the ranges~(\ref{eq:rangexy}). 
If $H(x,y)>0$ and $D>0$ everywhere in the ranges~(\ref{eq:rangexy}), curvature singularities do not appear on and outside the horizon. 
This depends on the value of $n$, and hence we classify the analysis into the following three cases: (i) $n=0$ (black ring), (ii) $n=1$ (black hole), (iii) $n\ge 2$ (black lens).
For this purpose, instead of using $H(x,y)$, it is more convenient to use the condition~(\ref{eq:d1neg}), which can be expressed from Eqs.~(\ref{eq:conifree}) and (\ref{eq:lenscon}) as
\begin{align}
d_1 = (1-\gamma)(1-\nu)^2(n-1-a)(n+1-a)<0.\label{eq:nescond1}
\end{align}
This provides a necessary condition for the absence of the surface $H(x,y)=0$, and 
curvature singularities exist if this condition is violated.

\subsubsection{$n=0$ $($a black ring$)$}

For $n=0$, the horizon cross-section has the topology of $S^2 \times S^1$.
In this case, we can show from Eq.~(\ref{eq:nescond1}) that  $d_1=-(1-\gamma)(1-\nu)^2(1-a^2)<0$ is always satisfied in the ranges of $\gamma,\nu,a$~(\ref{eq:paramrange}).
Moreover, we can solve Eqs.~(\ref{eq:nodms}), (\ref{eq:conifree}) and  (\ref{eq:lenscon}) as
\begin{align}
b =0,\quad  \gamma=\frac{\nu(3-\nu)}{1+\nu},\quad \tanh^3 \alpha = a,
\end{align}
which describes the charged dipole black ring as a two-soliton solution obtained in Ref.~\cite{Suzuki:2024coe}. Further details regarding the solution generation and analysis can be found in the reference. Therefore, we do not explore this case further in this paper.

\subsubsection{$n=1$ $($ a capped black hole with $S^3$-horizon topology $)$}

The $n=1$ case describes a regular solution called a ``capped black hole", characterized by a horizon cross-section  with a trivial topology of $S^3$ but the domain of outer communication with a non-trivial topology~\cite{Suzuki:2023nqf}.
In this case, we can show from Eq.~(\ref{eq:nescond1}) that under the condition~(\ref{eq:paramrange}),
\begin{align}
d_1 =- (1-\gamma)(1-\nu)^2a(2-a)<0  \quad \Longleftrightarrow \quad 0<a <1.
\end{align}
For this range of $a$, Eqs.~(\ref{eq:conifree}) and (\ref{eq:lenscon}) can be solved in terms of $\nu$ and $\gamma$ as 
\begin{align}
\begin{split}
&\nu = 1- \frac{2b(1-a^2)^2}{b\left(2a^2 (a-1)^2+1\right) +a\left((a-1)^3-1\right) },\\
&\gamma =\nu+\frac{(1-\nu)  \left(1-a+a^2 \right)}{1-(1+2b)a+(1+b)a^2},
\end{split}
\label{eq:Bd-sol}
\end{align}
from which we can show 
\begin{align}
0<\nu<\gamma<1 \Longleftrightarrow
\left\{ \begin{array}{ll} 
\displaystyle \frac{a(a-2)(1-a+a^2)}{(1-2a)(1+2a-2a^2)}<b<0 & (0<a<a_*)\\
-1<b<0 &(a_* \leq a <1) 
\end{array},\right.
\label{eq:a,b}
\end{align}
where $a_*=0.347\ldots$ is a root of $a^3-3a+1=0$.
From Eq.~(\ref{eq:nodms}), this also restricts the range of $\alpha$ to be positive.

\medskip

From Eq.~(\ref{eq:Bd-sol}), it is straightforward to show $H(x,y)>0$ in the coordinate ranges~(\ref{eq:rangexy}) in the parameter range~(\ref{eq:a,b})
by writing $H(x,y)$ in the following form~\cite{Suzuki:2023nqf}
\begin{align}
H(x,y)&=\biggl[\left\{ \nu  b^2 c_1^2  (1+\nu )^2(\gamma -\nu ) (1-\gamma )  (1-x^2)
+ \nu(1-\gamma )   (\gamma -\nu ) \left(b (1-\nu )^2 (1-x)-2 c_1 (1+\nu  x)\right)^2\right.\nonum
&+ b^2 c_1^2 (1+\nu )^3 (x+1) (\gamma -\nu )^2 \bigr\}(1+y)^2\biggr]
+\biggl[\bigl\{d_5  (1-x)^2+d_6(1-x^2)+ d_7(1+x) (1+\nu  x) \bigr\}(-1-y)\biggr]\nonum
& + \left[4 \left(1-a^2\right) (1-\gamma )^3 (1-\nu )^4 (1-x)
 +2c_2^2  (1-\gamma )  (1-\nu )^2   \left(1-x^2\right)
-4d_1  (1-\gamma ) (1-\nu)^2 (1+\nu ) (1+x)\right],\label{eq:H2}
\end{align}
with three extra auxiliary parameters:
\begin{align}
&d_5 := (1-\gamma ) (1-\nu )^3 [(\gamma -3 \nu ) \left(b^2 (\nu -1) (\gamma -\nu )-c_1^2\right)-2 b c_1 (3 \nu -1) (\gamma -\nu )],\nonum
&\frac{d_6}{1-\nu}:=\frac{d_7}{2}  := c_1(1-\gamma ) \left(1-\nu ^2\right) \left(c_2-(1-\gamma ) (a-b) (\gamma -\nu )\right).\label{eq:param-extra}
\end{align}
It is evident that the first and third square brackets in Eq.~(\ref{eq:H2})  are non-negative.
It can be shown from Appendix~\ref{app:hxy} that $d_5,d_6$ and $d_7$ in the second square bracket  are positive, hence as a result,  the second square bracket is also non-negative.
Thus, we can show that all three terms enclosed in a square bracket in Eq.~(\ref{eq:H2}) are nonnegative, and hence $H(x,y)$ is non-negative. 
Moreover, we can prove a stronger statement, $H(x,y)>0$, because 
none of the three square brackets can be zero simultaneously.
Having shown $H(x,y)>0$, 
the positivity of $D$ follows from $D=1+s^2(H(x,y)-H(y,x))/H(x,y)$ and
\begin{align}
&H(x,y)-H(y,x) = \frac{(1-\nu)(\gamma+\nu)(x-y)}{1+\nu}
\biggr(c_3(1-\gamma)(1-\nu)^3(1-x)(1-y)\nonum
&\quad + c_1^2 (1+\nu)^2(1+x)(-1-y)+ 2(-d_1)(1-\gamma)(1-\nu)[(1+\nu)(1+x)+(1+\nu x)(1-x)]\biggr)\geq 0,
\end{align}
where $c_3>0$ is obvious from the definition~(\ref{eq:def-cfs}).
\medskip

\begin{figure}
\includegraphics[width=7cm]{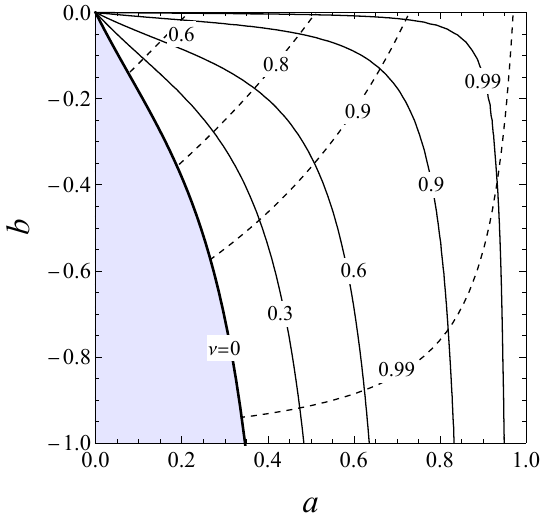}
\caption{
The parameter region for a regular capped black hole ($n=1$) given in Eq.~(\ref{eq:a,b}). 
The thick and dashed curves correspond to $\nu = {\rm constant}$ and $\tanh\alpha\ = {\rm constant}$, respectively. 
There is no regular black hole in the blue-colored region below $\nu=0$. \label{fig:paramregion}}
\end{figure}

Moreover, we can demonstrate that this regular solution does not permit the presence of CTCs both on and outside the horizon. To show this  explicitly, we need to ensure that the two-dimensional part $g_{IJ}$ (where $I,J=\psi,\phi$) of the metric~(\ref{eq:charged-metric}) is positive definite on and outside the horizon, except on the axes at $x=\pm 1$ and $y=-1$, i.e., ${\rm det}(g_{IJ})>0$ and ${\rm tr}(g_{IJ})>0$ for $-1 < x<1$ and $-1/\nu \leq y <-1$. Following the same reasoning as in the case of the charged dipole black ring discussed in Ref.~\cite{Suzuki:2024coe}, it suffices to demonstrate ${\rm det}(g_{IJ})>0$ for the ranges $-1 < x<1$ and $-1/\nu \leq y <-1$. This can be reduced to proving the positivity of $\Delta(x,y)$ defined by:
\begin{align}
\Delta(x,y) := -\frac{(1+\nu x) D H(x,y)}{\ell^4(1-x^2)(1+y)} {\rm det}(g_{IJ}).
\end{align}

\begin{figure}
\includegraphics[width=6cm]{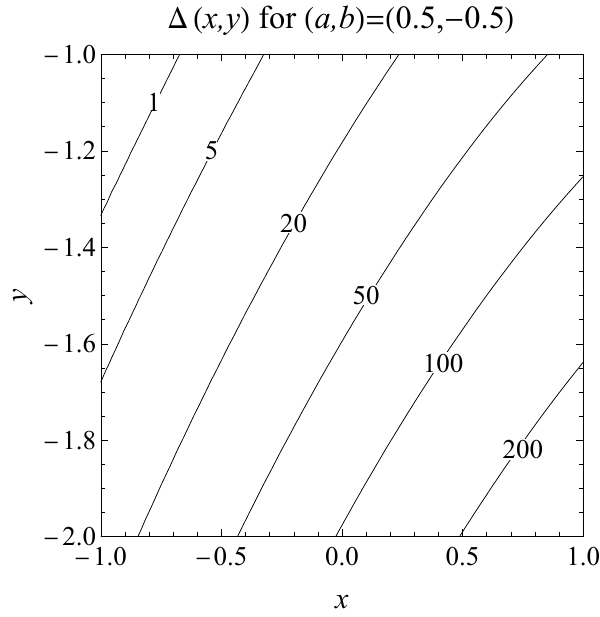}
\caption{Profile of ${\rm det}(g_{IJ})$ for parameters given by Eqs.~(\ref{eq:nodms}) and  (\ref{eq:Bd-sol}) with $(a,b)=(0.5,-0.5)$.
 One can obtain similar profiles for other sets of $(a,b)$ in the range~(\ref{eq:a,b}).\label{fig:ctcs}}
\end{figure}

One can easily observe the positivity of this quantity at infinity $(x,y)=(-1,-1)$ since the spacetime approaches the Minkowski metric in that limit. 
Additionally, around $(x,y)=(1,-1)$, one can show the positivity from the condition $d_1<0$, as expressed by
\begin{align}
\Delta(1,-1)=-4 d_1 (1-\nu )^3 (1+\nu ) \left(1+c^2 \nu +s^2\gamma \right)^3,
\end{align}
where we have also used Eq.~(\ref{eq:nodms}). The positivity on the horizon follows from
\begin{align}
\Delta(x,-1/\nu) = \frac{4 c_3^2 (1-\nu )^3 (1+\nu ) (\gamma +\nu ) \left(c^3
   \left(a c_1+\nu -1\right)-s^3 (b-a) (\gamma -\nu
   )\right){}^2}{\left(1-a^2\right)^2 (1-\gamma )}>0.
\end{align}
For other regions, proving the positivity analytically is challenging. Instead, we have numerically verified the positivity for several values in the parameter region~(\ref{eq:a,b}) (Fig.~\ref{fig:ctcs}).

\subsubsection{$n\geq 2$ {\rm (a black lens with $L(n;1)$-horizon topology) }}

For $n\geq 2$, we find $d_1>0$ from Eq.~(\ref{eq:nescond1}) in the range~(\ref{eq:paramrange}), which is contrary to the condition~(\ref{eq:d1neg}).
Therefore, this rules out the possibility of a black lens with a horizon of lens topology $L(n;1)$ for $n\geq 2$ as a regular solution. However, if we relax the condition~(\ref{eq:conifree}) and allow for a conical singularity at $x=1$, we can find parameter ranges without curvature singularities and CTCs, resembling the vacuum case discussed in Ref.~\cite{Chen:2008fa}.

\section{Capped black hole}\label{sec:phase}

\begin{figure}
\includegraphics[height=4cm]{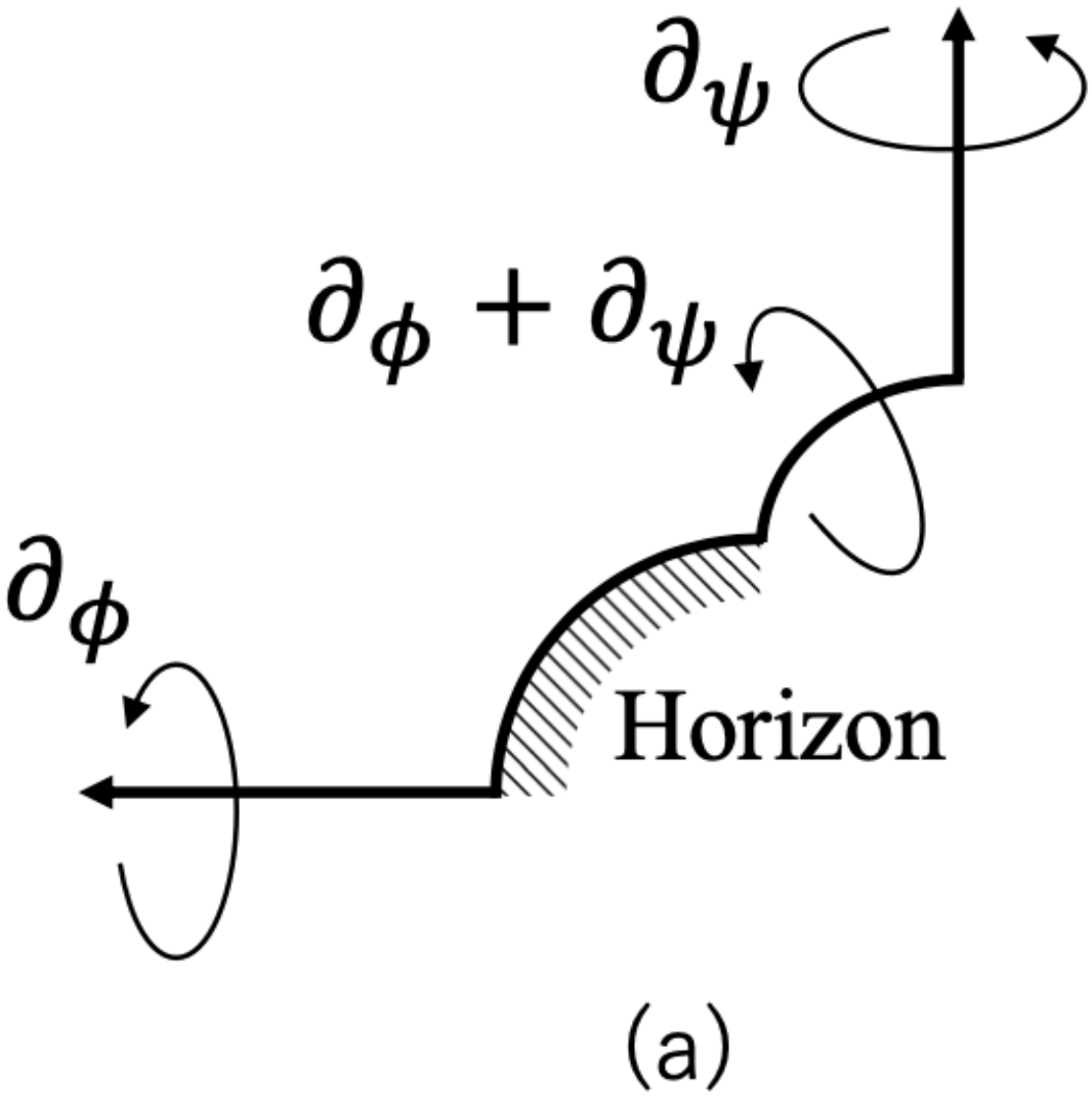}
\hspace{1cm}
\includegraphics[height=4cm]{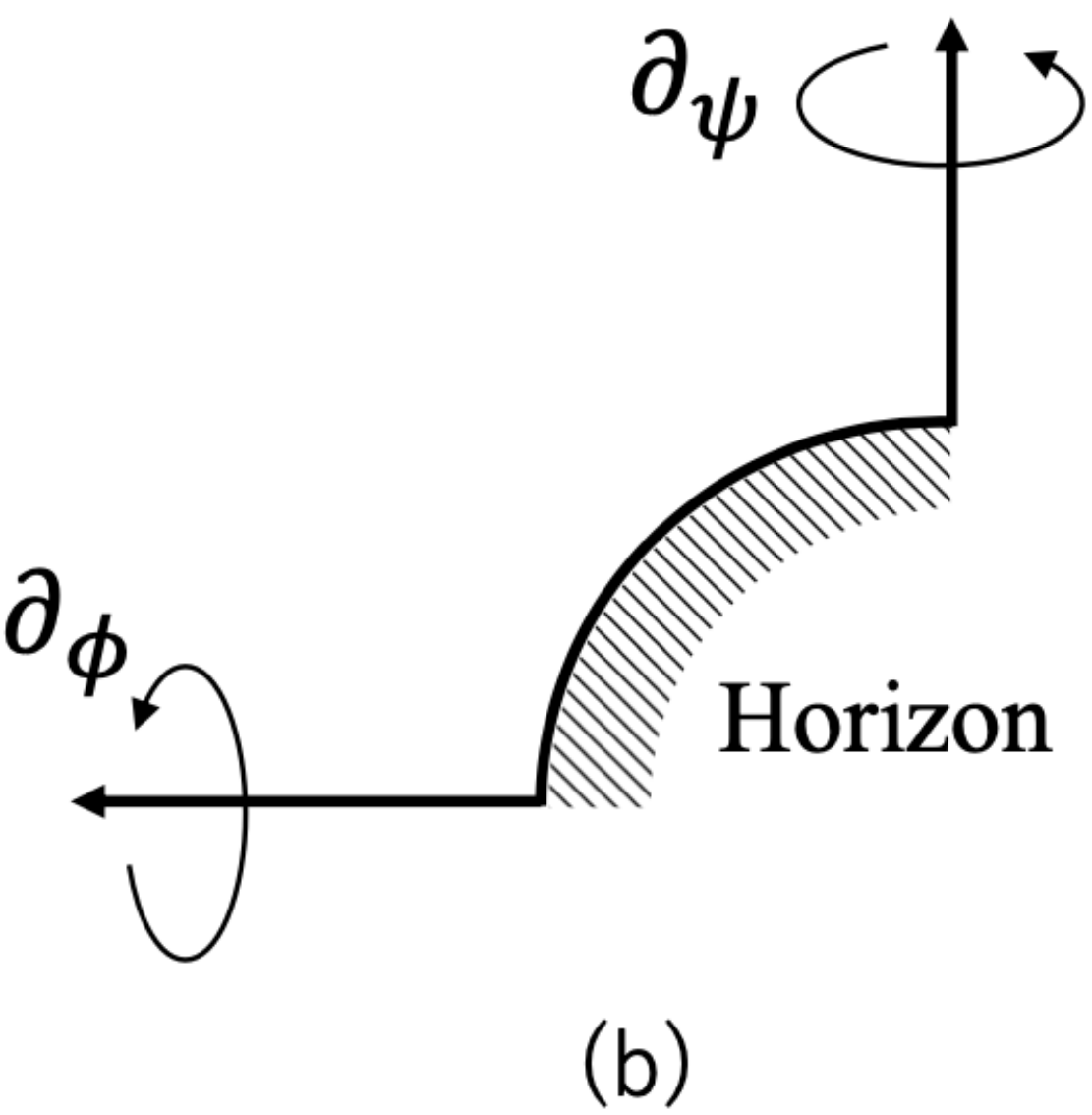}
\caption{The orbit space ${\cal M}^{(1,4)}/[{\mathbb R} \times U(1)\times U(1)]$ of (a) the capped black hole and (b) the Cveti\v{c}-Youm black hole.\label{eq:orbits}}
\end{figure}

For $n=1$, we obtain a spherical black hole that has a nontrivial external structure for the parameter range in Eqs.~(\ref{eq:Bd-sol}) and (\ref{eq:a,b})~\cite{Suzuki:2023nqf}. In Fig.~\ref{eq:orbits}, we illustrate the orbit spaces of our spherical black hole and known spherical black hole (Cveti\v{c}-Youm black hole). 
The inner axis at $x=1$ for $-1/\nu \leq y \leq -1$ has a disc topology since a $S^1$ generated by $\partial/\partial \psi$ has a fixed point at $y=-1$ but not at $y=-1/\nu$. Hence, the horizon is capped by a disc-shaped bubble at a pole and the solution is called `capped black hole'.
Below, we study the physical properties of the capped black hole.

\medskip

 As one can see from Fig.~\ref{fig:paramregion}, 
 besides the scale parameter $\ell$, the regular solution is 
 characterized by two independent parameters $(\nu,\tanh\alpha)\in (0,1) \times (0,1)$. We will see that $\nu$ controls the relative size of the disc-shaped bubble and horizon and $\tanh\alpha$ indicates the amount of the electric charge. Note that the metric is no longer Lorentzian on the boundary of the parameter region.

\subsection{Physical quantities}
As shown previously  in Eq.~(\ref{eq:aslimds}), the spacetime is asymptotically flat, and hence  
the ADM mass $M$ and two ADM angular momenta $J_\psi,J_\phi$ can be read off from the asymptotic of the metric at $r\to\infty$ in the coordinates~(\ref{eq:aslim}), 
\begin{align}
& ds^2 = -\left(1-\frac{8G_5 M}{3\pi r^2}\right)dt^2 - \frac{8G_5J_\psi \sin^2\theta}{\pi r^2}dt d\psi- \frac{8G_5J_\phi \cos^2\theta}{\pi r^2}dt d\phi \nonum
&\hspace{1cm}+dr^2+ r^2 \sin^2\theta d\psi^2+r^2\cos^2\theta d\phi^2 +r^2 d\theta^2,
\label{eq:aslimit}
\end{align}
with 
\begin{align}
 &M=\frac{3 \pi  \ell^2 (1+2 s^2) (\gamma +\nu ) (c_3
   (1-\nu )-d_1)}{4 G_5 \left(1-a^2\right) (1-\gamma )^2 (1+\nu)},\\
   & J_\psi = c^3 J_1 + s^3 J_2,\\
   & J_\phi = c^3 J_2 + s^3 J_1,
\end{align}
where $J_1$ and $J_2$ correspond to the angular momenta in the neutral case around the $\psi$-rotational axis and $\phi$-rotational axis, respectively, given by
\begin{align}
   &J_1=\frac{\pi  \ell^3 v_0 \left(c_3 d_2-c_1   c_2 c_3-(c_3-b c_2) d_1\right)}{4G_5 \left(1-a^2\right) (1-\gamma )^3 \nu },\\
&J_2 =\frac{\pi  \ell^3 v_0 (a-b) \left(2 c_3 \nu +d_1\right)}{2G_5
   \left(1-a^2\right) (1-\gamma ) (1+\nu )}.
 \end{align}
Additionally, the electric charge is defined by the integration over a three-dimensional closed surface $S$ surrounding a horizon and a bubble:
 \begin{align}
&Q := \fr{8\pi G_5} \int_{S} \left(\star F+\frac{1}{\sqrt{3}}F\wedge A \right)\nonum
&= \fr{8\pi G_5} \int_{S_\infty} \star F\nonum
&=- \frac{2\tanh\alpha}{\sqrt{3}}M, 
\end{align}
where in the second equality, we have used the fact that the Chern-Simons term falls much faster at the asymptotic limit.
This obviously
follows the Bogomol'nyi bound $M \geq \frac{\sqrt{3}}{2} Q$, which is saturated at the limit $\alpha\to\infty$.

\medskip

It is worth emphasizing that the electric charge evaluated over the three-dimensional surface $S_\infty$ at infinity does not coincide with one evaluated over the spatial cross section of the horizon $S_{\cal H}$. The rest of the contribution comes from the three-dimensional surface surrounding the disc-shaped bubble $\mathcal{D}$. By direct computation, one can confirm that
\begin{align}
Q &:= \fr{8\pi G_5} \int_{S_\infty} \left(*F + \fr{\sqrt{3}} F\wedge A\right)\\
&= \fr{8\pi G_5} \int_{S_{\cal H}} \left(*F + \fr{\sqrt{3}} F\wedge A\right)+\fr{8\pi G_5} \int_{{\cal D}_{\epsilon \to 0}} \left(*F + \fr{\sqrt{3}} F\wedge A\right),
\end{align}
where $\cD_\epsilon$ denotes the three-dimensional surface at $x=1-\epsilon$ for $-1/\nu\leq y\leq -1$.
This is same as the electric charge for the charged dipole black ring~\cite{Suzuki:2024coe}.

\medskip
In addition to these conserved charges, one can define the magnetic flux (this is not  a conserved charge) over the disc-shaped bubble $\cD$ at $x=1$ as
 \begin{align}
  q :&= \fr{4\pi} \int_{\cal D} F = \fr{2} A_\psi(x=1,y=-1/\nu)\nonum
  &=\frac{\sqrt{3} \ell s c d_1  v_0 (b d_2 s (\gamma -\nu )-2 c \nu 
   \tilde{d}_2)}{2
   (c^2 d_2^2 (\gamma -\nu )+2 \nu  s^2 {\tilde{d}_2}^2)}.\label{eq:magnetic-f-q}
\end{align}
where $\tilde{d}_2:=d_2-(\gamma -1) (\nu
   -1) (\gamma +\nu)$.
The area of a constant timeslice through the horizon is written as
\begin{align}
  A_H = 8 \pi^2\ell^2 \nu \kappa^{-1},
\end{align}
where the surface gravity $\kappa$ is given by Eq.~(\ref{eq:kappa}).

\begin{figure}
\begin{center}
\includegraphics[width=7cm]{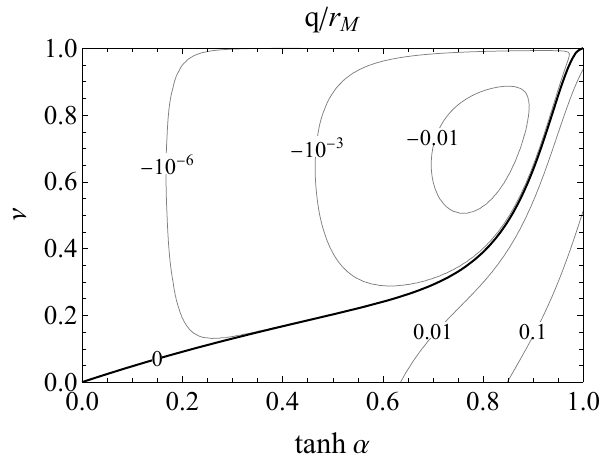}
\includegraphics[width=7cm]{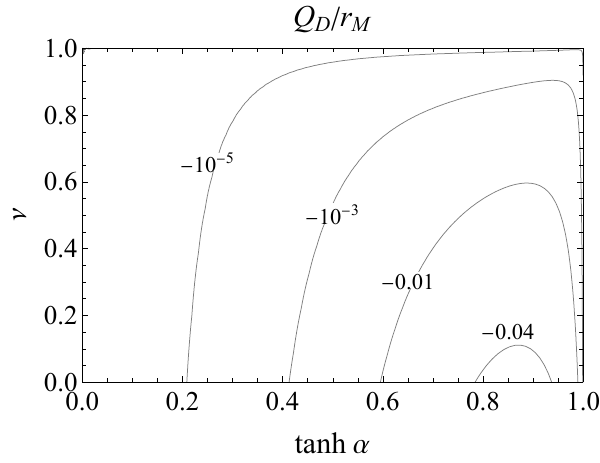}
\caption{Profiles for magnetic fluxes $q$ and ${\cal Q}_\cD$ in the $(\nu,\tanh\alpha)$-plane. \label{fig:fluxplot}}
\end{center}
\end{figure}

\medskip
In Ref.~\cite{Kunduri:2013vka}, it is shown that the black hole with a disc-shaped bubble must follow the first law
\begin{align}
 \delta M = \frac{\kappa }{8\pi} \delta A_H  + V_\psi \delta J_\psi + V_\phi \delta J_\phi
 + \fr{2} \Phi_H \delta Q + {\cal Q}_\cD \delta \Phi_\cD,
\end{align}
and the Smarr formula
\begin{align}
 M = \frac{3\kappa A_H}{16\pi}  + \frac{3}{2} V_\psi  J_\psi + \frac{3}{2}V_\phi J_\phi
 + \fr{2} \Phi_H Q  + \fr{2} {\cal Q}_\cD\Phi_\cD,
\end{align}
where $\Phi_H$ is the electric potential~(\ref{eq:phi-e}), $\Phi_\cD$ and ${\cal Q}_\cD$ are the magnetic potential and another type of a magnetic flux  on the bubble $\cD$, which are defined as
\begin{align}
 \Phi_\cD& := -(A_\phi+n A_\psi)\biggr|_{x=1}  = -\sqrt{3} c  s \ell v_0 (c (a-b)-s (1-a b)),\\
 {\cal Q}_\cD &:= \fr{4}\int_\cD \left[\iota_{v_H} (\star F)- \fr{\sqrt{3}} (\Phi-\Phi_H) F\right]  = \frac{\sqrt{3} \pi  \left(1-a^2\right) b c d_1 \ell s v_0}{4 c_3 \left(c^3 \left(a c_1+\nu -1\right)-s^3
   (b-a) (\gamma -\nu )\right)}.
\end{align}
As depicted in Fig.\ref{fig:fluxplot}, the two magnetic fluxes $q$ and ${\cal Q}_\cD$ differ in general.
Specifically, $q$ can vanish even with the existence of the bubble, while ${\cal Q}_\cD$ is negative definite.
When considering the horizon velocity in Eq.(\ref{eq:omegai}), one can verify that the capped black hole adheres to the aforementioned first law and Smarr formula by expressing each variable as a function of $(\ell,a,b)$ using Eqs.~(\ref{eq:nodms}) and (\ref{eq:Bd-sol}).

\begin{figure}
\includegraphics[width=5.8cm]{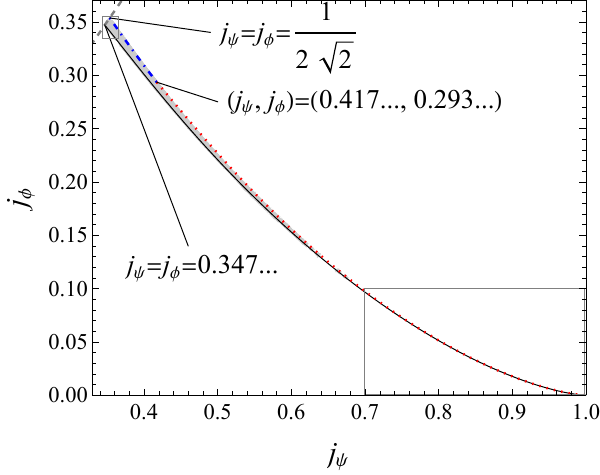}
\includegraphics[width=6cm]{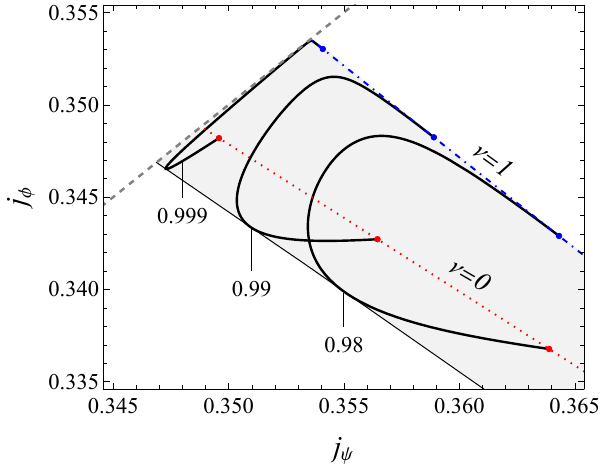}
\includegraphics[width=5.8cm]{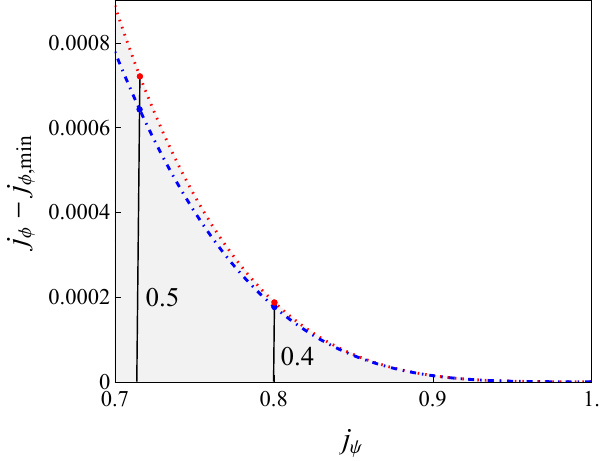}
\caption{
The allowed region for $j_\psi$ and $j_\phi$ and curves of constant $\tanh \alpha$ are displayed in the $(j_\psi,j_\phi)$ plane.
The allowed region is illustrated by the colored region. 
Each curve of constant $\tanh\alpha$ starts at the $\nu=0$ curve (blue dot-dashed) and ends at the $\nu=1$ curve (red dotted). 
In the middle panel, we show a close-up of the phases for $\tanh\alpha \approx 1$.
At the limit $\alpha \to 0$, the curves of constant $\tanh \alpha={\rm const.}$ converge around $(j_\psi,j_\phi)=(1,0)$, 
and the allowed region exists in a very narrow range. 
From the right panel, one can observe that this range becomes narrower and narrower as $j_\psi$ approaches $1$.
The plot depicts two angular momenta and several curves of constant $\tanh\alpha$ accompanied by the value of $\tanh\alpha$.\label{fig:jjplot}
 }
\end{figure}

\medskip
To discuss the phase space of the capped black hole, let us introduce the angular momenta and horizon area normalized by the  mass scale $r_M :=\sqrt{8 G_5 M/3\pi}$  as:
\begin{align}
j_\psi := \frac{4G_5}{\pi r_M^3} J_\psi,\quad
j_\phi := \frac{4G_5}{\pi r_M^3} J_\phi,\quad
a_H := \frac{\sqrt{2}}{\pi^2 r_M^3}A_H.\label{eq:capped-charges}
\end{align}
Figure~\ref{fig:jjplot} provides insight into the allowed region for the angular momenta in the $(j_\psi,j_\phi)$ plane. 
This figure reveals several important physical characteristics of the capped black hole:
\begin{itemize}
\item[(i)] 

The allowed range for angular momenta is constrained such that $0 < j_\phi < 1/(2\sqrt{2}) )\approx  0.353...$ and $0.347... < j_\psi < 1$. Compared to the Cveti\v{c}-Youm black hole~\cite{Cvetic:1996kv}, this region is notably narrower.

\item[(ii)] 

The solid curves represent different values of $\tanh\alpha$ (the electric charge), with each curve having endpoints at $\nu=0$ and $\nu=1$.

\item[(iii)]

The allowed region is bounded by the dashed line $j_\psi=j_\phi$, which can only be reached at the BPS limit $M=\sqrt{3}Q/2$ as $\alpha\to \infty$. However, it is important to note that the metric is not Lorentzian at the BPS limit. Therefore, the capped black hole does not admit equal angular momenta $j_\psi=j_\phi$.

\item[(iv)]

 Since each $\tanh\alpha={\rm const.}$ curve, as seen in the middle panel, is not closed, the capped black hole is uniquely specified by the conserved charges of its mass, two angular momenta, and electric charge. This implies that there is no continuous family of solutions parametrized by the magnetic flux $q$, highlighting the uniqueness of the capped black hole configuration.

\end{itemize}

\medskip

\subsection{Non-uniqueness of spherical black holes}
Now, we compare the capped black hole with the Cveti\v{c}-Youm solution~\cite{Cvetic:1996xz,Cvetic:1996kv} having the same conserved charges, the mass,  angular momenta and  electric charge. 
 The normalized angular momenta and the normalized horizon area for the Cveti\v{c}-Youm solution are give by, respectively, 
 \begin{align}
     j_\psi^{\rm CY} = \frac{c^3 j_1+ s^3 j_2}{(1+2s^2)^\frac{3}{2}},
     \quad 
          j_\phi^{\rm CY} = \frac{c^3 j_2+ s^3 j_1}{(1+2s^2)^\frac{3}{2}}, \quad 
a_H^{\rm CY} = \frac{\sqrt{2}\left[\sqrt{1-(j_1+j_2)^2}(c^3+s^3)+\sqrt{1-(j_1-j_2)^2}(c^3 -s^3)\right]}{(1+2s^2)^\frac{3}{2}},
\end{align}
where $\alpha$ in $(c,s)=(\cosh\alpha,\sinh\alpha)$ is the same parameter as in the capped black hole, and $j_1,j_2$ are the dimensionless parameters for the angular momenta.
To match the angular momenta of the Cveti\v{c}-Youm black hole with those of the capped black hole in Eq.~(\ref{eq:capped-charges}),
we set
\begin{align}
 j_1 = (1+2s^2)^{3/2}\frac{J_1}{r_M^3},\quad j_2 = (1+2s^2)^{3/2} \frac{J_2}{r_M^3}. \label{eq:match-j1j2}
\end{align}
Here we note that the Cveti\v{c}-Youm black hole reaches the extremal limit, which does not coincide with the BPS limit when $|j_1+j_2|=1$ or $|j_1-j_2|=1$~\cite{Cvetic:1996kv}.
Hence, the Cveti\v{c}-Youm black hole cannot match the capped black hole if $j_1$ and $j_2$ in Eq.~(\ref{eq:match-j1j2}) excess the bounds $|j_1+j_2|\leq 1$ and $|j_1-j_2|\leq1$.

\medskip

\begin{figure}
\includegraphics[width=7cm]{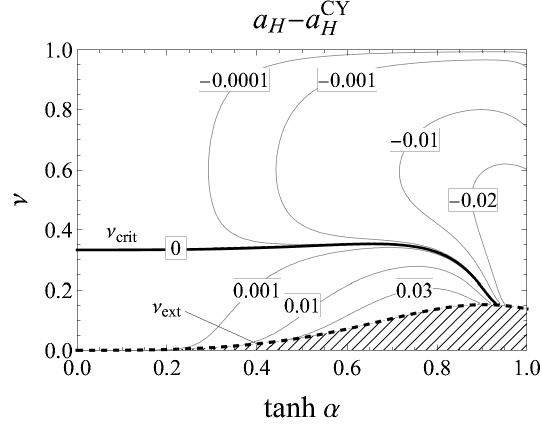}
\caption{The area of the horizon cross-section of the capped black hole  compared with that of  the Cveti\v{c}-Youm black hole of the same $(j_\psi,j_\phi)$ in the $(\nu,\tanh\alpha)$. \label{fig:areaplot}}
\end{figure}

Here are the observations from Fig.~\ref{fig:areaplot} regarding the difference in the horizon area of the two phases in the ($\nu$,$\tanh\alpha$)-plane:

\begin{itemize}
\item[(a)] For a fixed $\alpha$, the corresponding Cveti\v{c}-Youm black hole becomes extremal at $\nu=\nu_{\rm ext}(\alpha)$ for $|j_1+j_2|=0$, and has the same horizon area at $\nu=\nu_{\rm crit}(\alpha)$.
\item[(b)]   For $0<\nu<\nu_{\rm ext}(\alpha)$ (the shaded region) and a fixed $\alpha$, there is no Cveti\v{c}-Youm black hole corresponding to the capped black hole with the same angular momenta because of $|j_1+j_2|>1$,
\item[(c)] For $\nu_{\rm ext}(\alpha) < \nu < \nu_{\rm crit}(\alpha)$ and a fixed $\alpha$ ($\tanh\alpha < 0.940\ldots$), the capped black hole has larger entropy and hence thermodynamically more stable than the Cveti\v{c}-Youm black hole, and for $\nu > \nu_{\rm crit}(\alpha)$, on the other hand, the Cveti\v{c}-Youm black hole is more stable.
 \end{itemize}

\begin{figure}[h]
\begin{center}
\includegraphics[width=7cm]{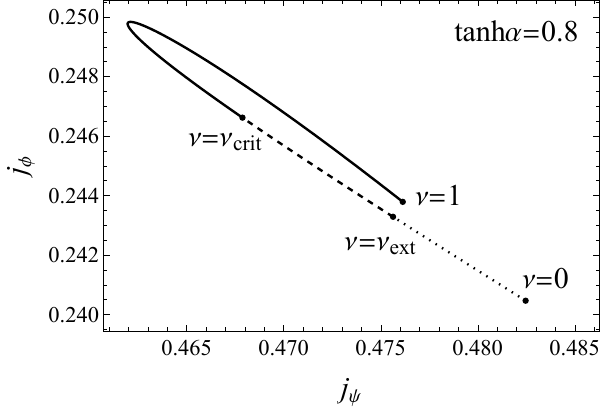}
\includegraphics[width=7cm]{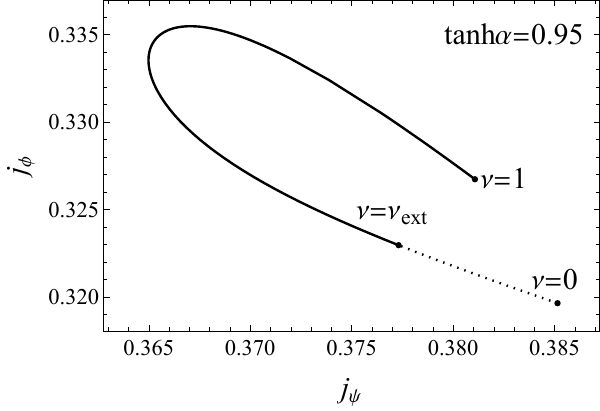}
\caption{The area of the horizon cross-section of the capped black hole  compared with that of the Cveti\v{c}-Youm black hole on the contour in $(j_\psi,j_\phi)$ plane for some fixed $\alpha$ ($\tanh\alpha=0.8,0.95$). The Cveti\v{c}-Youm black hole is favored on the thick curve, while ours is favored on the dashed. On the dotted curve, the Cveti\v{c}-Youm black hole does not exist.  \label{fig:jjplot2}}
\end{center}
\end{figure}

As depicted in Fig.~\ref{fig:jjplot2}, 
for $\tanh \alpha < 0.940...$, each curve of constant $\tanh\alpha$ can be segmented into three parts:  (i) $0<\nu< \nu_{\rm ext}(\alpha)$, (ii) $\nu_{\rm ext}(\alpha)<\nu< \nu_{\rm crit}(\alpha)$ and (iii) $\nu_{\rm crit}(\alpha)<\nu< 1$, while for $\tanh \alpha > 0.940...$, there are only two segments without the range $\nu_{\rm ext}(\alpha)<\nu<\nu_{\rm crit}(\alpha)$.

\subsection{Size of horizon and bubble}

\begin{figure}

\begin{minipage}{0.45\columnwidth}
\includegraphics[width=6.5cm]{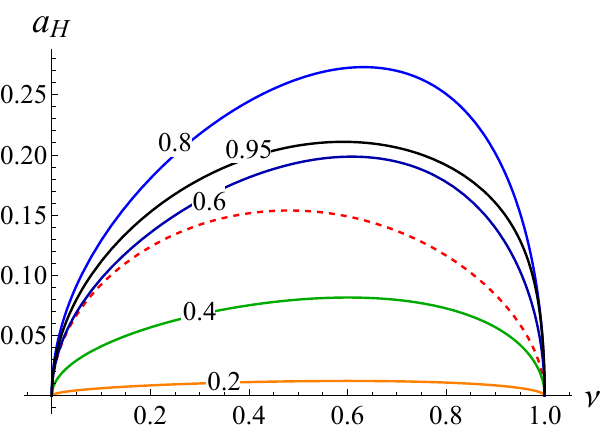}
\includegraphics[width=6.5cm]{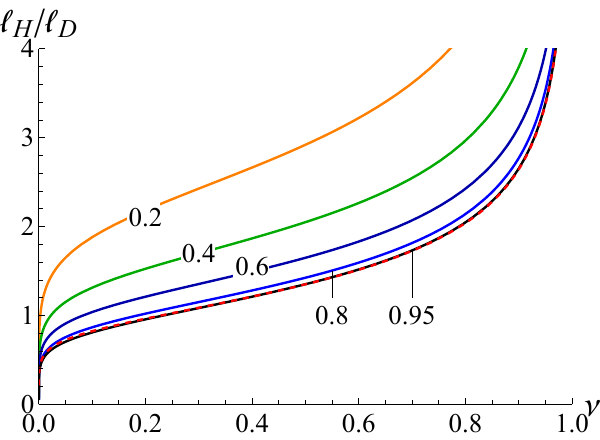}
\end{minipage}
\begin{minipage}{0.45\columnwidth}
\includegraphics[width=6.5cm]{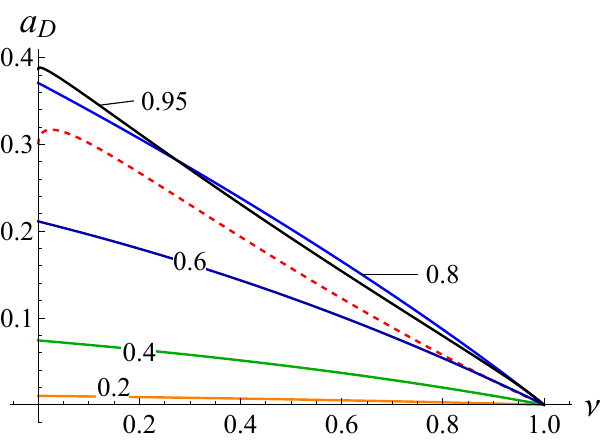}
\includegraphics[width=6.5cm]{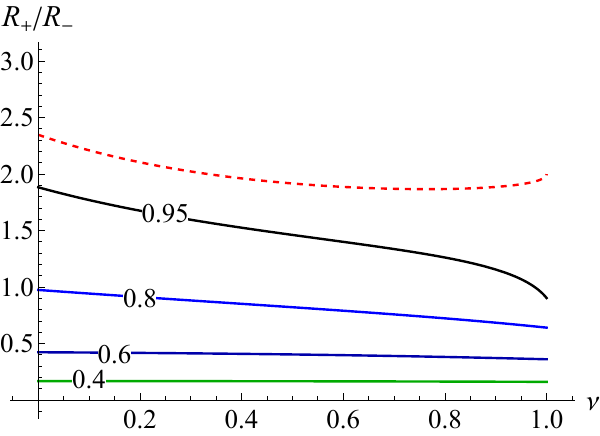}
\end{minipage}
\caption{
$\nu$-dependence of horizon and bubble scales for each $\tanh\alpha$. The value of $\tanh\alpha$ is shown with each curve. The red dashed curves correspond to the limit curve at $\tanh\alpha \to 1$. \label{fig:scales}
}
\end{figure}

As one might expect from the rod structure depicted in Fig.\ref{fig:rodtrans}, the parameter $\nu$ indicates the size of the bubble, with smaller values of $\nu$ indicating larger bubbles. 
The horizon disappears as $\nu$ approaches 0, while the bubble disappears as $\nu$ tends to 1. 
To illustrate this characteristic, we introduce two scales that represent the size of the horizon and the bubble. 
Analogous to the horizon area, one can calculate the two-dimensional area of the bubble as:
\begin{align}
a_D:= \frac{A_D}{\pi r_M^2} = \frac{2}{r_M^2} \int_{-1/\nu}^{-1} \sqrt{g_{yy}g_{\psi\psi}}\biggr|_{x=1} dy,
\end{align}
where $a_D$ is the mass-normalized bubble area.
As shown in the top panels of Fig.~\ref{fig:scales}, the horizon area vanishes as $\nu\to0$, while the bubble area remains finite. Although both areas vanish as $\nu\to1$, a comparison of the area scales defined by $\ell_H := a_H^{1/3}$ and $\ell_D := a_D^{1/2}$ reveals that the bubble scale decreases more rapidly, with $\ell_H/\ell_D \sim (1-\nu)^{-1/3}$ (see also the bottom left panel in Fig.~\ref{fig:scales}).

\medskip

To investigate the distortion of the $S^3$-horizon, we also compare $R_\pm$ defined in Eqs.~(\ref{eq:defRplus}) and (\ref{eq:defRminus}), which estimate the radial scales of the horizon at each pole. From the bottom right panel of Fig.~\ref{fig:scales}, it is evident that the ratio between $R_+$ and $R_-$ remains finite even as $\nu\to1$, indicating that the horizon shape does not collapse in the limit. If the electric charge is sufficiently small, we observe $R_+/R_- \to 0$, suggesting that the horizon shape becomes elongated along the $\psi$-rotation plane.

\subsection{Ergoregion}\label{sec:ergo}

\begin{figure}
\includegraphics[height=5cm]{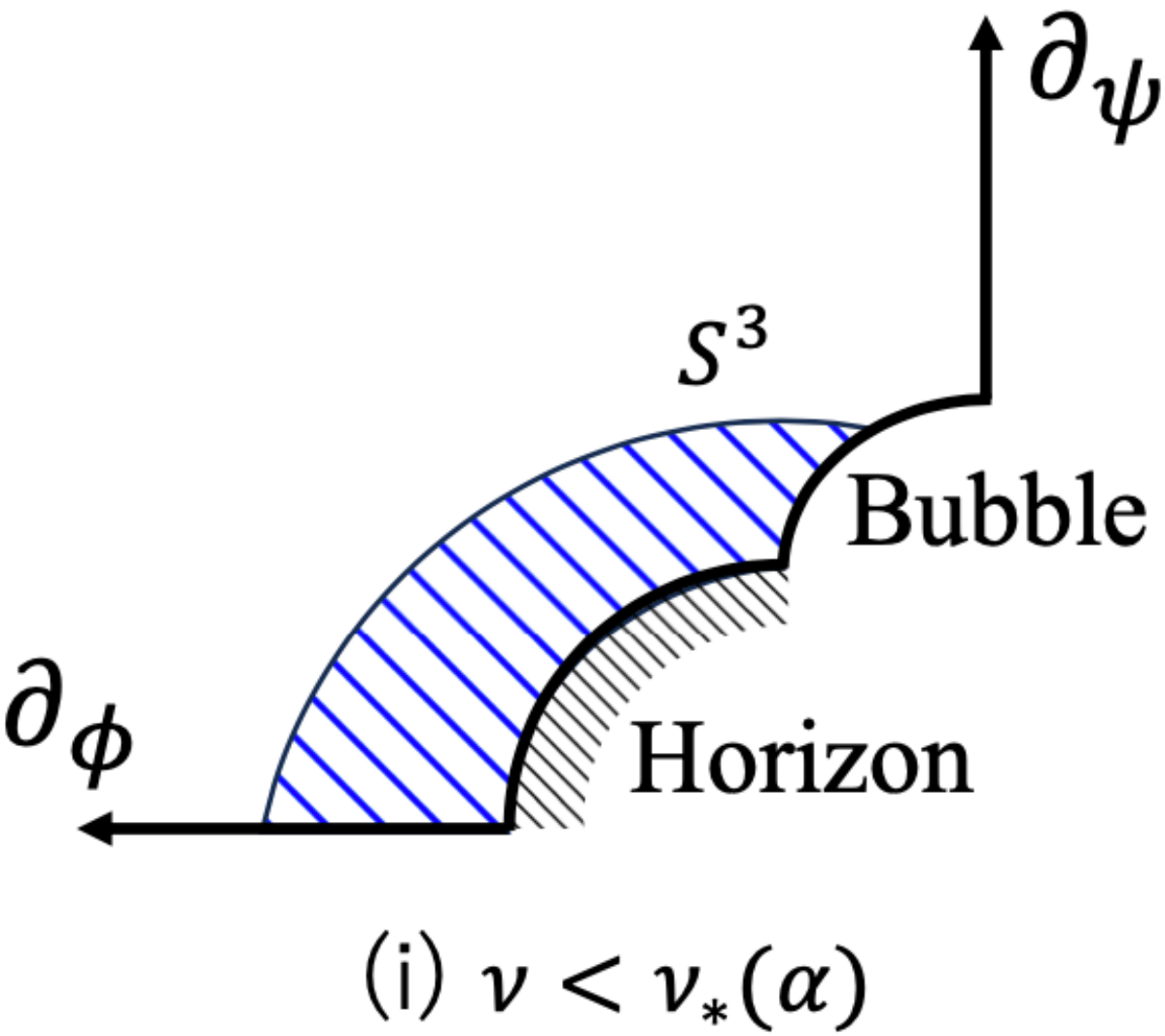}
\includegraphics[height=5cm]{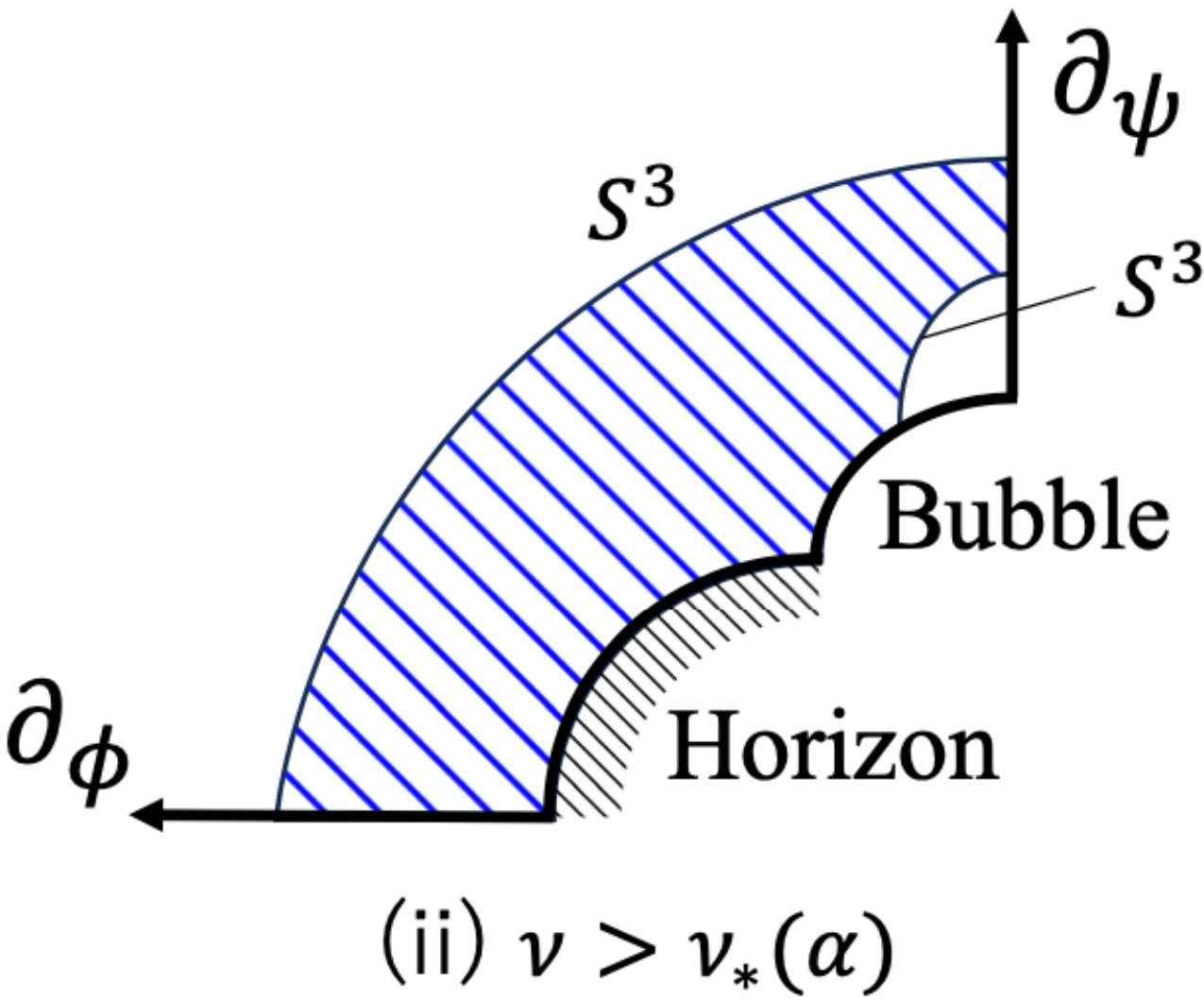}
\caption{Possible shapes for the ergoregion:  The ergoregions are illustrated by the blue hatched pattern in the orbit space for two cases, (a) $\nu<\nu_*(\alpha)$ and (b) $\nu>\nu_*(\alpha)$.\label{fig:ergo}}
\end{figure}

\begin{figure}
\includegraphics[width=7cm]{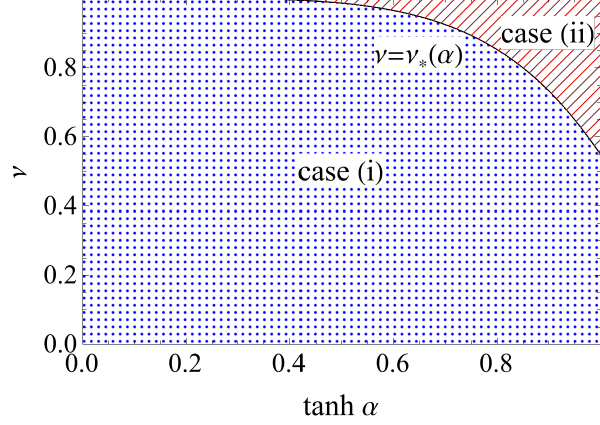}
\caption{The topology change of the ergoregion in the phase diagram. \label{fig:ergo-phase}}
\end{figure}

The ergoregion of the capped black hole~(\ref{eq:neutral-sol-cmetric}) is determined by the condition $H(y,x)<0$.
To comprehend the presence of the ergoregion,  it is convenient to use the following properties for the parameters within the range~(\ref{eq:a,b}) together with Eq.~(\ref{eq:Bd-sol}) (refer to Appendix \ref{app:ergo} for the proof):

\begin{enumerate}[(a)]
\item $H(y=-1,x=\pm 1)>0$,
\item $H(y=-1/\nu,x)<0 \ {\rm for}\ x\in [-1,1]$,
\item $\partial_x^2 H(y=-1,x)>0 \ {\rm for}\ x\in [-1,1]$,
\item $\partial_y^2 H(y,x)<0 \ {\rm for}\  (x,y) \in [-1,1]\times [-1/\nu,-1]$,
\item $\partial_y H(y=-1,x) + H(y=-1,x)>0\  {\rm for}\ x\in[-1,1]$.
\end{enumerate}
(a) indicates that both the asymptotic infinity and the intersection of the inner rotational axis and the $\psi$-rotational axis always lie outside the ergoregion, while (b) illustrates that the horizon is invariably situated inside the ergoregion.
(c) demonstrates that $H(y=-1,x)$ is a concave function of $x$, which, combined with (a),  leads to the following two potential behaviors on the $\psi$-rotational axis at $y=-1$:

\begin{itemize}
\item $H(y=-1,x) >0 \ {\rm for}\ x \in [-1,1]$,
\item $H(y=-1,x) > 0 \ {\rm for}\ x \in [-1,x_1)\cup (x_2,1]$ and $H(y=-1,x)<0\ {\rm for} \ x \in (x_1,x_2)$,
\end{itemize}
where $x_1$ and $x_2$ are certain constants such that $-1<x_1<x_2<1$.
Moreover, (d) and (e) ensure that for a given $x\in[-1,1]$:
\begin{itemize}
\item $H(y,x)=0$ has a single root for $y \in (-1/\nu,-1]$ if $H(y=-1,x)\geq 0$,
\item $\partial_y H(y=-1,x)=-H(y=-1,x)>0$ and then $H(y,x)$ is a monotonically increasing function of $y$ if $H(y=-1,x)<0$,
\end{itemize}
which excludes the case where $H(y,x)>0$ for $\exists y \in (-1/\nu,-1)$ but $H(y=-1,x)<0$ and $H(y=-1/\nu,x)<0$.
Therefore, we find that the capped black hole admits two types of ergosurfaces, whose topology changes across a value $\nu=\nu_*(\alpha)$ where a quadratic function $H(y=-1,x)$ exhibits a double root within the range $-1 \leq x \leq 1$~(Fig.~\ref{fig:ergo}):

\begin{enumerate}
\item[(i)] $0<\nu < \nu_*(\alpha)$: a single $S^3$-surface around the horizon, 
\item[(ii)] $\nu_*(\alpha)<\nu<1$: an outer $S^3$-surface that encompasses both the horizon and bubble, and an inner $S^3$-surface around the flat center at $(x,y)=(1,-1)$.
\end{enumerate}
In case (ii), the ergoregion extends to the rotation axis of $\psi$,
while the ball region around the center at $(x,y)=(1,-1)$ is excluded from the ergoregion due to being the fixed point of two rotations.
In Fig.~\ref{fig:ergo-phase}, we illustrate the threshold curve $\nu=\nu_*(\alpha)$ and the topology of the ergosurface in the phase diagram.

\medskip
One might understand the topology change of the ergoregion from the change in the bubble size discussed in the previous section. For small enough $\nu$, the bubble is sufficiently large compared to the horizon scale, and hence it protrudes out of the ergoregion as in case (i). As $\nu$ increases, the bubble becomes smaller, and for $\nu=\nu_*(\alpha)$, it is completely engulfed by the ergoregion. The inner ergosurface exists for $\nu_*(\alpha)\leq \nu <1$, but it vanishes as $\nu$ approaches 1, since the bubble shrinks to a point in the limit and the spacetime approaches the extremal (non-BPS) Cveti\v{c}-Youm black hole.
Note that in both cases, there does not exist a so-called evanescent ergosurface, which is a timelike surface outside the horizon where the timelike Killing vector field at infinity becomes null on the surface but remains timelike both inside and outside the surface~\cite{Tomizawa:2016kjh}.


\section{Summary and discussion}\label{sec:sum}.

In this paper, applying the electric Harrison transformation to the vacuum black lens solutions possessing a Dirac-Misner string singularity, we have constructed asymptotically flat, stationary, bi-axisymmetric black hole solutions within the bosonic sector of five-dimensional minimal supergravity.
Initially, we have obtained the vacuum black lens solutions, which inherently contained Dirac-Misner string singularities, using the inverse scattering method. 
Then, by implementing the Harrison transformation on the vacuum solution, we successfully eliminated the Dirac-Misner string singularities by appropriately adjusting the parameters involved.
The resulting black hole solutions exhibit horizon topologies of lens space $L(n;1)$, including the $n=0$ and $n=1$ cases. 
It has been demonstrated that these black hole solutions are singular for $n \ge 2$, but regular for $n=0,1$.
The $n=0$ case describes to a charged rotating black ring with a dipole charge constructed in Ref.~\cite{Suzuki:2024coe}, and the $n=1$ case describes a capped black hole constructed in Ref.~\cite{Suzuki:2023nqf}. 
In particular, the $n=1$ case is interesting because this solution describes an asymptotically flat, stationary, non-BPS black hole with a horizon cross-section of trivial topology $S^3$, while the domain of outer communication (DOC) exhibits a non-trivial topology.
This solution remains regular without any curvature singularities, conical singularities, Dirac-Misner string singularities, and orbifold singularities both on and outside the horizon.
It describes a charged rotating black hole capped by a disc-shaped bubble, which we call a “capped black hole”.
We have demonstrated that the spherical black hole carries mass, two angular momenta, an electric charge, and a magnetic flux, where only three of these quantities are independent.
Moreover, we have shown that this black hole can have identical conserved charges as the spherical black hole found by Cveti\v{c}-Youm, thus indicating a violation of black hole uniqueness even when assuming that the topology of the horizon cross-section is $S^3$.
Additionally, we have found that the capped black hole can possess a larger entropy compared to the Cveti\v{c}-Youm black hole, establishing the spherical black hole with a significant bubble as thermodynamically more stable.

\begin{figure}
\includegraphics[width=15cm]{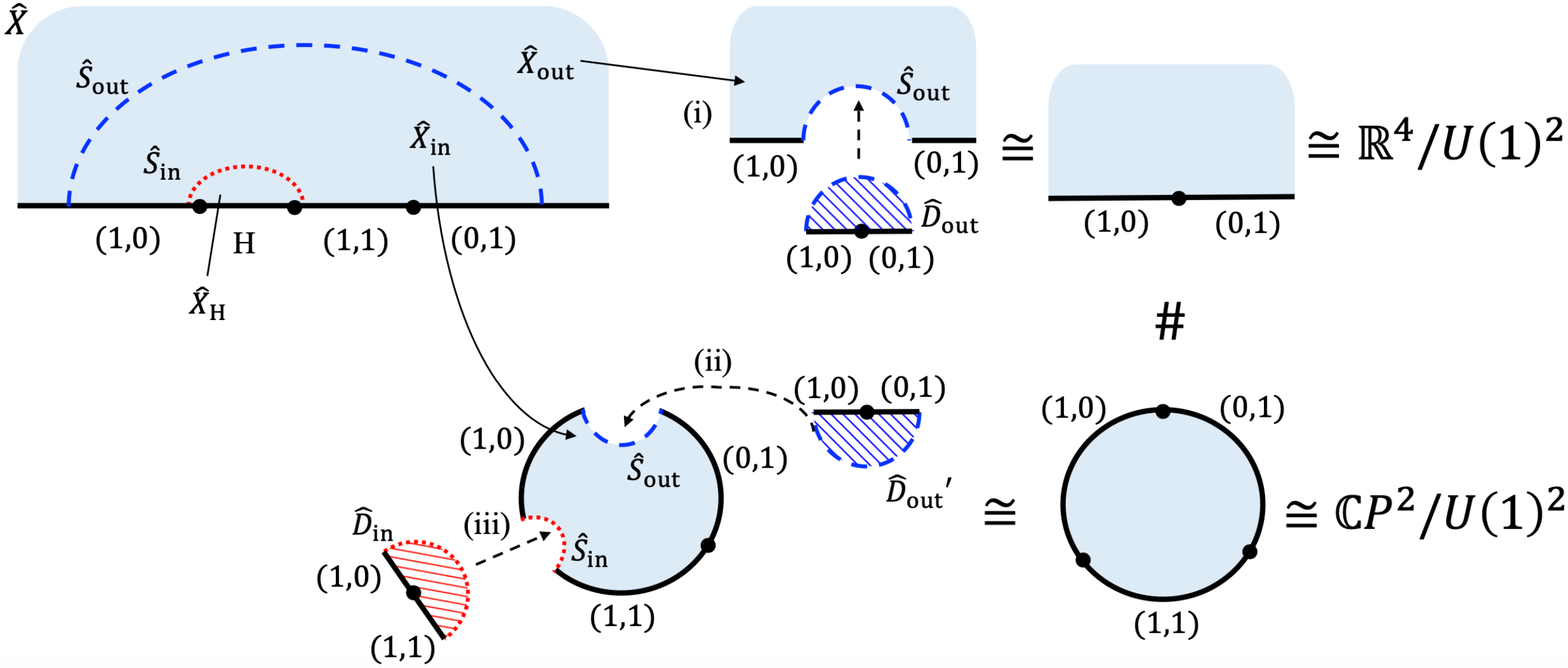}
\caption{The topology of the DOC can be seen from the rod diagram. \label{fig:DOC}}
\end{figure}

\medskip
The topology of the Domain of Outer Communication (DOC) on a time slice $\Sigma$ can easily be read off from the rod structure, as described in Ref. \cite{Hollands:2010qy}. 
According to the topology censorship theorem~\cite{Friedman:1993ty}, the intersection $X={\rm DOC} \cap \Sigma$ in an asymptotically flat spacetime must be simply connected.
Therefore, in a bi-axisymmetric spacetime, the orbit space $\hat X = X/ U(1)^2$  reduced to two dimensions by two $U(1)$-isometries is also simply connected. 
This results in the rod diagram representing the upper half-plane  in ${\mathbb R}^2$, as depicted in Fig.~\ref{fig:DOC}.
For simplicity, we assign only the spacial components of the rod vector to each rods, and the rod vectors on the semi-infinite rods are set to be $(0,1)$ and $(1,0)$, respectively. 
Note that, to match the orientation in Ref.~\cite{Hollands:2010qy}, we assign the rod vector $m_\phi \partial_\phi+m_\psi \psi$ to $(m_\phi,m_\psi)$ so that the semi-infinite rod with $(1,0)$ corresponds to the left side.
Following the procedure in Ref.~\cite{Hollands:2010qy}, to know the topological structure of  $X={\rm DOC} \cap \Sigma$ on a timeslice $\Sigma$ for the capped black hole, let us consider a sufficiently large outer sphere $S_{\rm out}$ and a sphere $S_{\rm in}$ sufficiently close  to the horizon, which divide $X$ into three regions: the asymptotic region $X_{\rm out}$ outside $S_{\rm out}$, including spatial infinity, the inner region $X_{\rm in}$ between $S_{\rm out}$ and $S_{\rm in}$, and the near-horizon region $X_{\rm H}$ between $S_{\rm in}$ and the horizon.
We denote the corresponding counterparts in the orbit space with hats.
As depicted in Fig.~\ref{fig:DOC}, the curve $\hat S_{\rm out}$, represented by the blue dashed curve, terminates at the two rods with $(1,0)$ and $(0,1)$, and the curve $\hat S_{\rm in}$, represented by the red dotted curve,  terminates at the two rods with $(1,0)$ and $(1,1)$, where as shown Ref.~\cite{Hollands:2007aj}, it is ensured from the orientations of these rod vectors that $S_{\rm out}$ and $S_{\rm in}$ are topologically  $S^3$.
The procedure in Ref.~\cite{Hollands:2010qy} is as follows:  
(i) First, by gluing $\hat X_{\rm out}$ and a half-disc $\hat{D}_{\rm out}$, which is such that $D_{\rm out} \cong {\mathbb B}^4$, along a semicircle $\hat S_{\rm out}$, one can obtain a upper half-plane [in other words, by gluing $X_{\rm out}$ and $D_{\rm out}$ along $S_{\rm out}$, one can obtain ${\mathbb R}^4$],
(ii) next, by gluing a half-disc $\hat{D}_{\rm out}'$ ($D_{\rm out}'\cong {\mathbb B}^4$) along a semicircle $\hat S_{\rm out}$ , and (iii) finally, by gluing another half-disc $\hat{D}_{\rm in}$ (${D}_{\rm in}\cong  {\mathbb B}^4$) along a semicircle $\hat S_{\rm in}$, 
one can obtain a compact two-dimensional space bounded by a closed curve with three endpoints 
[similarly, by gluing $D_{\rm out}'$ along $S_{\rm out}$ and $D_{\rm in}$ along $S_{\rm in}$ to $X_{\rm in}$, one can obtain a compact, simply connected $4$-manifold]. 
According to the classification in Ref.~\cite{Orlik1970}, the topology of the four-dimensional space with this type of rod diagram turns out to be ${\mathbb C}P^2$.
Thus, we find $X_{\rm out} \cup X_{\rm in} \cup D_{\rm in} \cong {\mathbb R}^4 \# {\mathbb C}P^2$.
Since $X\cong X_{\rm out}\cup X_{\rm in}$,
we can conclude that  the capped black hole constructed in this paper has the DOC such that  ${\rm DOC}\cap \Sigma \cong[ {\mathbb R}^4 \# {\mathbb C}P^2] \setminus {\mathbb B}^4$.

\medskip

The capped black hole derived in this paper is characterized by four conserved charges: the mass, two angular momenta, and an electric charge, along with a magnetic flux. However, these quantities are not all independent. It is expected that there may exist a more general capped black hole with five independent quantities. 
The construction of such a solution deserves further investigation.
Additionally, so far, an exact solution of a non-BPS black lens, even a vacuum solution, has not been found. 
In this paper, we also attempted the construction of black lenses applying the electric Harrison transformation on the Chen-Teo type configuration as described in Ref.~\cite{Chen:2008fa}. However, though we could obtain the capped black hole solution and charged dipole black ring solution as a by-product, our work does not yield any regular solutions for black lenses. Exploring this avenue is part of our future work.
It is conceivable that applying the Harrison transformation to the four-soliton solution referenced in Ref.~\cite{Tomizawa:2019acu} might enable us to uncover a regular charged black lens solution. 
Recently, a regular static black lens immersed in the external magnetic field--which is not asymptotically flat--was produced by combining the Harrison transformation and another type of transformation in the context of the five-dimensional Einstein-Maxwell theory~\cite{Tomlinson:2021wsp}. 
Hence, the presence of the magnetic field plays a significant role in the support of a black lens horizon, as suggested previously for supersymmetric black lenses~\cite{Tomizawa:2016kjh}.
In our forthcoming paper, we will discuss another construction of a non-BPS charged black lens solution in five-dimensional minimal supergravity.

\section*{Acknowledgement}
The authors are grateful to Marcus Khuri for useful comments and discussion on the topology of the DOC during the 32nd workshop on General Relativity and Gravitation in Japan (JGRG32).
R.S. was supported by JSPS KAKENHI Grant Number JP18K13541. 
S.T. was supported by JSPS KAKENHI Grant Number 21K03560.

\appendix

\section{Coefficients for $K(x,y)$}\label{sec:Kxy}

The coefficients in Eq.~(\ref{eq:def-Kxy}) are given by
\begin{align}
&k_1= 2 c_1 c_2 c_3 \nu  (\nu +1) (a-b) (\gamma -\nu ),\quad
k_2= -c_2 (1-\nu ) (a-b) (\gamma-\nu  ) \left(c_1 c_3-b d_1\right),\nonum
&k_3= \frac{c_2 d_1 (1-\nu ) \left(\gamma  c_3-c_3   \nu -d_2\right)}{(1-\gamma ) (\nu +1)},
k_4= \frac{(1-\gamma )^3 c_2 c_3 (1-\nu )^2 \left((1-\gamma ) (1-\nu ) (\gamma +\nu )+d_1-d_2\right)}{(1-\gamma )^4 (\nu +1)},\nonum
&k_5= 2 c_3 d_3   (1-\nu )^3 (\gamma -\nu ),\quad
   k_6= \frac{c_2 c_3 \left(\gamma ^2 (\nu -1)^2+\gamma  \left((\nu -1)^3-2 d_1\right)-(\nu -1) \left(d_2+(\nu -1) \nu \right)+d_1 (3 \nu -1)\right)}{(1-\gamma) (1+\nu)},\nonum
&   k_7= -\frac{8 d_1 \left[(1-\gamma ) (1-\nu )^2 (\gamma -\nu ) \left(b (\nu -1)+c_1\right)+2 c_1 c_3 (\gamma -\nu )-c_1 d_2 (1-\nu )\right]}{(1-\gamma )
   (\nu +1)}.
    \end{align}

\section{Proof of $H(x,y)>0$ and $D>0$ for $n=1$}\label{app:hxy}

Here we show the positivity of each term in Eq.~(\ref{eq:H2}).
With the negativity of $d_1$~(\ref{eq:d1neg}), it suffices to show the positivity of $d_5,d_6,d_7$ defined in Eq.~(\ref{eq:param-extra}).
For this, it is convenient to clarify the signature of $c_1$ by rewriting it with Eqs.~(\ref{eq:lenscon}) and (\ref{eq:conifree}) as
\begin{align}
c_1 =- (1-\gamma)^2(1-d)^2 (-d_1)^{-1} (1-a)(1-a^2)<0.
\end{align}
This also imply $c_2 <0$ due to the identity
\begin{align}
c_2 = c_1(1+\nu)-(1-\gamma)(1-\nu)a<0.
\end{align}
With $c_1,c_2<0$, the positivity of $d_6$ and $d_7$ is obvious from the definition.
The positivity of $d_5$ is a little less trivial. First, we consider the following quantity
\begin{align}
&\tilde{d}_5 = \frac{ab^2(2-a)(1-a^2)^2}{(1-\gamma)^4(1-\nu)^3}d_5 =
(2-a) a^3 \left(a^2-a+1\right)^2  
+b\left(a^2-a+1\right) \left(8 a^5-13 a^4+2 a^3-a^2+2 a-1\right) \nonum
&\quad+a b^2\left(2 a^6-14   a^5+22 a^4-17 a^3+16 a^2-8 a+2\right) 
+(2 a-1) b^3\left(a^2-a+1\right) \left(2 a^2-2 a-1\right) ,
\end{align}
where we used eq.~(\ref{eq:Bd-sol}).
One can see the positivity of $d_5$ by writing $\tilde{d}_5$  as the sum of positive definite terms
\begin{align}
& \tilde{d}_5 = (2-a)^3a^3(1-a+a^2)^2(1+b^3)+(-b)(1-b^2)(1-a+a^2)\left[\frac{5}{18}+\fr{2}(1-a)^4+\frac{9}{2}\left(a^2-\frac{2}{9}\right)^2+8 a^4(1-a)\right]\nonum
 &+b^2(1+b)a\left[2(1-a)^4+6a^2\left(a-\frac{3}{4}\right)^2+\frac{5}{8}a^2+2a^6+14a^4(1-a)\right]+(-b)^3 a^2(1+a)^3(1+a^2-a^3)>0.
\end{align}

\section{Properties of $H(y,x)$ for $n=1$}\label{app:ergo}
Here we prove several properties of $H(y,x)$ that is used to determine the topology of the ergoregion for the capped black hole in Sec.~\ref{sec:ergo}.
We use the following conditions proved in Sec.~\ref{sec:paramrange} and Appendix.~\ref{app:hxy}
\begin{align}
d_1<0,\quad c_1<0,\quad c_3>0.
\end{align}
\begin{enumerate}
\item Proof of $(a)\, H(y=-1,x=\pm 1)>0$ and $(c)\,\partial_x^2 H(y=-1,x)>0 \ {\rm for}\  x\in [-1,1]$\\
These can be shown directly from
\begin{align}
&H(y=-1,x=1) = - 8(1-\gamma)^2 d_1 (1-\nu)^2 > 0,\\
&H(y=-1,x=-1)=8(1-\gamma)^3(1-\nu)^4(1-a^2)>0,\\
&\partial_x^2 H(y=-1,x) =  8\nu(1-\gamma)^3(\gamma-\nu)(1-\nu)^2(a-b)^2>0.
\end{align}
\item Proof of $(b)\, H(y=-1/\nu,x)<0 \ {\rm for} \ x\in [-1,1]$\\
The positivity of $H(y=-1/\nu,x)$ is obvious by writing it as
\begin{align}
H(y=-1/\nu,x) = A_1 (1+x)^2+A_2(1-x)^2+A_3(1-x^2),
\end{align}
where
\begin{align}
&A_1 = \frac{(1-\nu)^2(1+\nu)(c_2-(1-\nu)(1-\gamma)(\gamma+\nu))^2}{\nu(\gamma-\nu)}>0,\\
&A_2 = \frac{(1-\gamma)c_3(\gamma-\nu)(1-\nu)^4(1+\nu)}{2\nu^2}>0,\\
&A_3 = \frac{(1-\gamma ) (\nu -1)^2 (\nu +1) (\gamma -\nu ) \left((\nu +1) \left(b (\nu -1)
   (\gamma +\nu )+2 c_1 \nu \right){}^2+4 (1-\gamma ) \gamma  \nu  (1-\nu )^2
   +4(-d_1) \nu
   ^2\right)}{2 \nu ^2 (\gamma +\nu )}>0.
\end{align}

\item Proof of $(d)\, \partial_y^2 H(y,x)<0\ {\rm for}\ (x,y) \in [-1,1]\times [-1/\nu,-1]$\\
$\partial_y^2 H(y,x)<0$ is obvious by writing it as
\begin{align}
\partial_y^2 H(y,x) =- B_1 (1+x)^2-B_2(1-x)^2-B_3(1-x^2),
\end{align}
where
\begin{align}
& B_1 = 2(1-\gamma)c_1^2 c_3 \nu(1+\nu)>0,\nonum
& B_2 = (1-\gamma)(1-\nu)^2(2\nu c_1+(\gamma-\nu)(1-\nu)b)^2>0,\nonum
& B_3 = (1-\gamma)(1-\nu)[4c_1^2 \nu^2(2-\gamma+\nu)+4b c_1 (\gamma-\nu)\nu(1-\nu^2)+(1-\nu)^3(\gamma^2-\nu^2)b^2]>0.
\end{align}

\item Proof of $(e)\, \partial_y H(y=-1,x) + H(y=-1,x)>0\ {\rm for} \ x\in[-1,1]$\\

$\partial_y H(y=-1,x) + H(y=-1,x)>0$ becomes obvious by writing it as
\begin{align}
\partial_y H(y=-1,x) + H(y=-1,x)  = C_1 (1+x)^2+C_2(1-x)^2+C_3(1-x^2),
\end{align}
where
\begin{align}
\begin{split}
C_1&=-c_1^2 (\nu +1) \left(1-\nu ^2\right) \left(2 (1-\gamma )^2-b^2 \left(\gamma ^2-\nu
   ^2\right)\right)+2 b c_1 (\gamma -1)  \left(\nu ^2-1\right)^2 (\gamma -\nu )\\
& +(1-\gamma   )^2 (\nu -1)^4 (2-\gamma +\nu ),\\
   C_2&=2 b c_1(\gamma -1)  \left(\nu ^2-1\right) (\nu -1)^2 (\gamma -\nu )-2c_1^2 (\gamma -1) 
   (\nu +1) (\nu -1)^3+(1-\gamma )^2 (\nu -1)^4 (2-\gamma +\nu ),\\
C_3& = 2 (1-\gamma )(- d_1) (\nu -1)^2 (2-\gamma +\nu )>0.
\end{split}
\end{align}
To see $C_1>0$ and $C_2>0$, one can write them by using Eq.~(\ref{eq:Bd-sol}) as
\begin{align}
&C_1=  \frac{a(2-a)(1-\nu)^4(1-\gamma)^3}{b^2(1-a)^2(1+a)^2} \left[-\frac{2b(1+b)^2(1-a+a^2)(1-a)^2\nu}{1-\nu}+(2-a)a(-b(1-a+a^2)+ab^2(2-a))\right]>0,
\end{align}
and
\begin{align}
C_2= \frac{(1-\gamma)^3(1-\nu)^4}{b^2} \left[ \frac{(1-a+a^2)^2(1+b)^2}{1-a^2}+ab^2(2-a)(1+2a-2a^2)-\frac{2b(1-a+a^2)(1+a(1-a)^2)}{1-a} \right]>0.
\end{align}

\end{enumerate}


\begin{thebibliography}{99}



\bibitem{Emparan:2008eg}
R.~Emparan and H.~S.~Reall,
``Black Holes in Higher Dimensions,''
Living Rev. Rel. \textbf{11}, 6 (2008)
[arXiv:0801.3471 [hep-th]].

\bibitem{Emparan:2006mm}
R.~Emparan and H.~S.~Reall,
``Black Rings,''
Class. Quant. Grav. \textbf{23}, R169 (2006)
[arXiv:hep-th/0608012 [hep-th]].



\bibitem{Mizoguchi:1998wv}
S.~Mizoguchi and N.~Ohta,
``More on the similarity between D = 5 simple supergravity and M theory,''
Phys. Lett. B \textbf{441}, 123-132 (1998)
[arXiv:hep-th/9807111 [hep-th]].




\bibitem{Mizoguchi:1999fu}
S.~Mizoguchi and G.~Schroder,
``On discrete U duality in M theory,''
Class. Quant. Grav. \textbf{17}, 835-870 (2000)
[arXiv:hep-th/9909150 [hep-th]].


\bibitem{Cremmer:1997ct}
E.~Cremmer, B.~Julia, H.~Lu and C.~N.~Pope,
``Dualization of dualities. 1.,''
Nucl. Phys. B \textbf{523}, 73-144 (1998)
[arXiv:hep-th/9710119 [hep-th]].

\bibitem{Cremmer:1998px}
E.~Cremmer, B.~Julia, H.~Lu and C.~N.~Pope,
``Dualization of dualities. 2. Twisted self-duality of doubled fields, and superdualities,''
Nucl. Phys. B \textbf{535}, 242-292 (1998)
[arXiv:hep-th/9806106 [hep-th]].


\bibitem{Ford:2007th}
J.~Ford, S.~Giusto, A.~Peet and A.~Saxena,
``Reduction without reduction: Adding KK-monopoles to five dimensional stationary axisymmetric solutions,''
Class. Quant. Grav. \textbf{25}, 075014 (2008)
[arXiv:0708.3823 [hep-th]].


\bibitem{Galtsov:2008pbf}
D.~V.~Gal'tsov and N.~G.~Scherbluk,
``Hidden symmetries in 5D supergravities and black rings,''
PoS \textbf{BHGRS}, 016 (2008)
[arXiv:0912.2770 [hep-th]].

\bibitem{Galtsov:2008bmt}
D.~V.~Gal'tsov and N.~G.~Scherbluk,
``Generating technique for U(1)**3 5D supergravity,''
Phys. Rev. D \textbf{78}, 064033 (2008)
[arXiv:0805.3924 [hep-th]].



\bibitem{Galtsov:2008jjb}
D.~V.~Gal'tsov and N.~G.~Scherbluk,
``Improved generating technique for D=5 supergravities and squashed Kaluza-Klein Black Holes,''
Phys. Rev. D \textbf{79}, 064020 (2009)
[arXiv:0812.2336 [hep-th]].


\bibitem{Compere:2009zh}
G.~Compere, S.~de Buyl, E.~Jamsin and A.~Virmani,
``G2 Dualities in D=5 Supergravity and Black Strings,''
Class. Quant. Grav. \textbf{26}, 125016 (2009)
[arXiv:0903.1645 [hep-th]].

\bibitem{Figueras:2009mc}
P.~Figueras, E.~Jamsin, J.~V.~Rocha and A.~Virmani,
``Integrability of Five Dimensional Minimal Supergravity and Charged Rotating Black Holes,''
Class. Quant. Grav. \textbf{27}, 135011 (2010)
[arXiv:0912.3199 [hep-th]].




\bibitem{Mizoguchi:2011zj}
S.~Mizoguchi and S.~Tomizawa,
``New approach to solution generation using SL(2,R)-duality of a dimensionally reduced space in five-dimensional minimal supergravity and new black holes,''
Phys. Rev. D \textbf{84}, 104009 (2011)
[arXiv:1106.3165 [hep-th]].


\bibitem{Bouchareb:2007ax}
A.~Bouchareb, G.~Clement, C.~M.~Chen, D.~V.~Gal'tsov, N.~G.~Scherbluk and T.~Wolf,
``G(2) generating technique for minimal D=5 supergravity and black rings,''
Phys. Rev. D \textbf{76}, 104032 (2007)
[erratum: Phys. Rev. D \textbf{78}, 029901 (2008)]
[arXiv:0708.2361 [hep-th]].




\bibitem{Myers:1986un}
R.~C.~Myers and M.~J.~Perry,
``Black Holes in Higher Dimensional Space-Times,''
Annals Phys. \textbf{172}, 304 (1986)



\bibitem{Emparan:2001wn}
R.~Emparan and H.~S.~Reall,
``A Rotating black ring solution in five-dimensions,''
Phys. Rev. Lett. \textbf{88}, 101101 (2002)
[arXiv:hep-th/0110260 [hep-th]].




 \bibitem{Hollands:2007aj} 
  S.~Hollands and S.~Yazadjiev,
  ``Uniqueness theorem for 5-dimensional black holes with two axial Killing fields,''
  Commun.\ Math.\ Phys.\  {\bf 283}, 749 (2008)
  [arXiv:0707.2775 [gr-qc]].

  
\bibitem{Mishima:2005id}
T.~Mishima and H.~Iguchi,
``New axisymmetric stationary solutions of five-dimensional vacuum Einstein equations with asymptotic flatness,''
Phys. Rev. D \textbf{73}, 044030 (2006)
[arXiv:hep-th/0504018 [hep-th]].

 %
  
\bibitem{Figueras:2005zp}
P.~Figueras,
``A Black ring with a rotating 2-sphere,''
JHEP \textbf{07}, 039 (2005)
[arXiv:hep-th/0505244 [hep-th]].


\bibitem{Pomeransky:2006bd}
A.~A.~Pomeransky and R.~A.~Sen'kov,
``Black ring with two angular momenta,''
[arXiv:hep-th/0612005 [hep-th]].

\bibitem{Evslin:2008gx}
J.~Evslin,
``Geometric Engineering 5d Black Holes with Rod Diagrams,''
JHEP \textbf{09}, 004 (2008)
[arXiv:0806.3389 [hep-th]].



\bibitem{Chen:2008fa}
Y.~Chen and E.~Teo, ``A Rotating black lens solution in five dimensions'', {} Phys.\ Rev.\ D {\bf 78}, 064062 (2008).




\bibitem{Tomizawa:2019acu}
S.~Tomizawa and T.~Mishima,
``Stationary and biaxisymmetric four-soliton solution in five dimensions,''
Phys. Rev. D \textbf{99}, no.10, 104053 (2019)
[arXiv:1902.10544 [hep-th]].



\bibitem{Lucietti:2020phh}
J.~Lucietti and F.~Tomlinson,
``On the nonexistence of a vacuum black lens,''
JHEP \textbf{21}, 005 (2020)
[arXiv:2012.00381 [gr-qc]].

\bibitem{Gauntlett:2002nw} 
  J.~P.~Gauntlett, J.~B.~Gutowski, C.~M.~Hull, S.~Pakis and H.~S.~Reall,
  ``All supersymmetric solutions of minimal supergravity in five- dimensions,''
  Class.\ Quant.\ Grav.\  {\bf 20}, 4587 (2003) 
  [hep-th/0209114].


\bibitem{Reall:2002bh}
H.~S.~Reall,
``Higher dimensional black holes and supersymmetry,''
Phys. Rev. D \textbf{68}, 024024 (2003)
[erratum: Phys. Rev. D \textbf{70}, 089902 (2004)]
[arXiv:hep-th/0211290 [hep-th]].


\bibitem{Breckenridge:1996is}
J.~C.~Breckenridge, R.~C.~Myers, A.~W.~Peet and C.~Vafa,
``D-branes and spinning black holes,''
Phys. Lett. B \textbf{391}, 93-98 (1997)
[arXiv:hep-th/9602065 [hep-th]].




\bibitem{Elvang:2004rt}
H.~Elvang, R.~Emparan, D.~Mateos and H.~S.~Reall,
``A Supersymmetric black ring,''
Phys. Rev. Lett. \textbf{93}, 211302 (2004)
[arXiv:hep-th/0407065 [hep-th]].



\bibitem{Kunduri:2014kja}
H.~K.~Kunduri and J.~Lucietti,
``Supersymmetric Black Holes with Lens-Space Topology,''
Phys. Rev. Lett. \textbf{113}, no.21, 211101 (2014)
[arXiv:1408.6083 [hep-th]].

\bibitem{Tomizawa:2016kjh}
S.~Tomizawa and M.~Nozawa,
``Supersymmetric black lenses in five dimensions,''
Phys. Rev. D \textbf{94}, no.4, 044037 (2016)
[arXiv:1606.06643 [hep-th]].


\bibitem{Breunholder:2017ubu}
V.~Breunh\"older and J.~Lucietti,
``Moduli space of supersymmetric solitons and black holes in five dimensions,''
Commun. Math. Phys. \textbf{365}, no.2, 471-513 (2019)
[arXiv:1712.07092 [hep-th]].



\bibitem{Tomizawa:2009ua}
S.~Tomizawa, Y.~Yasui and A.~Ishibashi,
``Uniqueness theorem for charged rotating black holes in five-dimensional minimal supergravity,''
Phys. Rev. D \textbf{79}, 124023 (2009)
[arXiv:0901.4724 [hep-th]].


\bibitem{Cvetic:1996xz}
M.~Cvetic and D.~Youm,
``General rotating five-dimensional black holes of toroidally compactified heterotic string,''
Nucl. Phys. B \textbf{476}, 118-132 (1996)
[arXiv:hep-th/9603100 [hep-th]].


\bibitem{Friedman:1993ty}
J.~L.~Friedman, K.~Schleich and D.~M.~Witt,
``Topological censorship,''
Phys. Rev. Lett. \textbf{71}, 1486-1489 (1993)
[erratum: Phys. Rev. Lett. \textbf{75}, 1872 (1995)]
[arXiv:gr-qc/9305017 [gr-qc]].

\bibitem{Hollands:2010qy}
S.~Hollands, J.~Holland and A.~Ishibashi,
Annales Henri Poincare \textbf{12}, 279-301 (2011)
doi:10.1007/s00023-011-0079-2
[arXiv:1002.0490 [gr-qc]].




\bibitem{Gibbons:2002bh}
G.~W.~Gibbons, D.~Ida and T.~Shiromizu,
``Uniqueness and nonuniqueness of static vacuum black holes in higher dimensions,''
Prog. Theor. Phys. Suppl. \textbf{148}, 284-290 (2003)
[arXiv:gr-qc/0203004 [gr-qc]].

\bibitem{Gibbons:2002av}
G.~W.~Gibbons, D.~Ida and T.~Shiromizu,
``Uniqueness and nonuniqueness of static black holes in higher dimensions,''
Phys. Rev. Lett. \textbf{89}, 041101 (2002)
[arXiv:hep-th/0206049 [hep-th]].



\bibitem{Tangherlini:1963bw} 
  F.~R.~Tangherlini,
  ``Schwarzschild field in n dimensions and the dimensionality of space problem,''
  Nuovo Cim.\  {\bf 27}, 636 (1963).
 

\bibitem{Kunduri:2014iga}
H.~K.~Kunduri and J.~Lucietti,
``Black hole non-uniqueness via spacetime topology in five dimensions,''
JHEP \textbf{10}, 082 (2014)
[arXiv:1407.8002 [hep-th]].


\bibitem{Ehlers}
J. Ehlers, Dissertation, Hamburg University.

\bibitem{Harrison}
B.K. Harrison, New solutions of the Einstein-Maxwell equations from old,  J. Math. Phys {\bf 9}, 1744 (1968).



\bibitem{exact}
H. Stephani, D. Kramer, M. A. H. MacCallum, C. Hoenselaers and E. Herlt, {\it Exact solutions of Einstein’s Field Equations}, 2nd ed. (Cambridge University Press, Cambridge, 2003).





\bibitem{Chamseddine:1980mpx}
A.~H.~Chamseddine and H.~Nicolai,
``Coupling the SO(2) Supergravity Through Dimensional Reduction,''
Phys. Lett. B \textbf{96}, 89-93 (1980)
[erratum: Phys. Lett. B \textbf{785}, 631-632 (2018)]
[arXiv:1808.08955 [hep-th]].


\bibitem{Cremmer:1979up}
E.~Cremmer and B.~Julia,
``The SO(8) Supergravity,''
Nucl. Phys. B \textbf{159}, 141-212 (1979)





\bibitem{Mizoguchi:2012vg}
S.~Mizoguchi and S.~Tomizawa,
``Flipped $SL(2,R)$ duality in five-dimensional supergravity,''
Phys. Rev. D \textbf{86}, 024022 (2012)
[arXiv:1201.3063 [hep-th]].

\bibitem{Tomizawa:2012nk}
S.~Tomizawa and S.~Mizoguchi,
``General Kaluza-Klein black holes with all six independent charges in five-dimensional minimal supergravity,''
Phys. Rev. D \textbf{87}, no.2, 024027 (2013)
[arXiv:1210.6723 [hep-th]].




\bibitem{Tomizawa:2008qr}
S.~Tomizawa, Y.~Yasui and Y.~Morisawa,
``Charged Rotating Kaluza-Klein Black Holes Generated by G2(2) Transformation,''
Class. Quant. Grav. \textbf{26}, 145006 (2009)
[arXiv:0809.2001 [hep-th]].

\bibitem{Rasheed:1995zv}
D.~Rasheed,
``The Rotating dyonic black holes of Kaluza-Klein theory,''
Nucl. Phys. B \textbf{454}, 379-401 (1995)
[arXiv:hep-th/9505038 [hep-th]].

\bibitem{Ishihara:2005dp}
H.~Ishihara and K.~Matsuno,
``Kaluza-Klein black holes with squashed horizons,''
Prog. Theor. Phys. \textbf{116}, 417-422 (2006)
[arXiv:hep-th/0510094 [hep-th]].







\bibitem{Suzuki:2024coe}
R.~Suzuki and S.~Tomizawa,
``New construction of a charged dipole black ring by Harrison transformation,''
[arXiv:2402.07589 [hep-th]].



\bibitem{Belinsky:1979mh}
  V.~A.~Belinsky and V.~E.~Sakharov,
  ``Stationary Gravitational Solitons with Axial Symmetry,''
  Sov.\ Phys.\ JETP {\bf 50}, 1 (1979).

\bibitem{Belinski:2001ph} 
  V.~A.~Belinski and E.~Verdaguer,
  \textit{Gravitational Solitons},
  (Cambridge University Press, Cambridge, 2001).



\bibitem{Pomeransky:2005sj}
  A.~A.~Pomeransky,
  ``Complete integrability of higher-dimensional Einstein equations with additional symmetry, and rotating black holes,''
  Phys.\ Rev.\ D {\bf 73} (2006) 044004
  [hep-th/0507250].




\bibitem{Emparan:2001wk}
R.~Emparan and H.~S.~Reall,
``Generalized Weyl solutions,''
Phys. Rev. D \textbf{65}, 084025 (2002)
[arXiv:hep-th/0110258 [hep-th]].



\bibitem{Harmark:2004rm}
T.~Harmark,
``Stationary and axisymmetric solutions of higher-dimensional general relativity,''
Phys. Rev. D \textbf{70}, 124002 (2004)
[arXiv:hep-th/0408141 [hep-th]].












































\bibitem{Iguchi:2006rd}
H.~Iguchi and T.~Mishima,
``Solitonic generation of five-dimensional black ring solution,''
Phys. Rev. D \textbf{73}, 121501 (2006)
[arXiv:hep-th/0604050 [hep-th]].


\bibitem{Tomizawa:2006vp}
S.~Tomizawa and M.~Nozawa,
``Vacuum solutions of five-dimensional Einstein equations generated by inverse scattering method. II. Production of black ring solution,''
Phys. Rev. D \textbf{73}, 124034 (2006)
[arXiv:hep-th/0604067 [hep-th]].








\bibitem{Tomizawa:2005wv}
S.~Tomizawa, Y.~Morisawa and Y.~Yasui,
``Vacuum solutions of five dimensional Einstein equations generated by inverse scattering method,''
Phys. Rev. D \textbf{73}, 064009 (2006)
[arXiv:hep-th/0512252 [hep-th]].



\bibitem{Tomizawa:2006jz}
S.~Tomizawa, H.~Iguchi and T.~Mishima,
``Relationship between solitonic solutions of five-dimensional Einstein equations,''
Phys. Rev. D \textbf{74}, 104004 (2006)
[arXiv:hep-th/0608169 [hep-th]].



\bibitem{Elvang:2007rd}
H.~Elvang and P.~Figueras,
``Black Saturn,''
JHEP \textbf{05}, 050 (2007)
[arXiv:hep-th/0701035 [hep-th]].

\bibitem{Izumi:2007qx}
K.~Izumi,
``Orthogonal black di-ring solution,''
Prog. Theor. Phys. \textbf{119}, 757-774 (2008)
[arXiv:0712.0902 [hep-th]].

\bibitem{Elvang:2007hs}
H.~Elvang and M.~J.~Rodriguez,
``Bicycling Black Rings,''
JHEP \textbf{04}, 045 (2008)
[arXiv:0712.2425 [hep-th]].

\bibitem{Iguchi:2007is}
H.~Iguchi and T.~Mishima,
``Black di-ring and infinite nonuniqueness,''
Phys. Rev. D \textbf{75}, 064018 (2007)
[erratum: Phys. Rev. D \textbf{78}, 069903 (2008)]
[arXiv:hep-th/0701043 [hep-th]].

\bibitem{Evslin:2007fv}
J.~Evslin and C.~Krishnan,
``The Black Di-Ring: An Inverse Scattering Construction,''
Class. Quant. Grav. \textbf{26}, 125018 (2009)
[arXiv:0706.1231 [hep-th]].








\bibitem{Tomizawa:2007mz}
S.~Tomizawa, H.~Iguchi and T.~Mishima,
``Rotating Black Holes on Kaluza-Klein Bubbles,''
Phys. Rev. D \textbf{78}, 084001 (2008)
[arXiv:hep-th/0702207 [hep-th]].

\bibitem{Iguchi:2007xs}
H.~Iguchi, T.~Mishima and S.~Tomizawa,
``Boosted black holes on Kaluza-Klein bubbles,''
Phys. Rev. D \textbf{76}, 124019 (2007)
[erratum: Phys. Rev. D \textbf{78}, 109903 (2008)]
[arXiv:0705.2520 [hep-th]].




\bibitem{Morisawa:2007di}
Y.~Morisawa, S.~Tomizawa and Y.~Yasui,
``Boundary Value Problem for Black Rings,''
Phys. Rev. D \textbf{77}, 064019 (2008)
[arXiv:0710.4600 [hep-th]].





















\bibitem{Chen:2010ih}
Y.~Chen and E.~Teo,
``Black holes on gravitational instantons,''
Nucl. Phys. B \textbf{850}, 253-272 (2011)
[arXiv:1011.6464 [hep-th]].



\bibitem{Chen:2011jb}
Y.~Chen, K.~Hong and E.~Teo,
``Unbalanced Pomeransky-Sen'kov black ring,''
Phys. Rev. D \textbf{84}, 084030 (2011)
[arXiv:1108.1849 [hep-th]].



\bibitem{Iguchi:2011qi}
H.~Iguchi, K.~Izumi and T.~Mishima,
``Systematic solution-generation of five-dimensional black holes,''
Prog. Theor. Phys. Suppl. \textbf{189}, 93-125 (2011)
[arXiv:1106.0387 [gr-qc]].


\bibitem{Rocha:2011vv}
J.~V.~Rocha, M.~J.~Rodriguez and A.~Virmani,
``Inverse Scattering Construction of a Dipole Black Ring,''
JHEP \textbf{11}, 008 (2011)
[arXiv:1108.3527 [hep-th]].



\bibitem{Chen:2012kd}
Y.~Chen, K.~Hong and E.~Teo,
``A Doubly rotating black ring with dipole charge,''
JHEP \textbf{06}, 148 (2012)
[arXiv:1204.5785 [hep-th]].

\bibitem{Chen:2012zb}
Y.~Chen and E.~Teo,
``Rotating black rings on Taub-NUT,''
JHEP \textbf{06}, 068 (2012)
[arXiv:1204.3116 [hep-th]].




\bibitem{Rocha:2012vs}
J.~V.~Rocha, M.~J.~Rodriguez and O.~Varela,
``An Electrically charged doubly spinning dipole black ring,''
JHEP \textbf{12}, 121 (2012)
[arXiv:1205.0527 [hep-th]].

\bibitem{Feldman:2012vd}
A.~Feldman and A.~A.~Pomeransky,
``Charged black rings in supergravity with a single non-zero gauge field,''
JHEP \textbf{07}, 141 (2012)
[arXiv:1206.1026 [hep-th]].


\bibitem{Chen:2015iex}
Y.~Chen,
``Gravitational multisoliton solutions on flat space,''
Phys. Rev. D \textbf{93}, no.4, 044021 (2016)
[arXiv:1512.00032 [gr-qc]].







\bibitem{Lucietti:2020ltw}
J.~Lucietti and F.~Tomlinson,
``Moduli space of stationary vacuum black holes from integrability,''
Adv. Theor. Math. Phys. \textbf{26}, no.2, 371 (2022)
[arXiv:2008.12761 [gr-qc]].



\bibitem{Tomizawa:2022qyd}
S.~Tomizawa and R.~Suzuki,
``Black lens in a bubble of nothing,''
Phys. Rev. D \textbf{106}, no.12, 124029 (2022)
[arXiv:2209.11640 [hep-th]].






\bibitem{Suzuki:2023nqf}
R.~Suzuki and S.~Tomizawa,
``A Capped Black Hole in Five Dimensions,''
[arXiv:2311.11653 [hep-th]].







\bibitem{Misner:1963fr}
C.~W.~Misner,
``The Flatter regions of Newman, Unti and Tamburino's generalized Schwarzschild space,''
J. Math. Phys. \textbf{4}, 924-938 (1963)




\bibitem{Kunduri:2013vka}
H.~K.~Kunduri and J.~Lucietti,
``The first law of soliton and black hole mechanics in five dimensions,''
Class. Quant. Grav. \textbf{31}, no.3, 032001 (2014)
[arXiv:1310.4810 [hep-th]].


\bibitem{Cvetic:1996kv}
M.~Cvetic and D.~Youm,
``Entropy of nonextreme charged rotating black holes in string theory,''
Phys. Rev. D \textbf{54}, 2612-2620 (1996)
[arXiv:hep-th/9603147 [hep-th]].


\bibitem{Orlik1970}
P.~Orlik and F.~Raymond,
 ``Actions of the Torus on 4-Manifolds. I.'',
Transactions of the American Mathematical Society 152, no. 2 (1970): 531-559.



\bibitem{Tomlinson:2021wsp}
F.~Tomlinson,
``Five-dimensional electrostatic black holes in a background field,''
Class. Quant. Grav. \textbf{39}, no.13, 135008 (2022)
[arXiv:2111.14809 [gr-qc]].

\end{thebibliography}
\end{document}